\documentclass[aps,pre,twocolumn,superscriptaddress,nofootinbib,longbibliography,floatfix]{revtex4-2}

\usepackage{amsmath,amssymb,bm}
\usepackage{graphicx}
\usepackage{tikz}
\usepackage{xcolor}
\usepackage[normalem]{ulem}

\newcommand{\dd}{\mathrm{d}}
\newcommand{\ee}{\mathrm{e}}
\newcommand{\ii}{\mathrm{i}}

\newcommand{\phid}{\phi^\dagger}
\newcommand{\Fop}{\mathcal F}

\newcommand{\mean}[1]{\langle #1\rangle}

\begin{document}

\title{Branching stochastic mechanics: Relative localization and collective poles from Bohm/Fisher feedback}

\author{Beno\^it Bischoff}
\affiliation{Université Paris-Saclay, CEA\\ Institut de Recherche sur les Lois Fondamentales de l'Univers, Gif-sur-Yvette, France}
\affiliation{Universit\'e Paris-Saclay, \'Ecole Normale Supérieure Paris-Saclay,
Gif-sur-Yvette, France}

\author{Eric Dumonteil}
\email{eric.dumonteil@cea.fr}
\affiliation{Université Paris-Saclay, CEA\\ Institut de Recherche sur les Lois Fondamentales de l'Univers, Gif-sur-Yvette, France}

\begin{abstract}
Branching stochastic mechanics (BSM) provides a reciprocal branching representation of the Schr\"odinger--Nagasawa pair. Its centered forward--backward kernel $C_{\rm FB}=\mathbb E_\omega[\psi_F\psi_B]$ resolves an organized connected sector, with weight $\rho_{\rm BSM}=-C_{\rm FB}(x,x)$ on the anticorrelated branch.

Here we investigate how this sector forms, localizes, and propagates under Bohm/Fisher feedback. We retain the branching covariance on the prescribed background as the bare noise kernel and truncate the nonlinear interaction to Bohm/Fisher drift vertices. A Martin--Siggia--Rose--Janssen--de Dominicis formulation and a causal two-loop two-particle-irreducible (2PI) closure determine response and correlation functions self-consistently. While the free connected theory exhibits secular growth and ultraviolet accumulation, the dressed theory develops a finite relative screening length. A reduced numerical evolution shows bounded formation of the localized sector, and a self-similar Fisher construction defines a saturated information velocity $c_\star$.

A Born--Oppenheimer separation connects this internal organization to collective propagation. The instantaneous adiabatic kernel has two pole families at fixed internal momentum: a gapless difference branch and a gapped sum branch. Projecting both inverse response kernels onto the same localized internal profile defines their collective coefficients. If both propagation speeds match $c_\star$ and the projected gap matches $mc_\star^2/\hbar$, the gapped branch takes the infrared Klein--Gordon form. These matching conditions define a candidate relativistic fixed point whose dynamical realization remains to be tested.
\end{abstract}


\maketitle

\section{Introduction}
\label{sec:introduction}

The Schr\"odinger wave function describes a quantum state that may extend over a large region of space. How a localized structure might arise while preserving this extended quantum description motivates branching stochastic mechanics (BSM), introduced in Ref.~\cite{dumonteil_branching_2026}. In this framework, branching superprocesses reproduce Schr\"odinger dynamics at the mean-field level while retaining genealogical information about the shared ancestry of stochastic histories. The Bohm potential, related to Fisher information, contributes in this context to the local branching dynamics and thereby influences the formation of these genealogical correlations.

Here we investigate whether Bohm/Fisher feedback can generate a finite correlation length, how the resulting connected sector forms, and what collective response it supports. Throughout this work, localization refers to the relative coordinate of the connected kernel. Its interpretation as a localized measurement outcome would require an explicit detector coupling.

We recall only the elements of Ref.~\cite{dumonteil_branching_2026} that are required to formulate this
problem. Writing the Schr\"odinger field as
$\Psi_{\rm S}(x,t)=R(x,t)\exp[\ii S(x,t)/\hbar]$, the
Schr\"odinger--Nagasawa transformation~\cite{Nagasawa1993} is
\begin{equation}
\phi=R\ee^{-S/\hbar},
\qquad
\phid=R\ee^{S/\hbar},
\qquad
\rho=\phid\phi=R^2,
\label{eq:intro-SN-fields}
\end{equation}
and the two real fields obey
\begin{equation}
\partial_t\phi=\Fop[\rho]\phi,
\qquad
\partial_t\phid=-\Fop[\rho]\phid,
\label{eq:SN-pair}
\end{equation}
with
\begin{equation}
\Fop[\rho]
=
\frac{\hbar}{2m}\nabla^2
+
\frac{V+2Q[\rho]}{\hbar},
\qquad
Q[\rho]
=
-\frac{\hbar^2}{2m}
\frac{\nabla^2\sqrt{\rho}}{\sqrt{\rho}}.
\label{eq:Fop-Q}
\end{equation}
Together with the phase and boundary conditions, this pair is an exact
real-variable representation of Schr\"odinger dynamics. It belongs to the
same broad line of real and stochastic reformulations developed by Madelung,
Bohm, Nelson, and Nagasawa
~\cite{Madelung1927,Bohm1952,Nelson1966,Nagasawa1993}.

Ref.~\cite{dumonteil_branching_2026} then uses the Nagasawa pair as the deterministic backbone of a
reciprocal branching construction in the measure-valued setting
~\cite{Dawson1993,Etheridge2000}. In physicists' notation, the two marginal
branching equations are
\begin{equation}
\begin{aligned}
\partial_t\Phi_F
&=
\Fop[\rho] \,\Phi_F
+
\sqrt{2\nu_F\Phi_F}\,\xi_F,
\\
\partial_s\check\Phi_B
&=
\Fop[\rho](x,t_f-s)\,\check\Phi_B
+
\sqrt{2\nu_B\check\Phi_B}\,\check\xi_B,
\end{aligned}
\label{eq:intro-BSM-SPDE}
\end{equation}
where $t_f$ is the terminal time of the reciprocal interval,
$s=t_f-t$, and $\check\Phi_B(x,s)=\Phi_B(x,t_f-s)$. The coefficients
$\nu_F$ and $\nu_B$ fix the forward and backward branching normalizations,
while $\xi_F$ and $\check\xi_B$ are the local space--time white-noise drivers.
The stochastic lift, its normalization, the reverse-time
construction of the backward sector, and the regulated treatment required when
the Bohm/Fisher functional acts on rough or nodal stochastic realizations are
established in Ref.~\cite{dumonteil_branching_2026}.  We use that
construction here as the starting stochastic dynamics and do not rederive
those ingredients.

The quantity carried into the present work is the centered reciprocal sector.
In the common reciprocal basis of Ref.~\cite{dumonteil_branching_2026},
$\Phi_{S,F}=\ee^{S/\hbar}\Phi_F=R+\psi_F$ and
$\Phi_{S,B}=\ee^{-S/\hbar}\Phi_B=R+\psi_B$. The signed connected
forward--backward kernel and the associated positive organized connected weight on the
anticorrelated branch are
\begin{align}
C_{\rm FB}(x,y,t)
&=
\mathbb E_\omega[\psi_F(x,t)\psi_B(y,t)],
\nonumber\\
\rho_{\rm BSM}(x,t)
&=
-C_{\rm FB}(x,x,t).
\label{eq:intro-paperI-pair}
\end{align}
Here and below, $\mathbb E_\omega$ denotes expectation over
realizations of the branching-noise process.

Within the centered reference used here,
\begin{equation*}
\begin{aligned}
\underbrace{\mathbb E_\omega[\Phi_F\Phi_B]}_
  {\text{mean reciprocal product}}
&=
\underbrace{R^2}_{\text{reference}}
+
\underbrace{C_{\rm FB}(x,x,t)}_
  {\text{signed connected correction}}
\\
&=R^2-\rho_{\rm BSM}.
\end{aligned}
\end{equation*}
The nonnegative mean product, normalized by its spatial integral
when finite and strictly positive, defines a candidate probability
density, but this interpretation will not be pursued here.
By contrast, $\rho_{\rm BSM}$ is the organized connected weight:
on the anticorrelated branch, it measures the deficit of the mean
product relative to $R^2$, rather than a probability density.
The diagonal and relative dependence of the same kernel play
different roles.

The diagonal
measures the organized connected weight, with the Born matching $\rho_{\rm BSM}=R^2$ imposed as a stationary fixed-point condition,
whereas the dependence on $x-y$ resolves the internal range
of the forward--backward organization. This separation is the
structural result of Ref.~\cite{dumonteil_branching_2026} that this work develops dynamically.

The dynamical problem involves feedback. In the frozen-background construction the
Bohm/Fisher term entering the drift is evaluated on the prescribed quantum
state. Once the stochastic fields themselves are allowed to contribute to the
state-dependent Bohm term, a fluctuation modifies the operator that transports
subsequent fluctuations. The response and covariance can no longer be treated
as two independent layers. The first objective of this work is therefore to
formulate the BSM pair as a stochastic field theory in which this closed
propagator--covariance loop can be followed explicitly. We use the
Martin--Siggia--Rose--Janssen--de~Dominicis response formalism
~\cite{Martin1973,Janssen1976,DeDominicis1976} and then resum the resulting
nonequilibrium dynamics with a two-particle-irreducible effective action
~\cite{Cornwall1974,Berges2004,AartsBerges2001,Bode2022}.

A self-consistent treatment is needed because the continuously driven connected covariance accumulates secularly even though the bare response remains oscillatory. Because the Bohm/Fisher interaction carries
spatial derivatives, the corresponding perturbative corrections also become
increasingly sensitive to high internal modes. A fixed-order expansion can
therefore diagnose the breakdown of the free theory but cannot provide the
late-time connected state. To isolate this feedback mechanism, the elementary branching covariance is
evaluated on the prescribed Schr\"odinger--Nagasawa background and retained as
the bare stochastic seed, while the nonlinear 2PI interaction is restricted to
the Bohm/Fisher drift vertices.  The causal two-loop closure then feeds the
dressed response and dressed covariance back into those vertices.  The cubic
contractions separate higher-gradient, diagonal, and non-derivative response
families, while the quartic Bohm vertex adds a local Hartree contribution and the cubic vertices also dress the effective noise.  Their combined
infrared scale produces a screened component in the relative connected
spectrum.  Writing $\psi\equiv\psi_F$ and $\psi^\dagger\equiv\psi_B$, we introduce the reciprocal combinations $U\equiv\psi+\psi^\dagger$ and $W\equiv\psi^\dagger-\psi$, which are respectively density-like and phase-like at linear order. The complete
connected $U/W$ covariance matrix carries the two channel projections of the
same reciprocal information.  The physical forward--backward observable is reconstructed from the complete
covariance matrix,
$C_{\rm FB}=(C_{UU}+C_{UW}-C_{WU}-C_{WW})/4$.
At equal time the covariance symmetry
$C_{UW}(x,y)=C_{WU}(y,x)$ makes the crossed contribution antisymmetric under
exchange of the relative coordinate.  It therefore drops out of the parity-even
relative projection used for the Born--Oppenheimer profiles, without requiring
the crossed covariance blocks themselves to vanish.  The numerical 2PI
evolution retains those blocks in the self-consistent feedback.  The two
diagonal channel residues then separate an extended reference-mode contribution
from a screened relative contribution, while their common stochastic part
drops out of the even reciprocal combination.

The formation process also has a temporal component. The 2PI equations retain
 their central-time memory and determine how the instantaneous response and
 covariance dress one another.  The quantity followed kinetically is the
 physical reciprocal kernel $C_{\rm FB}$ itself, rather than its separate
 $U/W$ projections.  We use an effective Kolmogorov--Johnson--Mehl--Avrami
 construction
 ~\cite{Kolmogorov1937,JohnsonMehl1939,Avrami1939,Avrami1940,Avrami1941}
 only as a post-processing description of the bounded formation of the
 normalized forward--backward weight $f_{\rm FB}(T)$ extracted from this kernel, where $T$ denotes the central evolution time.  The $U/W$ covariances remain the internal 2PI variables entering the
 self-energies and Hartree contractions, but they are not assigned independent
 Avrami order parameters.

The Avrami exponent is then interpreted through the geometry of the evolving
 reciprocal sector.  During formation, the localized part of $C_{\rm FB}$ is
 approximately characterized by one evolving correlation length while its
 normalized reduced shape changes more slowly.  This motivates a homogeneous
 capture law for the fraction of reciprocal weight that has become organized,
 without identifying that fraction with either $C_{UU}$ or $C_{WW}$ separately.

We also formulate a self-similar Fisher argument at the level of the intrinsic Schr\"odinger--Nagasawa dynamics, before the 2PI
 Hartree dressing is applied.  At a chosen internal resolution, each admissible
 possibility carries a normalized local spatial weight $p_a$ and the associated
 Fisher score $\nabla\ln p_a$.  These scores are linked to the intrinsic Schr\"odinger--Nagasawa operator through the Fisher functional that generates its Bohm term.  The reciprocal fraction
 $f_{\rm FB}(T)$ then weights the pair-score covariance: only possibilities
 belonging to the forward--backward overlap contribute to the organized pair
 information. In this model, reciprocal saturation gives a pair-score contribution that scales quadratically with the number of internal resolution elements.  Together with the bare inverse-mass diffusion scaling
 and an invariant elementary mass--length product, this defines the
 resolution-independent information velocity $c_\star$. The Hartree correction dresses the formed relative sector and enters its localization and renewal scales without redefining $c_\star$~\cite{HallReginatto2002}.

The final part of the construction separates internal organization from
collective propagation. A Born--Oppenheimer (BO) hierarchy
~\cite{BornOppenheimer1927} treats the localized relative coordinate as the
fast sector and the center-of-mass coordinate as the slow sector. At fixed central time, retaining both frequency factors of the instantaneous adiabatic kernel reveals two pole families of the
same connected field at fixed internal momentum: a gapless difference branch
and a sum branch carrying the internal frequency of the organized relative
sector. The localized relative state has a finite internal spectral width,
so neither branch is identified with a single internal momentum component.
The collective theory is obtained by projecting the two inverse pole kernels
onto the same normalized localized internal profile.  This gives one common
Born--Oppenheimer definition of the two spatial coefficients and of the
sum-sector gap. We then ask whether the two projected propagation speeds
can match $c_\star$ and the gap can match the mass-dependent frequency
$mc_\star^2/\hbar$. These are additional fixed-point conditions, under
which the gapped collective sector takes the infrared Klein--Gordon
form~\cite{Klein1926,Gordon1926}. This conditional correspondence motivates
the relativistic interpretation examined in the discussion. The construction
thus separates what is generated dynamically from what is imposed at the final
collective fixed point: relative localization and the two fixed-$q$ pole
families follow from the connected field theory, the bounded formation is
observed in the reduced 2PI evolution and summarized by Avrami kinetics, the
self-similar Fisher model supplies the scale $c_\star$, and the final common
cone and gap matching complete the collective interpretation.

The paper is organized as follows. Section~\ref{sec:msrjd} formulates the BSM
connected sector as a stochastic field theory and defines the MSRJD action.
Section~\ref{sec:quadratic} analyzes the secular and ultraviolet breakdown of
the free connected dynamics. Section~\ref{sec:2pi} develops the causal
two-loop closure and the localizing relative sector. Section~\ref{sec:avrami}
studies the bounded reciprocal conversion and its effective Avrami kinetics,
while Sec.~\ref{sec:selfsimilar} develops the self-similar Fisher scaling and
the resolution-independent information velocity. Section~\ref{sec:global-collective-poles}
reconstructs the two collective pole families, and Sec.~\ref{sec:discussion}
separates the derived infrared results from the additional fixed-point matching
conditions.

\section{Stochastic field theory for the BSM connected sector}
\label{sec:msrjd}

\subsection{From the BSM equations to the field-theory problem}

The starting point of this section is the pair of BSM stochastic equations
written in Eq.~\eqref{eq:intro-BSM-SPDE}.  Their branching-process derivation,
normalization, and reciprocal interpretation are established in Ref.~\cite{dumonteil_branching_2026} and
we take that stochastic construction as given here.  The dynamical variables are the positive forward and backward
branching fields $\Phi_F$ and $\check\Phi_B$. Their deterministic drift operators are inherited from the
Schr\"odinger--Nagasawa pair, while their square-root terms generate the local
branching fluctuations. The relation between the marginal branching construction
and the one-sector compatibility fields is also established in Ref.~\cite{dumonteil_branching_2026}. The aim of the present section is to construct the stochastic field theory of the connected fluctuations generated by this pair, following the broader field-theoretic treatment of branching and reaction--diffusion processes \cite{Doi1976,Peliti1985}, while allowing the Bohm/Fisher operator to respond self-consistently to the fluctuating state.

For that purpose it is convenient to return the backward field to the physical
time label $t$ and express the same two BSM fields in the common reciprocal basis
introduced in Ref.~\cite{dumonteil_branching_2026}.  The reverse-time construction, its endpoint convention, and its
rewriting with a physical-time label are part of the reciprocal stochastic
problem established in Ref.~\cite{dumonteil_branching_2026}. Below we use
that representation without introducing a second stochastic model.  We collect the reciprocal fields as
\begin{equation}
\Phi_S
=
\begin{pmatrix}
\Phi_{S,F}\\
\Phi_{S,B}
\end{pmatrix}
=
R
\begin{pmatrix}
1\\1
\end{pmatrix}
+
\begin{pmatrix}
\psi\\
\psi^\dagger
\end{pmatrix},
\label{eq:BSM-reciprocal-doublet}
\end{equation}
where from now on $\psi\equiv\psi_F$ and
$\psi^\dagger\equiv\psi_B$.  The dagger is a reciprocal label for the independent real backward field and does not denote complex conjugation.  In the common reciprocal normalization of
Ref.~\cite{dumonteil_branching_2026}, $\nu_2$ denotes the branching coefficient, while the original-basis
coefficients are $\nu_F=\nu_2\ee^{-S/\hbar}$ and
$\nu_B=\nu_2\ee^{S/\hbar}$.  The phase-reweighted stochastic sources are
collected in $\eta_S=(\eta_\psi,\eta_{\psi^\dagger})^T$.

The first statistical quantity required by the field theory is the conditional
quadratic variation of these sources.  The square-root branching law gives
\begin{equation}
\left\langle
\eta_S(x,t)\eta_S^T(y,t')
\mid\Phi_S
\right\rangle
=
2\Gamma_S^{(0)}[\Phi_S]\,
\delta^{(d)}(x-y)\delta(t-t'),
\label{eq:BSM-conditional-covariance}
\end{equation}
where the reciprocal half-kernel inherited from Ref.~\cite{dumonteil_branching_2026} is
\begin{equation}
\Gamma_S^{(0)}[\Phi_S]
=
\nu_2
\begin{pmatrix}
\Phi_{S,F}&0\\
0&\Phi_{S,B}
\end{pmatrix}.
\label{eq:Gamma-multiplicative}
\end{equation}
Here the superscript $(0)$ labels the undressed BSM source before the
Bohm/Fisher feedback is resummed.  The two diagonal entries describe the
elementary forward and backward branching genealogies.  Their interaction through
the state-dependent Bohm/Fisher operator generates the reciprocal connected
sector during the evolution.

Expanding Eq.~\eqref{eq:Gamma-multiplicative} around the smooth reference gives
\begin{equation}
\Gamma_S^{(0)}[\Phi_S]
=
\underbrace{\nu_2R\,\mathbb I}_{\widehat\Gamma_S^{(0)}}
+
\nu_2
\begin{pmatrix}
\psi&0\\
0&\psi^\dagger
\end{pmatrix}.
\label{eq:Gamma-background-vertex}
\end{equation}
Here $\mathbb I$ denotes the $2\times2$ identity in reciprocal field space.
Equation~\eqref{eq:Gamma-background-vertex} is the exact local expansion of Ref.~\cite{dumonteil_branching_2026} branching covariance around the prescribed one-point background.  Its
first term defines the quadratic stochastic kernel
\begin{equation}
\widehat\Gamma_S^{(0)}=\nu_2R\,\mathbb I.
\label{eq:Gamma-hat}
\end{equation}

The present paper adopts a reduced \emph{background-noise/Bohm-feedback}
closure.  The kernel $\widehat\Gamma_S^{(0)}$ is retained as the bare stochastic
source, whereas the fluctuation-dependent second term in
Eq.~\eqref{eq:Gamma-background-vertex} is not promoted to an additional
MSRJD interaction vertex.  This is a truncation of the interaction content,
not a statement that the multiplicative correction vanishes in the complete
branching process.  Its purpose is to isolate the nonlinear mechanism studied
here: the feedback of the Bohm/Fisher drift on the connected fluctuations
generated around the Schr\"odinger--Nagasawa background.

We use this separation for three reasons.  First,
$\widehat\Gamma_S^{(0)}$ is the complete source entering the quadratic
fluctuation theory and therefore fixes the bare secular connected growth.
Second, the omitted term would first enter as an additional cubic
response-field vertex and would mix a second nonlinear mechanism --the
fluctuation dependence of the elementary branching rate-- with the Bohm/Fisher
feedback resummed here.  Third, fixing the bare
source does not freeze the covariance: the Bohm/Fisher 2PI skeletons dress the
retarded response and generate a nontrivial effective noise kernel, as derived in Sec.~\ref{sec:2pi}.  The resulting $C$ is therefore fully
self-consistent even though its primitive stochastic seed is evaluated on
$R$.  Restoring the field-dependent branching vertex provides a systematic
extension of the reduced theory.

The analytic development below keeps this reduced interaction content.  The
numerical realization extends this closure by retaining the complete
$U/W$ covariance matrix and, in addition to the Bohm--Bohm contribution to the effective noise,
includes the leading mixed Bohm--branching-noise ($BN$) correction generated by
the fluctuation-dependent branching vertex.  This numerical extension is not
fed back into the closed analytic residue formulas. The analytic residue formulas thus refer to the reduced closure.

\subsection{Bohm/Fisher action and MSRJD representation}

The deterministic part of the stochastic equations is generated by the real
Schr\"odinger--Nagasawa functional
\begin{align}
S_{\rm SN}[\phid,\phi]
&=
\int\dd t\,\dd^dx\,
\left[
\phid
\left(
\partial_t
-\frac{\hbar}{2m}\nabla^2
-\frac{V}{\hbar}
\right)
\phi
\right.
\nonumber\\
&\hspace{3.0cm}\left.
-\frac{\hbar}{4m}
\frac{(\nabla\rho)^2}{\rho}
\right].
\label{eq:SN-action-v2}
\end{align}
The nonlinear term is the Fisher functional~\cite{HallReginatto2002}
\begin{equation}
I_F[\rho]
=
\int\dd^dx\,\frac{(\nabla\rho)^2}{\rho},
\end{equation}
whose first variation gives
\begin{equation}
\frac{2Q[\rho]}{\hbar}
=
\frac{\hbar}{4m}
\frac{\delta I_F[\rho]}{\delta\rho}.
\label{eq:Fisher-variation}
\end{equation}
Variation of $S_{\rm SN}$ reproduces the two deterministic equations
\eqref{eq:SN-pair}.  In BSM these equations are supplemented by the
multiplicative martingales whose covariance was fixed above.

Introduce the physical doublet, the response doublet, and the stochastic
source,
\begin{equation}
\Phi
=
\begin{pmatrix}
\phi\\
\phid
\end{pmatrix},
\qquad
\widetilde\Phi
=
\begin{pmatrix}
\widetilde\phi\\
\widetilde\phi^\dagger
\end{pmatrix},
\qquad
\eta
=
\begin{pmatrix}
\eta_\phi\\
\eta_{\phid}
\end{pmatrix}.
\label{eq:MSRJD-doublets}
\end{equation}
The stochastic constraints take the compact form
\begin{equation}
\mathcal A[\rho]\Phi=\eta,
\qquad
\mathcal A[\rho]
=
\begin{pmatrix}
\partial_t-\Fop[\rho]&0\\
0&\partial_t+\Fop[\rho]
\end{pmatrix}.
\label{eq:MSRJD-A}
\end{equation}
Here $\rho=\phid\phi$.  Let
$\widehat\Gamma^{(0)}\equiv\Gamma^{(0)}[\bar\Phi]$ denote the branching
half-kernel evaluated on the prescribed Schr\"odinger--Nagasawa background.
After the reciprocal similarity transformation it becomes precisely
Eq.~\eqref{eq:Gamma-hat}.  The It\^o prescription is used throughout, with
$\Theta(0)=0$ for equal-time response contractions.

Within the background-noise/Bohm-feedback truncation defined above,
integrating over the local martingale source gives
\begin{equation}
S_{\rm red}[\widetilde\Phi,\Phi]
=
\int\dd t\,\dd^dx\,
\left[
\widetilde\Phi^T\mathcal A[\rho]\Phi
-
\widetilde\Phi^T\widehat\Gamma^{(0)}\widetilde\Phi
\right].
\label{eq:MSRJD-action-v2}
\end{equation}
The stochastic source is therefore bare only in the sense that its elementary
quadratic variation is evaluated on the reference background.  The drift
operator retains its full state-dependent Bohm/Fisher functional and generates
the interaction vertices resummed below.  Consequently the physical response
and connected covariance are not Gaussian or fixed: both are dressed by the
same self-consistent 2PI dynamics.

The two basic two-point objects are then defined independently.  The retarded
and advanced responses are
\begin{align}
G_{ab}^{R}(x,t;y,t')
&=
\mean{\delta\Phi_a(x,t)\widetilde\Phi_b(y,t')},
\nonumber\\
G_{ab}^{A}(x,t;y,t')
&=
G_{ba}^{R}(y,t';x,t),
\label{eq:retarded-advanced-definition}
\end{align}
with the usual causal support, while the physical connected covariance is
\begin{equation}
C_{ab}(x,t;y,t')
=
\mean{\delta\Phi_a(x,t)\delta\Phi_b(y,t')}_c.
\label{eq:physical-connected-covariance}
\end{equation}
Here $a,b\in\{\phi,\phid\}$ label the two physical Nagasawa components. After
the reciprocal change of basis the same notation will label the $U/W$ components.
We impose the centered projection
\begin{equation}
\mean{\delta\Phi}=0,
\label{eq:centered-tadpole}
\end{equation}
as the definition of the connected sector around the prescribed
Schr\"odinger--Nagasawa background.  This condition is not asserted here to be
an autonomous solution of the nonlinear first-moment equation.  Rather, it
implements the diagonal-preserving separation established for the reciprocal
construction in Ref.~\cite{dumonteil_branching_2026}. The present 2PI closure evolves the connected response
 and covariance hierarchy subject to that background/tadpole condition,
without excluding first-moment feedback in a more general interacting theory.

\subsection{Expansion around the Schr\"odinger--Nagasawa background}

We expand about
\begin{equation}
\begin{aligned}
\phi&=\bar\phi+\delta\phi,
&
\phid&=\bar\phid+\delta\phid,
\\
\bar\phi&=R\ee^{-S/\hbar},
&
\bar\phid&=R\ee^{S/\hbar}.
\end{aligned}
\label{eq:background-expansion}
\end{equation}
Removing the opposite phase weights defines the reciprocal fluctuations
\begin{equation}
\begin{aligned}
\delta\phi&=\ee^{-S/\hbar}\psi,
&
\delta\phid&=\ee^{S/\hbar}\psi^\dagger,
\\
\widetilde\phi&=\ee^{S/\hbar}\widetilde\psi,
&
\widetilde\phi^\dagger&=\ee^{-S/\hbar}\widetilde\psi^\dagger.
\end{aligned}
\label{eq:similarity-fields}
\end{equation}
With
\begin{equation}
\Psi=
\begin{pmatrix}\psi\\\psi^\dagger\end{pmatrix},
\qquad
\widetilde\Psi=
\begin{pmatrix}\widetilde\psi\\\widetilde\psi^\dagger\end{pmatrix},
\label{eq:Psi-doublets}
\end{equation}
the quadratic action is
\begin{equation}
S^{(2)}
=
\int\dd t\,\dd^dx\,
\left[
\widetilde\Psi^T\widehat{\mathcal K}_S\Psi
-
\widetilde\Psi^T\widehat\Gamma_S^{(0)}\widetilde\Psi
\right].
\label{eq:S2-psi}
\end{equation}
Using the Hamilton--Jacobi equation of the mean background,
\begin{equation}
\widehat{\mathcal K}_S
=
\begin{pmatrix}
\mathcal D_t&\mathcal B_0\\
-\mathcal B_0&\mathcal D_t
\end{pmatrix},
\label{eq:KS-hat}
\end{equation}
where
\begin{align}
\mathcal D_t
&=
\partial_t
+
\frac{\nabla S}{m}\cdot\nabla
+
\frac{\nabla^2S}{2m},
\nonumber\\
\mathcal B_0
&=
\frac{\hbar}{2m}\nabla^2
+
\frac{Q_0}{\hbar},
\qquad Q_0=Q[R^2].
\label{eq:Dt-B0}
\end{align}
The transformed source has the quadratic covariance
\begin{equation}
\mean{\eta_S(x,t)\eta_S^T(y,t')}
=
2\widehat\Gamma_S^{(0)}
\delta^{(d)}(x-y)\delta(t-t').
\label{eq:transformed-noise-covariance}
\end{equation}

For the 2PI construction we collect the physical and response fields in
\begin{equation}
\Phi^T
=
(\psi,\psi^\dagger,\widetilde\psi,\widetilde\psi^\dagger)
\label{eq:four-component-field}
\end{equation}
and write
\begin{equation}
S^{(2)}
=
\int\dd t\,\dd^dx\,
\Phi^T\mathcal K_{2PI}\Phi,
\end{equation}
with
\begin{equation}
\mathcal K_{2PI}
=
\frac{1}{2}
\begin{pmatrix}
0&\widehat{\mathcal K}_S^T\\
\widehat{\mathcal K}_S&-2\widehat\Gamma_S^{(0)}
\end{pmatrix}.
\label{eq:K2PI-matrix}
\end{equation}
The normalization of this symmetrized form and the related normalization used
below in the $U/W$ basis are summarized in
Appendix~\ref{app:UW-normalization}.

\subsection{Density-like and reciprocal phase-like channels}

The combinations
\begin{equation}
\begin{aligned}
U&=\psi+\psi^\dagger,
&
W&=\psi^\dagger-\psi,
\\
\widetilde U&=\widetilde\psi+\widetilde\psi^\dagger,
&
\widetilde W&=\widetilde\psi^\dagger-\widetilde\psi
\end{aligned}
\label{eq:UW-fields}
\end{equation}
separate a density-like channel from the reciprocal phase-like channel.  We
use
\begin{equation}
h_0\equiv\frac{\hbar}{4m}
\label{eq:h0-definition}
\end{equation}
in this reduced basis. Appendix~\ref{app:UW-normalization} gives the
normalization derivation and the associated Wigner-frequency convention.

The two reduced channels can be identified directly by linearizing the
Schr\"odinger--Nagasawa fields. For \(R\rightarrow R+\delta R\) and
\(S\rightarrow S+\delta S\),
\begin{align}
\psi&=\delta R-\frac{R}{\hbar}\delta S,
&
\psi^\dagger&=\delta R+\frac{R}{\hbar}\delta S,
\end{align}
so that
\begin{equation}
U=2\delta R,
\qquad
W=\frac{2R}{\hbar}\delta S.
\end{equation}
Thus \(U\) is the linear amplitude fluctuation and directly generates the
physical density fluctuation, \(\delta\rho^{(1)}=RU\), while \(W\) is the
conjugate phase fluctuation. The static Bohm/Fisher restoring kernel is generated
by variations of \(Q[\rho]\) and acts directly on density-shape fluctuations, whereas a uniform phase variation leaves \(\rho\) and \(Q[\rho]\) unchanged.

Although $U$ is the density-like fluctuation, the relative organization is
fixed by the residue structure of the dressed covariance rather than by this
linear identification alone.  The $UU$ and $WW$ projections share the same
response poles but carry different residues.  In the leading reduction below,
the non-common $UU$ residue retains the extended reference-mode structure,
whereas the non-common $WW$ residue contains the screened contribution.  The
physical reciprocal kernel is reconstructed from their combination through
Eq.~\eqref{eq:FB-overlap-UW-even}.

Let $Z=(x,t)$ and $Z'=(y,t')$ denote space--time points, and let
$X^T=(U,W)$ denote the reciprocal physical-field doublet.  The connected
correlator matrix is
\begin{equation}
C_{ij}(Z,Z')
=
\mean{X_i(Z)X_j(Z')}_c,
\qquad i,j\in\{U,W\},
\label{eq:UW-connected-correlators}
\end{equation}
with
\begin{equation}
C_{UU}(Z,Z')=\mean{U(Z)U(Z')}_c.
\label{eq:CUU-definition-msrjd}
\end{equation}
At linear order the physical density covariance is therefore
\begin{equation}
C_{\rho\rho}^{(1)}(Z,Z')
=
R(Z)R(Z')C_{UU}(Z,Z'),
\label{eq:Crhorho-CUU}
\end{equation}
where the complete covariance
\begin{equation}
C_{\rho\rho}(Z,Z')
=
\mean{\delta\rho(Z)\delta\rho(Z')}_c
\label{eq:Crhorho-definition}
\end{equation}
also contains the composite terms generated by $\psi^\dagger\psi$.

The forward--backward connected kernel of Ref.~\cite{dumonteil_branching_2026} is reconstructed from the
same matrix:
\begin{align}
\mathcal C_{\rm FB}(Z,Z')
&\equiv
\mean{\psi(Z)\psi^\dagger(Z')}_c
\nonumber\\
&=
\frac{1}{4}
\left[
C_{UU}+C_{UW}-C_{WU}-C_{WW}
\right](Z,Z').
\label{eq:FB-overlap-UW}
\end{align}
Its equal-time restriction is the $C_{\rm FB}$ of
Eq.~\eqref{eq:intro-paperI-pair}.  At equal time,
$C_{UW}(x,y)=C_{WU}(y,x)$.  Hence the crossed term
$C_{UW}-C_{WU}$ is odd under $r=x-y\rightarrow-r$, whereas the relative
profiles and cosine projections used below select the parity-even component.
For that even reciprocal sector,
\begin{equation}
\mathcal C_{\rm FB}^{(+)}
=
\frac{1}{4}\left(C_{UU}^{(+)}-C_{WW}^{(+)}\right),
\label{eq:FB-overlap-UW-even}
\end{equation}
where $X=(x+y)/2$ is the collective spatial coordinate and $F^{(+)}(X,r)=[F(X,r)+F(X,-r)]/2$.  This projection does not set
$C_{UW}$ or $C_{WU}$ to zero in the interacting dynamics.  In the analytic
Born--Oppenheimer and collective-pole sections below, $C_{\rm FB}$ denotes
this parity-even component unless the complete matrix combination is written
explicitly.

The bare reciprocal covariance of Eq.~\eqref{eq:Gamma-hat} becomes diagonal in
the $U/W$ channels.  With
$\eta_U=\eta_\psi+\eta_{\psi^\dagger}$ and
$\eta_W=\eta_{\psi^\dagger}-\eta_\psi$,
\begin{equation}
\mean{\eta_U\eta_W}=0,
\end{equation}
and
\begin{align}
\mean{\eta_U(x,t)\eta_U(y,t')}
&=
\mean{\eta_W(x,t)\eta_W(y,t')}
\nonumber\\
&=
4\nu_2R(x,t)\,
\delta^{(d)}(x-y)\delta(t-t').
\label{eq:bare-UW-noise}
\end{align}
The interacting theory dresses this source together with the response.  We
write the resulting effective noise kernel as
\begin{equation}
\mathcal N
=
\widehat\Gamma_S^{(0)}+\Sigma_C,
\label{eq:dressed-noise-kernel}
\end{equation}
where $\Sigma_C$ is the correlation self-energy introduced by the 2PI
construction.  Equations~\eqref{eq:K2PI-matrix} and
\eqref{eq:dressed-noise-kernel} define the propagator--covariance problem that
will be solved self-consistently below. We denote by $N_{ij}$ with $i,j\in\{U,W\}$ the coefficients of $\mathcal{N}$.

\section{Quadratic theory, secular growth, and the nonequilibrium problem}
\label{sec:quadratic}

\subsection{Stationary modal kernel in the infinite-well benchmark}

Consider a confined stationary mode $u_n$ satisfying
\begin{equation}
-\nabla^2u_n=k_n^2u_n.
\label{eq:stationary-mode}
\end{equation}
The stochastic construction is understood on the regular part of each nodal domain, where the Schr\"odinger--Nagasawa weight has fixed sign. We therefore take
\begin{equation}
R(x)=R_n(x)=|u_n(x)|
\end{equation}
locally on that domain (equivalently, one may choose the positive representative of $u_n$ there). The modal formulas below remain valid for generic $n$, but the nodes themselves require the phase/boundary matching and regularization discussed in Ref.~\cite{dumonteil_branching_2026} and Appendix~\ref{app:operators}.
For an absorbing one-dimensional well $[-L,L]$, we use the
orthonormal basis
\begin{equation}
u_p(x)
=
\frac{1}{\sqrt L}
\sin\left[
\frac{p\pi}{2}
\left(
1+\frac{x}{L}
\right)
\right]
\label{eq:well-basis-v2}
\end{equation}
with
\begin{equation}
\int_{-L}^{L}\dd x\,
u_p(x)u_q(x)
=
\delta_{pq}.
\label{eq:well-orthogonality}
\end{equation}

Writing
\begin{equation}
\frac{\hbar}{2m}\nabla^2u_p
=
-\Omega_pu_p
\end{equation}
the transformed free response kernel is diagonal in the spatial mode
index,
\begin{equation}
\widehat{\mathcal K}_{n,p}
=
\begin{pmatrix}
\partial_t&-\Lambda_{n,p}\\
\Lambda_{n,p}&\partial_t
\end{pmatrix}
\qquad
\Lambda_{n,p}
=
\Omega_p-\Omega_n.
\label{eq:modal-kernel}
\end{equation}
Its retarded propagator is
\begin{equation}
G^R_{n,p}(t,t')
=
\Theta(\tau)
\begin{pmatrix}
\cos(\Lambda_{n,p}\tau)&
\sin(\Lambda_{n,p}\tau)
\\
-\sin(\Lambda_{n,p}\tau)&
\cos(\Lambda_{n,p}\tau)
\end{pmatrix}
\label{eq:modal-retarded}
\end{equation}
with
\begin{equation}
\tau=t-t'.
\end{equation}
The free deterministic sector therefore consists entirely of bounded
oscillatory blocks, with the resonant case $\Lambda_{n,p}=0$ reducing
to the identity response.

The stochastic sector inherits the background weight of Ref.~\cite{dumonteil_branching_2026} branching source.  From Sec.~\ref{sec:msrjd},
\begin{equation}
\widehat\Gamma_S^{(0)}(x)
=
\nu_2R(x)\mathbb I.
\label{eq:free-reduced-noise}
\end{equation}
Projecting the source onto the well modes,
\begin{equation}
\eta_p(t)
=
\int_{-L}^{L}\dd x\,
u_p(x)\eta_S(x,t)
\end{equation}
introduces the real symmetric overlap matrix
\begin{equation}
\mathcal R_{pq}^{(n)}
\equiv
\int_{-L}^{L}\dd x\,
u_p(x)R(x)u_q(x).
\label{eq:noise-overlap-matrix}
\end{equation}
The modal covariance is therefore
\begin{equation}
\mean{
\eta_p(t)
\eta_q^T(t')
}
=
2\nu_2\,
\mathcal R_{pq}^{(n)}
\mathbb I\,
\delta(t-t').
\label{eq:modal-noise-covariance}
\end{equation}
Unlike a spatially constant source, the background-weighted
kernel is not diagonal in the well basis.  This does not alter the
secular conclusion.  For vanishing initial fluctuations,
\begin{equation}
\Psi_p(t)
=
\int_0^t\dd s\,
G^R_{n,p}(t,s)\eta_p(s)
\label{eq:free-modal-solution}
\end{equation}
and hence
\begin{align}
\mean{
\Psi_p(t)\Psi_q^T(t')
}_0
&=
2\nu_2\mathcal R_{pq}^{(n)}
\int_0^{\min(t,t')}
\dd s\,
\nonumber\\
&\qquad\times
G^R_{n,p}(t,s)
G^R_{n,q}(t',s)^T.
\label{eq:free-covariance}
\end{align}
The deterministic background still enters through the modal
frequencies \(\Lambda_{n,p}\), while its amplitude enters only through
the bounded overlap matrix \(\mathcal R_{pq}^{(n)}\).

\subsection{Exact free secular growth}

For a diagonal modal block \(p=q\), the free response is an orthogonal
rotation,
\begin{equation}
G^R_{n,p}(t,s)
G^R_{n,p}(t,s)^T
=
\mathbb I.
\label{eq:rotation-identity}
\end{equation}
Equation~\eqref{eq:free-covariance} therefore gives the exact secular
part
\begin{equation}
\mean{
\Psi_p(t)\Psi_p^T(t)
}_0
=
2\nu_2
\mathcal R_{pp}^{(n)}
t\,
\mathbb I.
\label{eq:free-equal-time-covariance}
\end{equation}
For \(p\neq q\), the integrand contains the relative rotation with
frequency \(\Lambda_{n,p}-\Lambda_{n,q}\). Away from accidental degeneracies, its contribution is bounded or oscillatory rather than
linear in time.  The background-weighted source therefore changes the modal
weights, but not the hierarchy between secular diagonal blocks and
nonsecular off-diagonal blocks.

For the secular diagonal pieces,
\begin{align}
C_{\psi\psi}^{pp}(t)
&=
2\nu_2
\mathcal R_{pp}^{(n)}
t,
\nonumber\\
C_{\psi^\dagger\psi^\dagger}^{pp}(t)
&=
2\nu_2
\mathcal R_{pp}^{(n)}
t,
\nonumber\\
C_{\psi\psi^\dagger}^{pp}(t)
&=
C_{\psi^\dagger\psi}^{pp}(t)
=
0.
\label{eq:free-psi-correlators}
\end{align}
Passing to
\(U=\psi+\psi^\dagger\) and
\(W=\psi^\dagger-\psi\) gives
\begin{equation}
\begin{aligned}
C_{UU}^{pp}(t)
&=
C_{WW}^{pp}(t)
=
4\nu_2\mathcal R_{pp}^{(n)}t,
\\
C_{UW}^{pp}(t)
&=
C_{WU}^{pp}(t)
=
0.
\end{aligned}
\label{eq:free-UW-secular}
\end{equation}
The linear growth comes from the secular accumulation of continuously driven connected covariance, while the free response remains bounded.  The factor \(R\) required by the underlying branching
law modifies the source residues but leaves the existence of the
secular growth unchanged.

These are correlators of the reduced fluctuation fields. The physical
density fluctuation is instead, at linear order,
\begin{equation}
\delta\rho^{(1)}(x,t)
=
R_n(x)U(x,t)
\end{equation}
and therefore
\begin{equation}
C_{\rho\rho}^{(1)}(x,y;t,t')
=
R_n(x)R_n(y)
C_{UU}(x,y;t,t').
\label{eq:physical-density-covariance-well}
\end{equation}
The physical observable carries an additional factor of the background
through Eq.~\eqref{eq:physical-density-covariance-well}, while the bare connected source already contains the single factor of \(R\)
required by the multiplicative genealogy.

The quadratic theory therefore exhibits the following nonequilibrium behavior: the deterministic response remains bounded, while the
connected covariance grows without bound,
\begin{equation}
C^{(0)}(t)\propto\nu_2t.
\label{eq:free-secular-scaling}
\end{equation}
A fixed-order treatment of the Bohm/Fisher feedback can therefore
only remain valid for finite times.

\subsection{Secular breakdown and ultraviolet accumulation}

The free theory already exhibits a linear secular growth of the
connected covariance,
\begin{equation}
C^{(0)}(t)\propto \nu_2 t.
\label{eq:free-secular-summary}
\end{equation}
The perturbative Bohm/Fisher correction therefore inherits an explicit
time dependence when this covariance is inserted into the first
self-energy diagrams.

A second effect follows from the differential structure of the
Bohm/Fisher interaction. In the infinite-well basis,
\begin{equation}
\nabla^2u_q=-k_q^2u_q
\end{equation}
so that high internal modes are weighted quadratically,
\begin{equation}
\mathcal V_{pq}
\propto q^2
\qquad
q\rightarrow\infty,
\label{eq:Bohm-vertex-UV}
\end{equation}
where $\mathcal V_{pq}$ denotes the corresponding projected
Bohm/Fisher vertex. The explicit modal projection is given in
Appendix~\ref{app:secular}.

 Denoting by $\Pi_p^{(1)}$ the first fixed-order correction to the projected
response kernel in well mode $p$, and by $N_{\rm UV}$ the retained modal
ultraviolet cutoff, combining the secular covariance with this derivative
enhancement gives the asymptotic perturbative behavior
\begin{equation}
\Pi_p^{(1)}(t;N_{\rm UV})
=
\mathcal O
\left(
\nu_2\,p\,N_{\rm UV}^3\,t
\right).
\label{eq:secular-UV-scaling}
\end{equation}
The precise prefactor depends on the channel and on the finite modal
projection, but the two relevant scalings are universal within the
present benchmark:
\begin{equation}
\Pi^{(1)}\propto t,
\qquad
\Pi^{(1)}\propto N_{\rm UV}^3.
\end{equation}

The first scaling reflects the continuous stochastic accumulation of
connected fluctuations. The second reflects the short-distance
enhancement carried by the spatial derivatives of the Bohm/Fisher
vertex.

The fixed-order expansion is therefore only meaningful while
\begin{equation}
\left\|
\Pi_p^{(1)}
\right\|
\ll
\left\|
\widehat{\mathcal K}_{n,p}
\right\|.
\end{equation}
Since the perturbative correction grows secularly, this condition
cannot remain valid at arbitrarily long times. The calculation should
therefore be interpreted as a diagnosis of the breakdown of the bare
expansion, rather than as a prediction of the dynamics beyond that
regime.

The ultraviolet dependence is kept explicit through a fixed physical
modal cutoff $N_{\rm UV}$. No continuum limit is assumed here.

The quadratic theory thus identifies the two limitations that motivate
the self-consistent treatment developed below: secular late-time growth
and ultraviolet modal accumulation. To address both effects, we use a nonequilibrium resummation in which the full response functions and full
connected correlators re-enter the loops that dress them.

The same free calculation also fixes the channel ordering used in the
analytic interacting reduction.  Equation~\eqref{eq:free-UW-secular} shows
that the secular equal-time contribution is carried by the diagonal $UU$ and
$WW$ blocks, whereas the crossed modal blocks have no secular diagonal term. Away from modal degeneracies, their free contributions are bounded and
oscillatory.  Moreover, the bare $U/W$ noise of
Eq.~\eqref{eq:bare-UW-noise} is exactly diagonal.  We therefore use, as the
leading channel hierarchy of the reduced analytic closure,
\begin{equation}
\begin{aligned}
|C_{UW}|,\ |C_{WU}|&\ll |C_{UU}|,\ |C_{WW}|,\\
|N_{UW}|,\ |N_{WU}|&\ll |N_{UU}|,\ |N_{WW}|.
\end{aligned}
\label{eq:secular-channel-hierarchy}
\end{equation}
This hierarchy organizes the analytic residue expansion without setting the crossed blocks to zero in the nonlinear numerical evolution.

\section{Two-loop 2PI resummation and the self-consistent overlap mass}
\label{sec:2pi}
\subsection{Effective action and causal Bohm/Fisher skeletons}

The quadratic kernel in Eq.~\eqref{eq:K2PI-matrix} defines the bare
four-component MSRJD theory. In the density--phase basis, we collect the
physical and response fields as
\begin{equation}
X=
\begin{pmatrix}U\\W\end{pmatrix},
\qquad
\widetilde X=
\begin{pmatrix}\widetilde U\\\widetilde W\end{pmatrix}.
\label{eq:2PI-UW-doublets}
\end{equation}
We keep the response--response block off shell during the functional variation and write
\begin{equation}
\mathbb G=
\begin{pmatrix}
C&G\\
G^T&H
\end{pmatrix}
\label{eq:full-MSRJD-propagator}
\end{equation}
where
\begin{align}
G_{ab}(1,2)&=\mean{X_a(1)\widetilde X_b(2)},
\label{eq:2PI-response-definition}\\
C_{ab}(1,2)&=\mean{X_a(1)X_b(2)}_c.
\label{eq:2PI-correlation-definition}
\end{align}
The physical MSRJD solution is obtained only after the variation by setting
\begin{equation}
H_{ab}(1,2)=\mean{\widetilde X_a(1)\widetilde X_b(2)}=0.
\label{eq:no-response-response-contraction}
\end{equation}
Here $GGC$ denotes a contraction with two response lines $G$ and one
covariance line $C$, while $HCC$ contains one auxiliary
response--response line $H$ and two covariance lines. Keeping $H$ during
the variation allows its line in an $HCC$ contraction to be opened,
producing a correction to the physical noise kernel.

The 2PI effective action~\cite{Cornwall1974,Berges2004,AartsBerges2001,Bode2022} is
used here in the classical MSR setting developed for nonequilibrium stochastic
processes by Bode~\cite{Bode2022}, with the interaction specialized to the
Bohm/Fisher vertices of BSM.
\begin{equation}
\Gamma[\mathbb G]
=
\frac12\operatorname{Tr}\ln\mathbb G^{-1}
+
\frac12\operatorname{Tr}(\mathbb G_0^{-1}\mathbb G)
+
\Gamma_2[\mathbb G]
+
\text{const.}
\label{eq:2PI-action}
\end{equation}
and stationarity gives
\begin{equation}
\mathbb G^{-1}=\mathbb G_0^{-1}+\Sigma[\mathbb G],
\qquad
\Sigma=2\frac{\delta\Gamma_2}{\delta\mathbb G}.
\label{eq:2PI-Dyson}
\end{equation}

For the closed analytic formulas below we now implement the hierarchy
motivated by the secular analysis, Eq.~\eqref{eq:secular-channel-hierarchy}:
the leading covariance and effective-noise residues are taken from the
diagonal $UU$ and $WW$ sectors, while crossed terms are retained only where
they are required by the response structure.  This fixes the reduced analytic
closure used to identify the cubic response coefficients and the leading Hartree coefficients defined below.  The numerical solver does not impose this projection and evolves
all four covariance blocks.

In this 2PI closure, the background branching kernel
$\widehat\Gamma_S^{(0)}=\nu_2R\mathbb I$ belongs to the quadratic part of
$\mathbb G_0^{-1}$ and provides the primitive stochastic seed.  The interaction
functional $\Gamma_2$ is built only from the cubic and quartic Bohm/Fisher drift
vertices.  Thus the fluctuation-dependent branching term displayed in
Eq.~\eqref{eq:Gamma-background-vertex}, which would generate an additional
$\widetilde X\widetilde X X$ vertex in the complete multiplicative theory, is
outside the present truncation.

The vertex is omitted as a truncation choice, not canceled by causality. This closure tests whether Bohm/Fisher feedback alone can reorganize the continuously generated branching covariance into a bounded, localized reciprocal sector.  The
truncation remains dynamically self-consistent because the Bohm cubic vertex
dresses the retarded response through the $GGC$ topology and dresses the
effective noise through the off-shell $HCC$ topology, while the quartic Bohm
vertex supplies the Hartree response correction.  Hence both $G$ and $C$
re-enter the loops that determine their subsequent evolution.  The bare source
is fixed on $R$, but the effective noise and the connected covariance are not.

The numerical closure retains the complete $2\times2$ physical covariance in
this Bohm/Fisher feedback, including the crossed $UW$ and $WU$ blocks, together
with the distinct crossed response functions.  In the analytic formulas below
we display the diagonal $U/W$ projection that identifies the coefficient
families entering the even reciprocal residue.  This is a projection of the
formulas, not a vanishing condition on the crossed covariance blocks of the
evolved theory.  The off-shell $HCC$ variation of the Bohm vertices is retained
and contributes to the dressed noise kernel before the physical condition
$H=0$ is imposed.
The reduced analytic formulas do not include the multiplicative-noise
vertex.  In the numerical closure, however, its leading mixed $BN$ contribution
to the noise self-energy is retained together with the complete crossed
covariance sector, while higher multiplicative-noise skeletons remain outside the
present truncation.

\paragraph{Cubic Bohm vertex and two-loop corrections}
The spatial structure of the cubic drift vertex is conveniently written with
\begin{align}
\mathcal A_R(\chi)
&=
\nabla^2\!\left(\frac{\chi}{R}\right)
+2\nabla\ln R\cdot\nabla\!\left(\frac{\chi}{R}\right),
\label{eq:2PI-AR}\\
\mathcal D_R(\chi,\zeta)
&=
\nabla^2\!\left(\frac{\chi\zeta}{R^2}\right)
+2\nabla\ln R\cdot\nabla\!\left(\frac{\chi\zeta}{R^2}\right).
\label{eq:2PI-DR}
\end{align}
With
\begin{equation}
\kappa=\frac{\hbar}{8m}=\frac{h_0}{2},
\label{eq:2PI-kappa}
\end{equation}
the cubic interaction in the response normalization used here is
\begin{equation}
\begin{aligned}
S_B^{(3)}
&=-2\kappa\int\dd t\,\dd^dx\;
\widetilde U\,W\,\mathcal A_R(U)
\\
&\quad
+\kappa\int\dd t\,\dd^dx\;
R\,\widetilde W\,\mathcal D_R(W,W).
\end{aligned}
\label{eq:Bohm-cubic-UW}
\end{equation}
Writing
\begin{align}
V_U(1)&=-2\kappa\,\widetilde U(1)W(1)\mathcal A_1U(1),
\label{eq:2PI-VU}\\
V_W(1)&=\kappa\,R(1)\widetilde W(1)\mathcal D_1[W(1),W(1)],
\label{eq:2PI-VW}
\end{align}
the two cubic vertices generate the response $GGC$ topology and, off shell,
the complementary $HCC$ topology.

\begin{figure}[!htbp]
\centering
\begin{tikzpicture}[scale=.92,transform shape,every node/.style={font=\small}]
\node[circle,fill=black,inner sep=1.5pt] (a1) at (-3.6,0) {};
\node[circle,fill=black,inner sep=1.5pt] (a2) at (-1.0,0) {};
\draw[thick,->] (a1) .. controls (-2.9,.75) and (-1.7,.75) .. (a2);
\draw[thick,->] (a2) .. controls (-1.7,-.75) and (-2.9,-.75) .. (a1);
\draw[thick] (a1) -- (a2);
\node at (-2.3,1.05) {$GGC$};
\node[circle,fill=black,inner sep=1.5pt] (b1) at (1.0,0) {};
\node[circle,fill=black,inner sep=1.5pt] (b2) at (3.6,0) {};
\draw[thick,dashed] (b1) .. controls (1.7,.85) and (2.9,.85) .. (b2);
\draw[thick] (b1) .. controls (1.7,.25) and (2.9,.25) .. (b2);
\draw[thick] (b1) .. controls (1.7,-.45) and (2.9,-.45) .. (b2);
\node at (2.3,1.05) {$HCC$};
\end{tikzpicture}
\caption{Two-cubic Bohm/Fisher contributions.  Opening a response line in
the $GGC$ topology dresses the retarded kernel.  Opening the auxiliary
$H$ line in the off-shell $HCC$ topology dresses the effective noise.
The physical condition $H=0$ is imposed only after these functional
variations.}
\label{fig:two-vertex-diagram}
\end{figure}
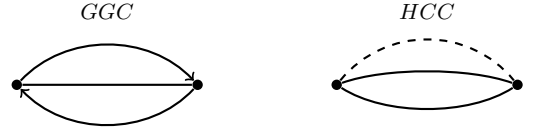

We denote the momenta conjugate to the relative coordinate $r=x-y$
and the collective coordinate $X=(x+y)/2$ by $q$ and $Q$, respectively.
For the two leg times $t$ and $t'$, the central time is $T=(t+t')/2$.
The reduced relative frequency is $\nu$, and $\Omega$ is conjugate to $T$.
Their leg assignments and normalization are given in
Sec.~\ref{sec:relative-localization} and Appendix~\ref{app:UW-normalization}.
At $Q=0$ let
\begin{equation}
s(q)=k_n^2-q^2.
\label{eq:s-relative-2pi}
\end{equation}
The nonlocal memory kernels are defined before the adiabatic projection,
which is applied only afterwards,
\begin{equation}
\Sigma^{\rm ad}(q,T)
=
\Sigma(q,Q=0;\nu=0,\Omega=0;T).
\label{eq:adiabatic-self-energy-projection}
\end{equation}
Under the instantaneous exchange reduction $E_U=E_W\equiv E$, opening the
$GGC$ topology gives the following three response families,
\begin{equation}
\Sigma_{3}^{R,\rm ad}(q,T)=
\begin{pmatrix}
E(q,T)s(q)&\mu_{2PI}(q,T)\\
D(q,T)s^2(q)&E(q,T)s(q)
\end{pmatrix}.
\label{eq:2PI-cubic-reduced-selfenergy}
\end{equation}
Here $D$ carries two external derivative factors, $E$ one, and
$\mu_{2PI}$ none.  Their explicit modal and Fourier loop expressions,
with the derivatives kept on the appropriate legs, are given in
Appendix~\ref{app:Sigma}.

The $HCC$ completion can be expressed using the two vertices $V_U$ and $V_W$.  Under the same diagonal-sector analytic projection
used for the residues, pairing two $V_U$ vertices and opening the auxiliary
$H_{UU}$ line leaves one $UU$ and one $WW$ covariance, while pairing two
$V_W$ vertices and opening $H_{WW}$ leaves two $WW$ covariances.  In a locally
uniform patch, with $g=h_0/R$ and $r=q-p$ denoting a loop momentum here rather than a spatial separation, this projection gives
\begin{align}
\delta N_{UU}^{BB}(q;t,t')
&=
g^2\int_p s^2(p)\,
C_{UU}(p;t,t')C_{WW}(r;t,t'),
\nonumber\\
\delta N_{WW}^{BB}(q;t,t')
&=
\frac{g^2s^2(q)}{2}
\int_p
C_{WW}(p;t,t')C_{WW}(r;t,t').
\label{eq:HCC-noise-main}
\end{align}
Mixed $V_UV_W$ pairings generate the crossed $UW$ and $WU$ noise blocks.
The full numerical $HCC$ contraction also retains the crossed covariance
contractions inside each block, although only the diagonal-sector expressions above are displayed for the analytic residue decomposition.  The derivation and modal
form are summarized in Appendix~\ref{app:Sigma}.
Thus the same cubic Bohm interaction dresses the retarded kernel through
$GGC$ and the effective noise through its off-shell $HCC$ completion.

\paragraph{Quartic Bohm vertex and Hartree correction}
The same Bohm/Fisher expansion contains a quartic drift vertex.  Introducing
\begin{equation}
u=\frac{U}{2R},\qquad w=\frac{W}{2R},
\label{eq:2PI-uw-relative}
\end{equation}
and
\begin{equation}
\mathcal L_R=\nabla^2+2\nabla\ln R\cdot\nabla,
\label{eq:2PI-LR}
\end{equation}
its complete form is
\begin{equation}
\begin{aligned}
S_B^{(4)}
=2h_0\int\dd t\,\dd^dx\,R\Big[
&\widetilde U\{w\mathcal L_R(w^2)+2uw\mathcal L_Ru\}
\\
-&\widetilde W\{\mathcal L_R(uw^2)+w^2\mathcal L_Ru\}
\Big].
\end{aligned}
\label{eq:Bohm-quartic-UW}
\end{equation}
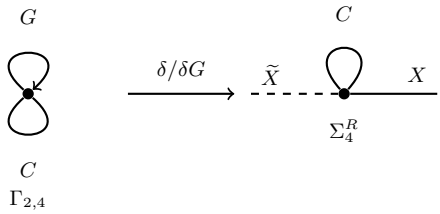
\begin{figure}[!htbp]
\centering
\begin{tikzpicture}[scale=.88,transform shape,every node/.style={font=\small}]
\node[circle,fill=black,inner sep=1.7pt] (v1) at (-2.4,0) {};
\draw[thick,->] (v1) .. controls (-3.3,.8) and (-1.5,.8) .. (v1);
\draw[thick] (v1) .. controls (-3.3,-.8) and (-1.5,-.8) .. (v1);
\node at (-2.4,1.15) {$G$};
\node at (-2.4,-1.15) {$C$};
\node at (-2.4,-1.6) {$\Gamma_{2,4}$};
\draw[->,thick] (-.9,0) -- (.7,0);
\node at (-.1,.35) {$\delta/\delta G$};
\node[circle,fill=black,inner sep=1.7pt] (v2) at (2.35,0) {};
\draw[thick,dashed] (.95,0) -- (v2);
\draw[thick] (v2) -- (3.75,0);
\draw[thick] (v2) .. controls (1.55,.9) and (3.15,.9) .. (v2);
\node at (1.25,.28) {$\widetilde X$};
\node at (3.45,.28) {$X$};
\node at (2.35,1.2) {$C$};
\node at (2.35,-.55) {$\Sigma_4^R$};
\end{tikzpicture}
\caption{Quartic Bohm/Fisher contribution.  Left: the one-vertex off-shell
2PI term contains one response contraction and one physical covariance
contraction.  Functional differentiation opens the response contraction.
Right: the resulting Hartree response self-energy leaves one equal-time
$C$ loop and one external response--physical pair.}
\label{fig:quartic-hartree-diagram}
\end{figure}

For the local homogeneous reduction, the dominant diagonal covariance
moments give
\begin{align}
&a_4(T)=2h_0\langle w^2\rangle_c,
\nonumber\\
&M_4(T)=2h_0\left[
\langle|\nabla u|^2\rangle_c+
\langle|\nabla w|^2\rangle_c
\right]\ge0 .
\label{eq:a4-M4-main}
\end{align}
The same Hartree contraction also generates a traceless diagonal term
$d_4\propto\langle uw\rangle_c,\langle\nabla u\cdot\nabla w\rangle_c$.
Because it is built entirely from crossed $UW$ correlators, which remain
subleading with respect to the diagonal $UU$ and $WW$ covariances in the
localized regime, this contribution is neglected in the analytic reduction
below.  It is retained in the complete numerical 2PI evolution, and its explicit form is given in Appendix~\ref{app:quartic}.
In a uniform patch this mass-like term is equivalently
\begin{align}
M_4(T)&=
\frac{h_0}{2R^2}
\int\frac{\dd^dp}{(2\pi)^d}\,p^2
\nonumber\\
&\quad\times\left[
C_{UU}(p;T,T)+C_{WW}(p;T,T)
\right].
\label{eq:M4-spectral-main}
\end{align}
The quartic Hartree correction to the response kernel is therefore
\begin{equation}
\delta\mathcal K_4(q,T)=
\begin{pmatrix}
0&a_4(T)s(q)-M_4(T)\\
-a_4(T)s(q)&0
\end{pmatrix}.
\label{eq:quartic-Hartree-reduction}
\end{equation}
The quartic vertex renormalizes the relative stiffness and contributes the
positive non-derivative term $M_4$ to the infrared response.  For the normalized
field $w=W/(2R)$ we define the dimensionless correlation dressing
\begin{equation}
\zeta(T)
\equiv
1+2\langle w^2\rangle_c
=
1+2C_{ww}(r=0,T).
\label{eq:zeta-correlation}
\end{equation}
Since $a_4=2h_0\langle w^2\rangle_c$, the dressed stiffness is
\begin{equation}
h_{\rm eff}(T)
=
h_0+a_4(T)
=
h_0\zeta(T)
=
\frac{\hbar}{4m}\zeta(T).
\label{eq:heff-correlation-scaling}
\end{equation}
The Hartree dressing therefore preserves the inverse-mass scaling of the
Schr\"odinger--Nagasawa coefficient while renormalizing its dimensionless
amplitude through the local equal-time connected $ww$ covariance.  Equations~\eqref{eq:a4-M4-main}--\eqref{eq:heff-correlation-scaling}
define the reduced analytic Hartree contraction. The numerical realization evaluates the same equal-time Hartree moments from the full covariance, with the physical correlation cutoff specified in Appendix~\ref{app:numerical-2pi}.
The full inhomogeneous contraction, before the local reduction used here, is given in
Appendix~\ref{app:quartic}.

\paragraph{Combined reduced 2PI kernel}
The cubic and quartic corrections can be combined while keeping their
diagrammatic origin explicit.  We collect the complete zeroth-order coefficient
of the intrinsic relative kernel into the frequency coefficient
\begin{equation}
M(T)
=
M_4(T)-\mu_{2PI}(T)-h_{\rm eff}(T)k_n^2.
\label{eq:total-branching-mass}
\end{equation}
At $Q=0$ the instantaneous response kernel can then be written directly as
\begin{align}
&\widehat{\mathcal K}_{\rm 2PI}(q,\omega;T)=
\nonumber\\
&\begin{pmatrix}
-\ii\omega+Es & -\left[h_{\rm eff}q^2+M\right]
\\[1mm]
-\left[h_{\rm eff}s-Ds^2\right] & -\ii\omega+Es
\end{pmatrix}.
\label{eq:corrected-2PI-kernel}
\end{align}
Inside this matrix, $s=s(q)$, $D=D(q,T)$, $E=E(q,T)$,
$h_{\rm eff}=h_{\rm eff}(T)$, and $M=M(T)$.
The coefficient $D$ carries the higher-gradient response correction, $E$
the diagonal dynamical dressing, and $h_{\rm eff}$ the Hartree-renormalized
relative stiffness.  The quartic term $M_4$ is an equal-time covariance
contraction, whereas $\mu_{2PI}$ is generated by the mixed $GC$ topology
propagated through the Bohm/Fisher response.  Equation~\eqref{eq:total-branching-mass}
assembles these contributions with the stationary reference frequency into the
single coefficient that controls the intrinsic infrared kernel. Its interpretation as a local renewal frequency is introduced in Sec.~\ref{sec:selfsimilar}.  The dressed response and covariance determine all three
contributions self-consistently at each central time.

\subsection{Reciprocal relative sector and slow collective projection}
\label{sec:relative-localization}

We now use the two diagonal channel projections $C_{UU}$ and $C_{WW}$ to
separate the internal relative structure from its slow collective response.
The complete numerical observable remains
$C_{\rm FB}=(C_{UU}+C_{UW}-C_{WU}-C_{WW})/4$, while its parity-even equal-time
relative projection is Eq.~\eqref{eq:FB-overlap-UW-even}.  The crossed blocks
remain present in the 2PI feedback and therefore influence the dressed
coefficients indirectly without defining an additional even relative residue.  The two diagonal channels have the same response denominator and
different residues. Their combination contains both an extended reference-mode sector and the screened relative sector.

The screening length characterizes the localized contribution to the
relative kernel. Because an extended reference-mode contribution also
remains, this length does not describe the decay of the entire reciprocal
correlator.

The present section keeps the frequency factor that becomes slow in the
collective variables $(Q,\Omega)$.  The complementary factor is restored in
Sec.~\ref{sec:global-collective-poles}, where both global pole families are
retained simultaneously.

\paragraph{Connected correlator and adiabatic central-time reduction}

The connected covariance matrix follows from the dressed response
matrix and the full noise kernel,
\begin{align}
C
&=
G\,\mathcal N\,G^T,
&
\mathcal N
&=
\widehat\Gamma_S^{(0)}+\Sigma_C.
\label{eq:C-GNG-relative}
\end{align}
In the $U/W$ basis the dressed noise kernel is kept in the general form
\begin{equation}
\mathcal N=
\begin{pmatrix}
N_{UU}&N_{UW}\\
N_{WU}&N_{WW}
\end{pmatrix}.
\label{eq:full-UW-noise-kernel}
\end{equation}
The reciprocal construction of Ref.~\cite{dumonteil_branching_2026} singles out correlated and
anticorrelated combinations as the two privileged stochastic channels.  This
motivates, in the same basis, the hierarchy
\begin{equation}
|N_{UW}|,\ |N_{WU}|\ll |N_{UU}|,
\qquad
|N_{UW}|,\ |N_{WU}|\ll |N_{WW}|.
\label{eq:weak-crossed-noise-hierarchy}
\end{equation}
The hierarchy is used only for the analytic residue reduction below.  It is
not a strict vanishing condition and does not impose
$C_{UW}=C_{WU}=0$: crossed physical covariances are generated by the coupled
response and remain in the complete 2PI feedback.

To separate internal and collective motion, we introduce the two
Wigner legs
\begin{align}
q_x
&=
q+\frac{Q}{2},
&
q_y
&=
q-\frac{Q}{2},
\nonumber\\
\omega_x
&=
\nu+\frac{\Omega}{4},
&
\omega_y
&=
\nu-\frac{\Omega}{4},
\label{eq:relative-collective-legs}
\end{align}
Here $\Omega$ is the collective frequency conjugate to the central
time $T$.  The $\Omega/4$ splitting follows from the normalization of
the unnormalized $U/W$ basis and is derived in
Appendix~\ref{app:UW-normalization} and is used throughout the main text.

We define
\begin{equation}
s_\pm(q,Q)
=
k_n^2-
\left(
q\pm\frac{Q}{2}
\right)^2.
\label{eq:s-pm-relative-body}
\end{equation}
The variables $(q,\nu)$ resolve the internal relative dynamics,
whereas $(Q,\Omega)$ describe the slow collective deformation and the propagation.

The 2PI corrections remain nonlocal functionals of the full
nonequilibrium propagators and retain their central-time dependence.
The reduction below is therefore performed at leading adiabatic order in the
central time.  At each fixed $T$, the instantaneous coefficients $D(q,T)$,
$E(q,T)$, $h_{\rm eff}(T)$, and $M(T)$ are held fixed while the
relative-frequency integration and the expansion in $(Q,\Omega)$ are carried
out.  Their momentum and central-time dependence is retained between
successive slices.
For compactness, the explicit arguments $(q,T)$ are suppressed in the following
matrices.  With $q_x=q+Q/2$ and $q_y=q-Q/2$, the dressed response kernel of the
$x$ leg is
\begin{align}
\mathcal K_x
&=
\begin{pmatrix}
Es_+-\ii\omega_x
&
-\left(h_{\rm eff}q_x^2+M\right)
\\[1mm]
-\left(h_{\rm eff}s_+-Ds_+s_-\right)
&
Es_+-\ii\omega_x
\end{pmatrix}
\nonumber\\
&\equiv
\begin{pmatrix}
d_x & B_x\\
-C_x & d_x
\end{pmatrix}
\label{eq:Kx-relative-dressed}
\end{align}
with
\begin{align}
&d_x=Es_+-\ii\omega_x,
\qquad
B_x=-\left(h_{\rm eff}q_x^2+M\right),
\nonumber\\
&C_x=h_{\rm eff}s_+-Ds_+s_-.
\label{eq:x-blocks-relative}
\end{align}

The second leg entering the transposed propagator is obtained by exchanging the
two Wigner legs together with the conjugate frequency assignment,
\begin{equation}
\mathcal K_y^\ast
=
\begin{pmatrix}
d_y^\ast & B_y\\
-C_y & d_y^\ast
\end{pmatrix},
\end{equation}
where
\begin{align}
&d_y^\ast=Es_-+\ii\omega_y,
\qquad
B_y=-\left(h_{\rm eff}q_y^2+M\right),
\nonumber\\
&C_y=h_{\rm eff}s_--Ds_+s_-.
\label{eq:y-blocks-relative}
\end{align}

It is convenient to introduce the spectral products
\begin{equation}
P_x=B_xC_x,
\qquad
P_y=B_yC_y,
\label{eq:relative-Pxy-general}
\end{equation}
so that the two response determinants are simply
\begin{equation}
\Delta_x=d_x^2+P_x,
\qquad
\Delta_y^\ast=(d_y^\ast)^2+P_y.
\label{eq:relative-Deltaxy-general}
\end{equation}

The labels $x$ and $y$ thus refer to the two legs of the two-point
function. They are exchanged by $Q\rightarrow-Q$, while both are
evaluated at the same central time $T$.

Under the channel hierarchy of Eq.~\eqref{eq:weak-crossed-noise-hierarchy},
matrix inversion gives the leading analytic residue projection
\begin{align}
C_{UU}(q,Q;\nu,\Omega,T)
&\simeq
\frac{d_xd_y^\ast N_{UU}+B_xB_yN_{WW}}
{\Delta_x\Delta_y^\ast},
\label{eq:CUU-start-relative}\\
C_{WW}(q,Q;\nu,\Omega,T)
&\simeq
\frac{C_xC_yN_{UU}+d_xd_y^\ast N_{WW}}
{\Delta_x\Delta_y^\ast}.
\label{eq:CWW-start-relative}
\end{align}
These expressions retain the internal relative variables $(q,\nu)$, the slow
collective variables $(Q,\Omega)$, and the instantaneous nonequilibrium
dressing at central time $T$.  The hierarchy concerns only the crossed source
weights in this analytic residue projection, while crossed response and covariance blocks remain in the 2PI dynamics.  Consequently $C_{UU}$
and $C_{WW}$ have the same response denominator and therefore the same pole
locations, while their leading residue weights are interchanged between the
two diagonal sectors.

At the bare level the reduced branching source is diagonal.  In the residue
normalization used after the internal-frequency integration, common overall
normalization factors are absorbed into $N_{ii}$, so that
$N_{UU}^{(0)}=N_{WW}^{(0)}\propto\nu_2R$.  The unabsorbed numerical source is
the covariance of Eq.~\eqref{eq:bare-UW-noise}, namely
$4\nu_2R(x)\delta(x-y)$ in each diagonal $U/W$ block.  The $HCC$ variation
dresses the full noise matrix in the numerical evolution.  For the analytic Born--Oppenheimer
residues only the diagonal effective weights $N_{UU}^{\rm eff}$ and
$N_{WW}^{\rm eff}$ are written explicitly, while crossed blocks continue to feed back through the self-consistent numerical propagators and covariances.  No
additional source-anticorrelation parameter is imposed.

\paragraph{Reciprocal observable and relative-frequency integration}
The two diagonal channel correlators are intermediate projections of the same
reciprocal observable.  For the parity-even relative sector of
Eq.~\eqref{eq:FB-overlap-UW-even}, Eq.~\eqref{eq:weak-crossed-noise-hierarchy}
gives directly
\begin{align}
&C_{\rm FB}(q,Q;\nu,\Omega,T)
=\frac14\left(C_{UU}-C_{WW}\right)
\nonumber\\
&\qquad\simeq
\frac{
\left(d_xd_y^\ast-C_xC_y\right)N_{UU}
+
\left(B_xB_y-d_xd_y^\ast\right)N_{WW}
}{4\,\Delta_x\Delta_y^\ast}.
\label{eq:CFB-complete-two-frequency}
\end{align}
Crossed physical covariances remain dynamically active in the complete 2PI
feedback, since the simplification above concerns only the leading analytic source
weights entering the residue algebra.

The integration over the internal relative frequency $\nu$ is performed next,
before separating fast relative organization from slow collective motion.
From this point through the Born--Oppenheimer residue reduction, we work
in the conservative sector $E=0$. The general frequency factor retaining
$E$ is restored in Sec.~\ref{sec:global-collective-poles}.
We write $\varepsilon_x=\sqrt{P_x}$ and $\varepsilon_y=\sqrt{P_y}$ for
the frequency-like roots, so that they remain distinct from the momenta.
Summing the poles of either frequency leg gives
\begin{equation}
C_{\rm FB}^{\rm eq}(q,Q;\Omega,T)
\simeq
\sum_{\sigma=\pm1}
\frac{R_\sigma^{\rm FB}(q,Q;\Omega,T)}
{\mathcal D_\sigma(q,Q;\Omega,T)},
\label{eq:CFB-after-nu}
\end{equation}
with
\begin{equation}
R_\sigma^{\rm FB}
=\frac14\left(R_\sigma^{U}-R_\sigma^{W}\right)
\label{eq:CFB-residue-combination}
\end{equation}
where $R_\sigma^U$ and $R_\sigma^W$ are the residues obtained from the internal-frequency integration in the $U$ and $W$ sectors, respectively, and
\begin{equation}
\mathcal D_\sigma(q,Q;\Omega,T)
=
P_x(q,Q,T)
-
\left[
\frac{\Omega}{2}
+\sigma \varepsilon_y(q,Q,T)
\right]^2,
\label{eq:Dsigma-relative-body}
\end{equation}
The remaining denominator factorizes exactly as
\begin{align}
\mathcal D_\sigma
&=
\left[\varepsilon_x-\frac{\Omega}{2}-\sigma \varepsilon_y\right]
\left[\varepsilon_x+\frac{\Omega}{2}+\sigma \varepsilon_y\right].
\label{eq:Dsigma-factor-body}
\end{align}
Equation~\eqref{eq:CFB-after-nu} is the frequency-integrated reciprocal
correlator from which the Born--Oppenheimer hierarchy is constructed.

\paragraph{Born--Oppenheimer relative structure}
We now apply the Born--Oppenheimer hierarchy~\cite{BornOppenheimer1927}.  The
relative-frequency integration has already been performed, and the separation
concerns the two remaining scales contained in the factors of
Eq.~\eqref{eq:Dsigma-factor-body}.  The fast combination is of order
$\varepsilon_x+\varepsilon_y$, while the difference $\varepsilon_x-\varepsilon_y$ becomes slow in the collective
variables $(Q,\Omega)$.

For the relative residues we first retain the finite-well reference scale
$k_n$.  At $Q=0$ we define
\begin{align}
&B(q,T)=-\left[h_{\rm eff}(T)q^2+M(T)\right],
\nonumber\\
&C(q,T)=s(q)\left[h_{\rm eff}(T)-D(q,T)s(q)\right],
\label{eq:BO-finite-reference-BC}
\end{align}
and $P(q,T)=B(q,T)C(q,T)$, $\varepsilon(q,T)=\sqrt{P(q,T)}$ on the real-pole support, where $P(q,T)>0$.  The higher-gradient coefficient $D$ is retained in the denominator
of the finite-reference residue.
For the Born--Oppenheimer residue reduction we retain the conservative condition $E=0$ and further restrict to the regime in which the quartic derivative correction is subleading to the dressed localizing coefficient.  Operationally we require
\begin{equation}
E(q,T)=0,
\qquad
\left|D(q,T)s^2(q)\right|
\ll
\left|M(T)+h_{\rm eff}(T)k_n^2\right|.
\label{eq:infrared-derivative-hierarchy}
\end{equation}
The full $D$ dependence is nevertheless retained in the
finite-reference denominator, because it controls the detailed continuation
of the extended sector.  In the non-common numerator the exact correction is
proportional to $D(q,T)s^2(q)$ and is dropped only under
Eq.~\eqref{eq:infrared-derivative-hierarchy}.  After removing the
common causal normalization inherited from the $\nu$ integration, the slow
difference-family residues are then
\begin{align}
\widehat Z_-^{W}(q,T)
&=
N_{UU}^{\rm eff}+N_{WW}^{\rm eff}
-
\frac{\left[M+h_{\rm eff}k_n^2\right]N_{UU}^{\rm eff}}
{h_{\rm eff}q^2+M},
\label{eq:BO-ZW-finite}
\\
\widehat Z_-^{U}(q,T)
&=
N_{UU}^{\rm eff}+N_{WW}^{\rm eff}
\nonumber\\
&\qquad\qquad+\frac{\left[M+h_{\rm eff}k_n^2\right]N_{WW}^{\rm eff}}
{(q^2-k_n^2)\left[h_{\rm eff}-D(q,T)s(q)\right]}.
\label{eq:BO-ZU-finite}
\end{align}
Here $N_{ii}^{\rm eff}$ denotes the effective diagonal noise weight associated
with channel $i\in\{U,W\}$ after the internal-frequency reduction.  The term
$N_{UU}^{\rm eff}+N_{WW}^{\rm eff}$ is common to the two channel residues and
cancels from the even reciprocal projection.  The physical slow residue is
therefore
\begin{align}
\widehat Z_-^{\rm FB}(q,T)
&\equiv
\frac14\left(\widehat Z_-^{U}-\widehat Z_-^{W}\right)
\nonumber\\
&=
\frac{M+h_{\rm eff}k_n^2}{4}
\nonumber\\
&\qquad\times\left[
-\frac{N_{WW}^{\rm eff}}
{s(q)\left[h_{\rm eff}-Ds(q)\right]}
+
\frac{N_{UU}^{\rm eff}}
{h_{\rm eff}q^2+M}
\right].
\label{eq:BO-ZFB-two-relative-sectors}
\end{align}

\begin{figure*}[!htbp]
\centering
\includegraphics[width=\textwidth]{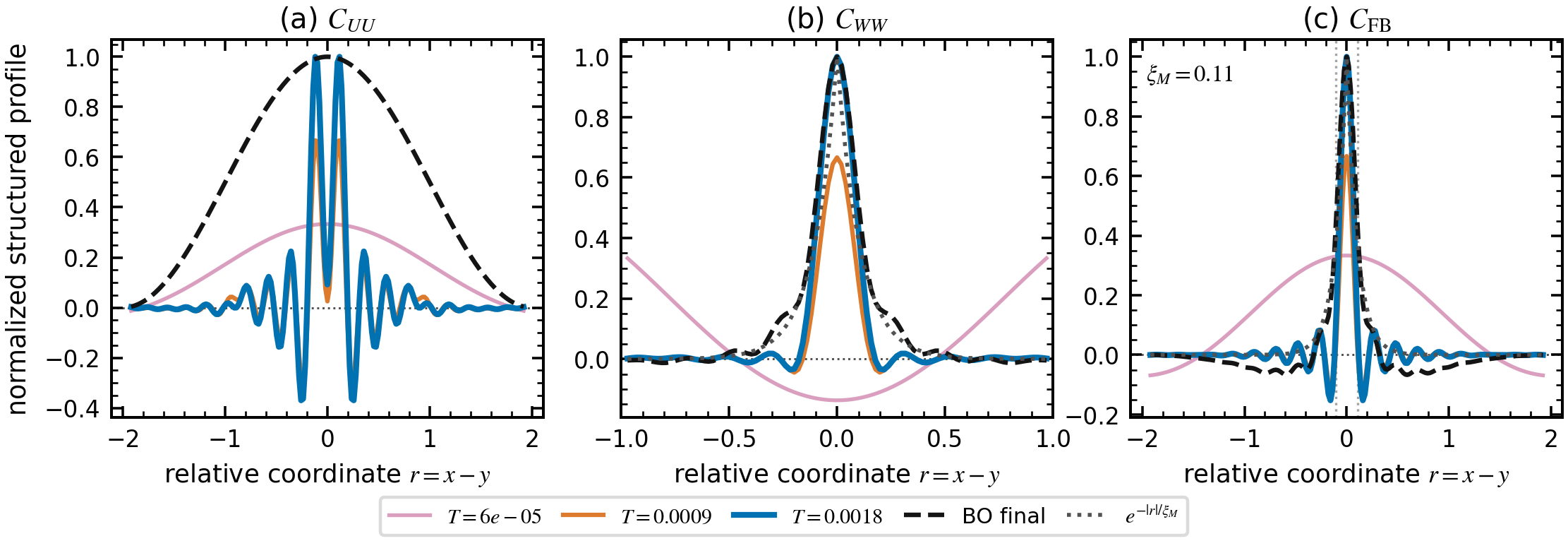}
\caption{
Relative-coordinate shape diagnostics at fixed collective coordinate
$X=x_{\rm ref}$: (a) $UU$, (b) $WW$, and (c) the full reciprocal combination
$C_{\rm FB}=(C_{UU}+C_{UW}-C_{WU}-C_{WW})/4$.
The profiles are processed with the cosine window and common-component
estimation specified in Appendix~\ref{app:numerical-2pi}.  The $U$-channel BO
comparison retains $D(q,T)$ in the finite-reference denominator of
Eq.~\eqref{eq:BO-ZU-finite}, while the $Ds^2$ term is neglected only in the non-common numerator as stated above.
The early, middle and final snapshots are independently rescaled to absolute
peaks $1/3$, $2/3$ and $1$ to compare shapes. These display levels do not represent measured growth amplitudes. Dashed curves show separately normalized final BO shape
comparisons. Dotted curves in the $WW$ and FB panels show the sign-matched
envelope $\exp(-|r|/\xi_M)$, where $\xi_M=\sqrt{h_{\rm eff}/M}$ is calculated
from the dressed coefficients, not fitted. Vertical guides in FB mark
$r=\pm\xi_M$. The envelope is unfiltered, whereas the numerical profiles
use $q\leq\min(q_{\rm grid,max},p/\xi_M)$, with $p=3$ in the supplied
reference settings and a smooth taper over the last $20\%$ of the band.
}
\label{fig:a-transition}
\end{figure*}

The two terms have distinct relative analytic structures. On the localizing branch $M(T)>0$, with $h_{\rm eff}(T)>0$, the first retains
the reference poles at $q=\pm k_n$ and gives an extended oscillatory component
after inverse Fourier transformation.  The second contains the screened poles
$q=\pm\ii\sqrt{M/h_{\rm eff}}$ and therefore introduces the finite relative
length
\begin{equation}
\xi(T)=\sqrt{\frac{h_{\rm eff}(T)}{M(T)}}.
\label{eq:Crel-Lorentzian-body-final}
\end{equation}
In one dimension its long-distance envelope is proportional to
$\exp[-|r|/\xi(T)]$.  The BO relative kernel thus contains an extended
reference-mode sector and an emergent localized sector within the same
forward--backward observable.

The slow factor follows from the same frequency-integrated denominator.  In
the collective regime
\begin{equation}
|\Omega|,\ |\varepsilon_x-\varepsilon_y|\ll \varepsilon(q,T),
\label{eq:slow-scale-separation}
\end{equation}
we define
\begin{equation}
\Omega_{\rm d}(q,Q,T)=2\left[\varepsilon_x-\varepsilon_y\right]
\label{eq:V-relative-body}
\end{equation}
and combine the conjugate difference poles to obtain
\begin{equation}
G_{\rm coll}^{\rm slow}(q,Q;\Omega,T)
=
\frac{2\Omega}
{\Omega^2-\Omega_{\rm d}^2(q,Q,T)}.
\label{eq:Gcoll-body-final}
\end{equation}
The resulting BO reciprocal correlator can therefore be written explicitly,
retaining the same finite-reference denominator as in
Eq.~\eqref{eq:BO-ZU-finite}, consistently with the hierarchy stated above.
\begin{align}
&C_{\rm FB}^{\rm slow,BO}(q,Q;\Omega,T)
\simeq
\widehat Z_-^{\rm FB}(q,T)
G_{\rm coll}^{\rm slow}(q,Q;\Omega,T)
\nonumber\\
&\quad=
\frac{M+h_{\rm eff}k_n^2}{4}\frac{2\Omega}
{\Omega^2-\Omega_{\rm d}^2}
\nonumber\\
&\qquad\times\left[
-\frac{N_{WW}^{\rm eff}}
{s(q)\left[h_{\rm eff}-Ds(q)\right]}
+
\frac{N_{UU}^{\rm eff}}
{h_{\rm eff}q^2+M}
\right]
\label{eq:CFB-BO-complete-slow}
\end{align}
up to the common causal normalization absorbed into the reduced residues.
Equation~\eqref{eq:CFB-BO-complete-slow} makes explicit that the extended and
screened relative sectors are transported by the same slow collective kernel.

The diagonal amplitude and the normalized relative shape play different roles
in this factorization.  We define
\begin{equation}
\rho_{\rm BSM}(X,T)\equiv-C_{\rm FB}(X,r=0,T)
\label{eq:BO-rhoBSM-definition}
\end{equation}
and, whenever this diagonal amplitude is nonzero, write
\begin{align}
&C_{\rm FB}(X,r,T)
=-\rho_{\rm BSM}(X,T)F_{\rm rel}(X,r,T),
\nonumber\\
&F_{\rm rel}(X,0,T)=1.
\label{eq:BO-CFB-collective-relative}
\end{align}
 This equation normalizes the relative BO shape, not the connected weight.
The stationary reciprocal construction of Ref.~\cite{dumonteil_branching_2026} imposes the matching
of $\rho_{\rm BSM}$ to the Born profile $R^2(X)$.  Taking that diagonal matching as given, the present 2PI analysis determines the dressed relative content of the same kernel, represented here by the coexisting $k_n$ and
$\xi^{-1}$ sectors.

\paragraph{Spectral definition and selection of the relative mass}

Equation~\eqref{eq:total-branching-mass} gives the complete zeroth-order
coefficient of the intrinsic relative kernel.  In the slow infrared sector,
\begin{equation}
\widehat{\mathcal K}_{\rm rel}^{\rm IR}(q,T)
=
h_{\rm eff}(T)q^2+M(T).
\label{eq:minimal-spectral-mass}
\end{equation}
The localizing branch is selected by
\begin{equation}
M(T)>0
\label{eq:localizing-sign-condition}
\end{equation}
Equation~\eqref{eq:localizing-sign-condition} is only the screening condition that produces a finite relative length.  It does not determine the sign of $C_{\rm FB}$ or impose the diagonal Born matching $\rho_{\rm BSM}=R^2$, because selection of the anticorrelated branch and the stationary amplitude matching are separate ingredients of the reciprocal construction.
The same coefficient enters the branching clock of
Sec.~\ref{sec:selfsimilar}.  Its decomposition
\begin{equation}
M(T)=M_4(T)-\mu_{2PI}(T)-h_{\rm eff}(T)k_n^2
\end{equation}
keeps the quartic covariance contraction, the cubic response--correlation
correction, and the stationary reference contribution explicit while the
intrinsic slow pole depends on their dressed sum.

The numerical implementation used for the self-consistent 2PI figures is
summarized in Appendix~\ref{app:numerical-2pi}.
\begin{figure*}[!htbp]
\centering
\includegraphics[width=\textwidth]{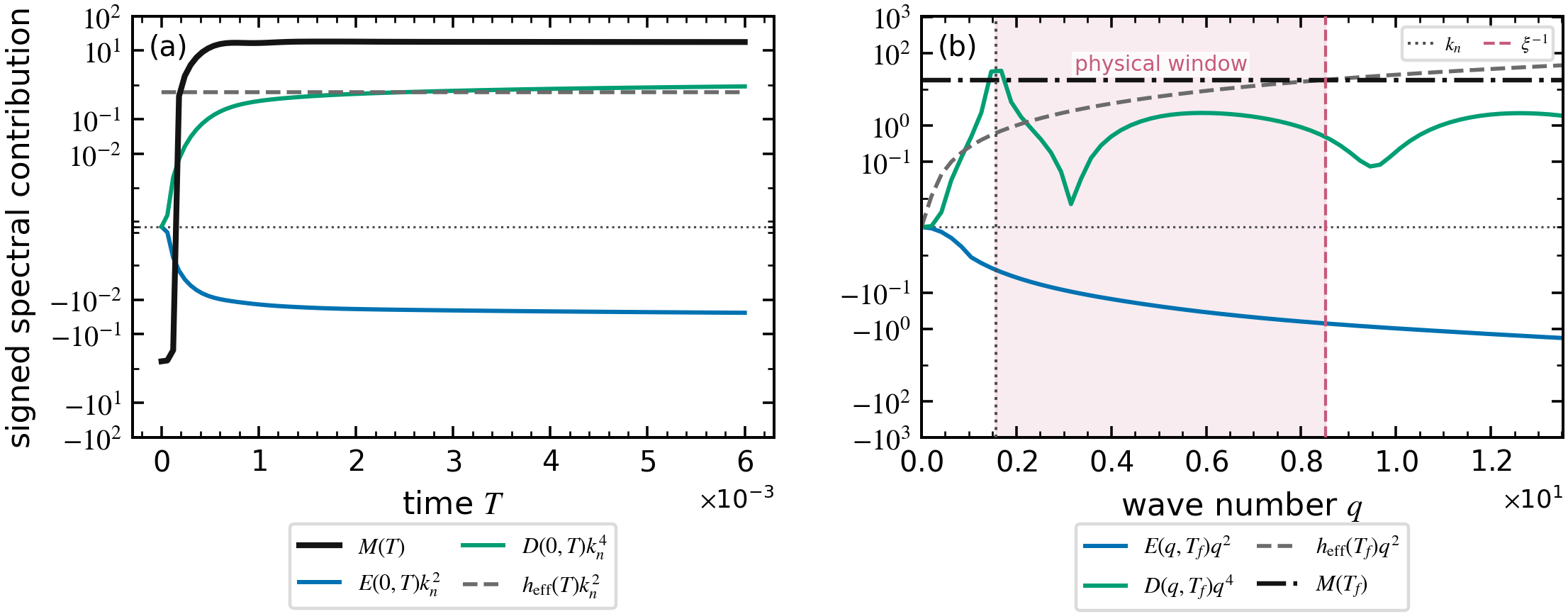}
\caption{
Signed contributions to the dressed relative kernel.
(a) $M(T)$, $E(q_0,T)k_n^2$, $D(q_0,T)k_n^4$, and
$h_{\rm eff}(T)k_n^2$, with $q_0$ the saved wave number nearest zero and
$M=M_4-\mu_{2PI}-h_{\rm eff}k_n^2$.
(b) $E(q,T_f)q^2$, $D(q,T_f)q^4$, and $h_{\rm eff}(T_f)q^2$ at the final
stored time, with a horizontal guide at $M(T_f)$.
The shaded comparison band lies between $k_n$ and
$\xi_\mu^{-1}=\sqrt{|\mu_{2PI}(q_0,T_f)|/h_{\rm eff}(T_f)}$.
This cubic-coefficient scale is distinct from the screened-pole length
$\xi_M$ and from a fitted spatial width. The shading alone does not establish the validity of an approximation. Both ordinates use a signed
symmetric-logarithmic scale. The signed $\mu_{2PI}$ values are retained in
the accompanying numerical exports.
}
\label{fig:2pi-coefficients}
\end{figure*}

\paragraph{Derivative corrections beyond the infrared form}
Retaining $D$ and $E$ changes the short-distance continuation of the screened
mode while preserving the definition of the localizing coefficient $M(T)$.
The instantaneous spectral product of the complete reduced kernel is
\begin{align}
P(q,T)
&=
-\left[h_{\rm eff}(T)q^2+M(T)\right]
\nonumber\\
&\qquad\qquad\times\left[h_{\rm eff}(T)s(q)-D(q,T)s^2(q)\right],
\label{eq:quartic-relative-kernel}
\end{align}
with the diagonal term $E^2s^2$ retained in the full response determinant.
The coefficient $D$ controls the higher-gradient continuation and $E$ the
diagonal dynamical dressing.  In the finite-$k_n$ Born--Oppenheimer residue of
Eq.~\eqref{eq:BO-ZU-finite}, $D$ is retained in the denominator and neglected
only in the non-common numerator under
Eq.~\eqref{eq:infrared-derivative-hierarchy}.  The Lorentzian screened pole
$h_{\rm eff}q^2+M$ and its length $\xi_M$ remain unchanged.  The additional
roots of the complete polynomial determine the short-distance continuation
resolved by the numerical 2PI evolution.
\label{sec:massive-closure}

\section{Avrami formation of the reciprocal forward--backward sector}
\label{sec:avrami}

\subsection{Bounded reciprocal formation variable}

The Avrami law was originally introduced to describe bounded conversion by
nucleation and growth
~\cite{Kolmogorov1937,JohnsonMehl1939,Avrami1939,Avrami1940,Avrami1941}.
Here its exponential structure is used only as a reduced description of the
central-time formation generated by the self-consistent 2PI dynamics.  It is
not inserted into the field equations.

The physical reciprocal observable is
\begin{equation}
C_{\rm FB}
\simeq
\frac14\left(C_{UU}-C_{WW}\right)
\label{eq:Avrami-CFB-observable}
\end{equation}
for the parity-even relative projection of Eq.~\eqref{eq:FB-overlap-UW-even}.
The Avrami construction is applied directly to this reconstructed forward--backward observable.  The separate $U/W$ covariances remain inside the complete 2PI feedback, but no independent kinetic order parameter is assigned to either channel.

Let $A_{\rm FB}(T)$ denote a positive amplitude extracted from the reciprocal
profile after the fixed sign convention used for the numerical comparison.  In
the numerical implementation below this is the spatial maximum of the stored
$C_{\rm FB}$ profile. Let $T_{\rm sel}$ mark the start of the resolved
conversion window and $A_{\rm FB}(\infty)$ denote its late-time plateau
amplitude. Over this window we define
\begin{equation}
f_{\rm FB}(T)
=
\frac{A_{\rm FB}(T)-A_{\rm FB}(T_{\rm sel})}
{A_{\rm FB}(\infty)-A_{\rm FB}(T_{\rm sel})}
\label{eq:Avrami-mu}
\end{equation}
with $0\leq f_{\rm FB}\leq1$ when the extracted reciprocal amplitude evolves
monotonically over the fitted interval.  The normalized quantity
$f_{\rm FB}$ is the reduced captured fraction used below in the Fisher
construction.  Its use as a pair-information weight is a closure of the
self-similar model, while the underlying $C_{\rm FB}$ is generated by the 2PI
evolution itself. The notation $f_{\rm FB}$ distinguishes this dimensionless
fraction from the reciprocal screening coefficient $\mu_{\rm FB}$ of Ref.~\cite{dumonteil_branching_2026}.

\subsection{Extended reciprocal activity and self-similar capture}

Introduce a single extended activity $X_{\rm FB}(T)$ by
\begin{equation}
f_{\rm FB}(T)=1-\exp[-X_{\rm FB}(T)]
\label{eq:Avrami-channel-mu}
\end{equation}
so that
\begin{equation}
\partial_Tf_{\rm FB}(T)
=
[1-f_{\rm FB}(T)]\,\partial_TX_{\rm FB}(T).
\label{eq:Avrami-channel-identity}
\end{equation}
We use the homogeneous capture closure
\begin{equation}
\partial_TX_{\rm FB}(T)
=
\Gamma_{\rm cap}\,\eta_{\rm FB}[X_{\rm FB}(T)],
\label{eq:X-growth-general}
\end{equation}
with
\begin{equation}
\eta_{\rm FB}(X_{\rm FB})
=
\eta_0
\left(\frac{X_{\rm FB}}{X_c}\right)^{d_{\rm cap}}.
\label{eq:eta-dcap}
\end{equation}
Here $\Gamma_{\rm cap}$ sets the capture-rate scale, $\eta_0$ normalizes
the capture factor, $X_c$ is a reference activity, and $d_{\rm cap}$
is the effective exponent in this homogeneous closure.
The localized part of the same reciprocal profile is approximately described by
\begin{equation}
F_{\rm loc}(r,T)
=
f\!\left(\frac{|r|}{\xi(T)}\right),
\label{eq:selfsimilar-relative-profile}
\end{equation}
with $f(s)\sim\exp(-s)$ on the infrared localizing branch.  This one-scale
structure motivates the effective capture exponent but does not replace the
complete extended-plus-screened Born--Oppenheimer residue of $C_{\rm FB}$.

For $0<d_{\rm cap}<1$,
\begin{equation}
X_{\rm FB}(T)
\simeq
\gamma_{\rm FB}(T-T_{\rm sel})^{\beta_{\rm FB}},
\qquad
\beta_{\rm FB}=\frac{1}{1-d_{\rm cap}}.
\label{eq:Avrami-beta}
\end{equation}
Hence the reduced reciprocal formation variable has the Avrami form
\begin{equation}
f_{\rm FB}(T)
=
1-\exp\left[-\gamma_{\rm FB}(T-T_{\rm sel})^{\beta_{\rm FB}}\right].
\label{eq:avrami-general}
\end{equation}
Only the forward--backward conversion exponent is used in the physical kinetic comparison.  Channel-resolved quantities remain diagnostics of the internal 2PI mechanism rather than separate Avrami observables.

For the numerical comparison, BSM-MC denotes the projected branching
Monte Carlo implementation of the BSM hierarchy used in
Ref.~\cite{dumonteil_branching_2026}. It provides the forward--backward
reference activity compared with the reduced 2PI evolution below.

\begin{figure}[!htbp]
\centering
\includegraphics[width=\columnwidth]{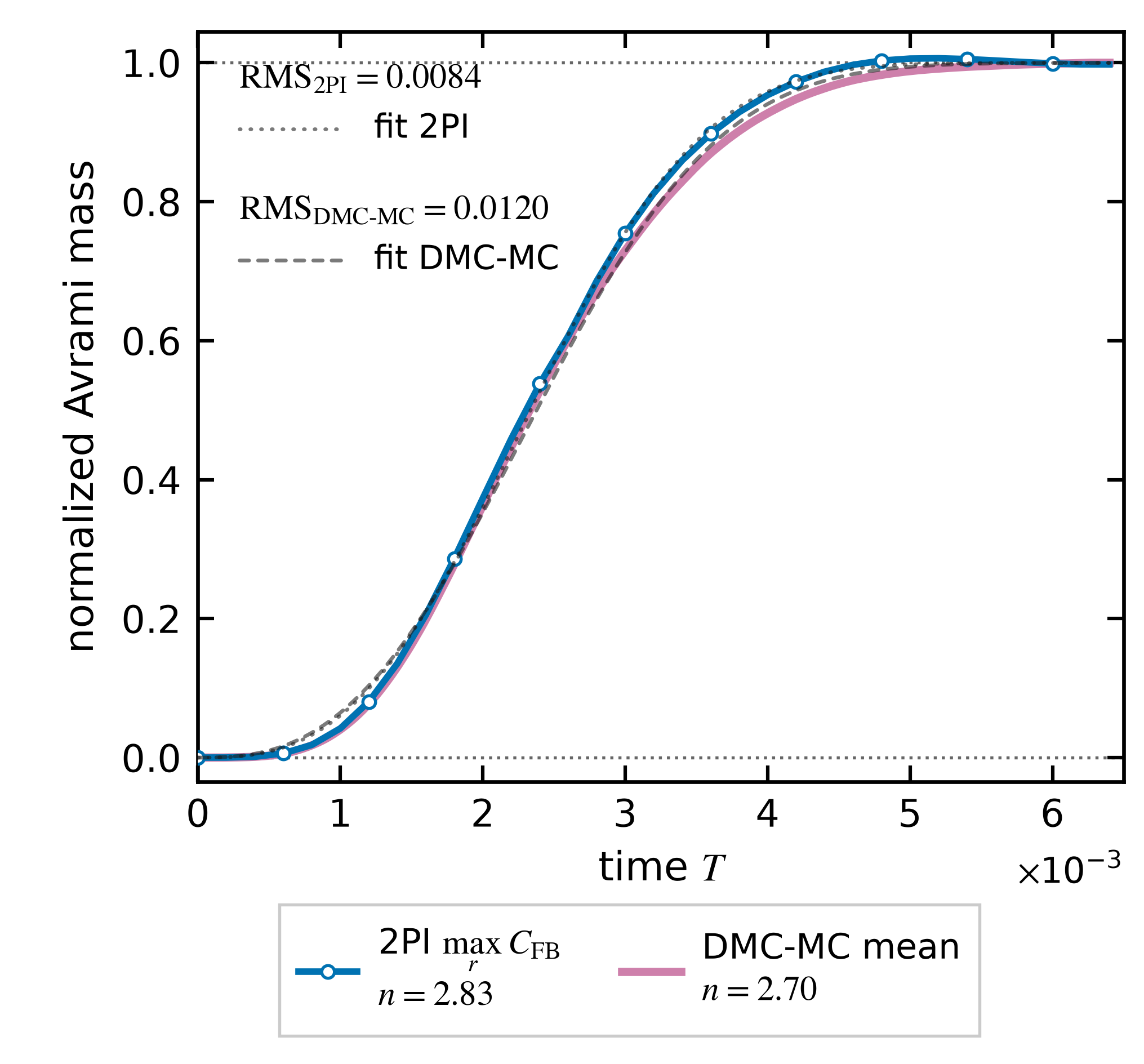}
\caption{
Normalized growth and saturation of the spatial maximum of the
forward--backward correlator in the reduced self-consistent 2PI evolution,
compared with the BSM-MC reference.
The 2PI observable is $A_{\rm FB}(T)=\max_r C_{\rm FB}(r,T)$ on the
exported spatial grid, with
$f_{\rm FB}(T)=[A_{\rm FB}(T)-A_{\rm FB}(0)]/
[A_{{\rm FB},{\rm tail}}-A_{\rm FB}(0)]$,
where $A_{{\rm FB},{\rm tail}}$ is the median of the final eight samples
on the saturated plateau.
The blue line shows $f_{\rm FB}$, with open circles marking every third
sample, while the mauve line shows the normalized BSM-MC forward--backward mean
activity over 100 realizations.
Thin, semi-transparent black dotted curves are nonlinear
Avrami fits $1-\exp[-\gamma T^n]$, with unit asymptote and
$T_0=0$ in the corresponding source clock.
Fits use $0.01<f_{\rm FB}<0.985$. The legend gives the exponents, and RMS residuals are evaluated over the displayed samples.
BSM-MC time is affinely rescaled to the 2PI conversion interval for a
comparison of normalized kinetic shapes.  The self-consistent 2PI dynamics
reaches the reciprocal saturation plateau described by the Avrami law.
Further numerical details are given in Appendix~\ref{app:numerical-2pi}.
}
\label{fig:avrami}
\end{figure}

\subsection{2PI saturation and BSM-MC comparison}

Figure~\ref{fig:avrami} tests directly the bounded formation of the reciprocal observable generated by the self-consistent evolution.  The solver evolves the dressed response and the full covariance matrix, and reconstructs
\begin{equation}
C_{\rm FB}
=\frac14\left(C_{UU}+C_{UW}-C_{WU}-C_{WW}\right).
\end{equation}
The crossed blocks remain in the numerical evolution and are not set to zero by the parity-even analytic reduction.  The Avrami fit is performed only after
this reconstruction and therefore summarizes, rather than imposes, the
central-time formation process.

The $U$ and $W$ projections still diagnose how the extended reference-mode and screened relative contributions are assembled inside the 2PI covariance, but the physical bounded conversion variable is their reciprocal combination $C_{\rm FB}$.  This avoids identifying the screened $W$ residue by itself with the captured physical weight.

\begin{figure}[!htbp]
\centering
\includegraphics[width=\columnwidth]{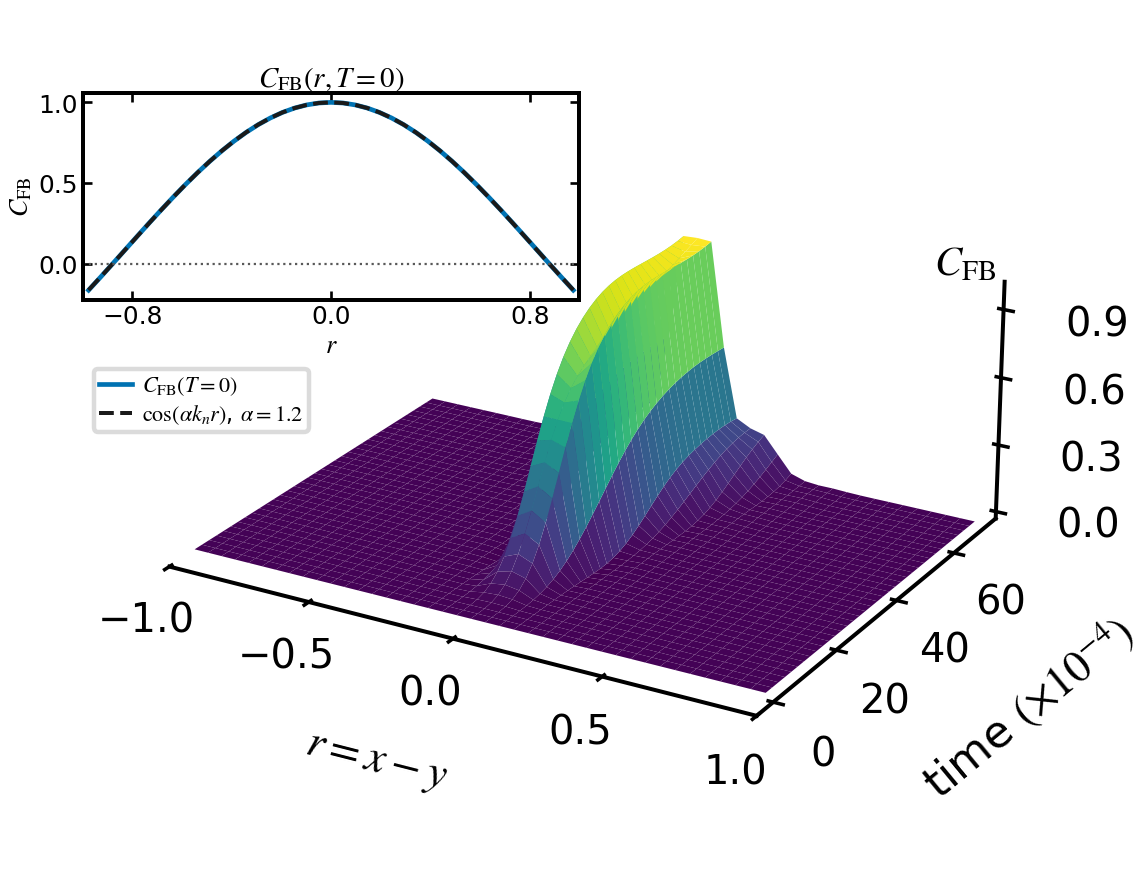}
\caption{
Formation of the relative reciprocal covariance at $X=x_{\rm ref}$.
The stored combination is $C_{\rm FB}=(C_{UU}+C_{UW}-C_{WU}-C_{WW})/4$.
For this surface, a constant far-tail median is removed at each time, and
one final-time absolute-peak scale normalizes all times. An overall sign
orients the final central value positively, so the ordinate
$\widetilde C_{\rm FB}$ denotes this display convention, not a probability
density. The initial profile is not subtracted.
The inset shows the earliest processed snapshot selected for the profile
figure at its actual stored time $T_i$, rather than the zero-covariance
initial condition, together with a fitted cosine reference
$a\cos(\alpha k_n r)+b$. See Appendix~\ref{app:numerical-2pi} for the
offset and normalization prescriptions.
}
\label{fig:cfb-cluster-formation}
\end{figure}

The forward--backward BSM-MC observable is normalized and its time coordinate
is affinely rescaled only to compare kinetic shape over the common conversion
window.  The comparison is therefore qualitative rather than a parameter-free
equality of amplitudes or microscopic time scales.  Figures~\ref{fig:avrami}
and \ref{fig:cfb-cluster-formation} provide complementary temporal and spatial
diagnostics of the same reciprocal organization.

\section{Self-similar information scaling and universal fixed point}
\label{sec:selfsimilar}

The preceding sections describe relative organization through the connected
kernel and its reduced numerical evolution. We now introduce a self-similar
information model to associate a velocity scale with reciprocal saturation.
This model uses the fraction $f_{\rm FB}$ extracted from the connected
kernel, together with assumptions about mass resolution, pair-score
alignment, and an invariant elementary mass--length product. The resulting
scale $c_\star$ is then used to formulate matching conditions for the
collective response.

\subsection{Mass self-similarity and internal time}

Following Ref.~\cite{dumonteil_branching_2026}, a \emph{possibility} denotes a locally admissible stochastic continuation in the branching representation. It serves as a bookkeeping degree of freedom and does not represent a hidden material particle. The theory propagates such possible continuations under the imposed dynamical and symmetry constraints. These continuations may equivalently be regarded as elementary carriers of information.

This subsection concerns the intrinsic Schr\"odinger--Nagasawa sector, $V=0$, before the correlation-dependent 2PI Hartree dressing is applied.  The self-similar construction is a statement about how the same underlying set of admissible Schr\"odinger--Nagasawa possibilities is represented at different internal resolutions.  The factor $\zeta(T)$ belongs instead to the correlated state produced by the 2PI dynamics and will re-enter only when the dressed renewal and localization scales are matched to the intrinsic information velocity.

At the coarse-grained mass scale $m$, the propagated field represents a structured collective information measure. Resolving this same measure into $N$ equivalent resolution elements defines the elementary mass weight $m_0$ through
\begin{equation}
m=Nm_0,
\qquad
N\in\mathbb N_{>0}.
\label{eq:mass-resolution}
\end{equation}
The integer $N$ is not interpreted as a fixed number of physical constituents. It specifies the internal resolution at which the same local information measure is represented. Equation~\eqref{eq:mass-resolution} is therefore a resolution identity rather than a constituent-number postulate.

For a uniform refinement into equivalent resolution elements, the collective field is represented as
\begin{equation}
\phi=N\phi_1.
\label{eq:selfsimilar-mass-v2}
\end{equation}
Here $\phi_1$ denotes the field carried by one elementary resolution element.  Let $\mathcal L_m^{(0)}$ denote the intrinsic Schr\"odinger--Nagasawa generator at mass $m$.  Its inverse-mass coefficient gives
\begin{equation}
\mathcal L_m^{(0)}
=
\frac{m_0}{m}
\mathcal L_{m_0}^{(0)}
=
\frac{1}{N}\mathcal L_{m_0}^{(0)}.
\label{eq:L-selfsimilar}
\end{equation}
The collective equation can consequently be written
\begin{equation}
N\partial_t\phi
=
\mathcal L_{m_0}^{(0)}[\phi].
\label{eq:N-time}
\end{equation}
Introducing the internal time
\begin{equation}
\dd\tau
=
\frac{\dd t}{N}
\label{eq:internal-time-v2}
\end{equation}
gives
\begin{equation}
\partial_\tau\phi
=
\mathcal L_{m_0}^{(0)}[\phi].
\label{eq:internal-evolution}
\end{equation}

Increasing the internal resolution contracts the internal time interval associated with one elementary possibility. For a fixed laboratory-time interval,
\begin{equation}
N\longrightarrow\infty
\qquad\Longrightarrow\qquad
\frac{\dd\tau}{\dd t}\longrightarrow0,
\end{equation}
and
\begin{equation}
\tau(t)=\frac{t}{N}\longrightarrow0.
\label{eq:time-freeze}
\end{equation}
At fixed total mass $m$, $m_0=m/N$, so the elementary intrinsic spatial generator grows proportionally to $N$ while the elementary internal-time interval decreases as $1/N$,
\begin{equation}
\mathcal L_{m_0}^{(0)}\,\dd\tau
\sim
N\,\frac{\dd t}{N}.
\end{equation}
The refinement therefore preserves the finite intrinsic Schr\"odinger--Nagasawa evolution without requiring a correlation-dependent factor.  Self-similarity fixes the kinematics of the possible continuations, whereas $\zeta$ and $M$ describe the dressing of the correlated sector.

\subsection{Fisher scores and reciprocal capture}

At resolution $m_0$, the $a$th internal possibility is represented by a normalized local spatial weight
\begin{equation}
p_a(x,T)\geq0,
\qquad
\int\dd^dx\,p_a(x,T)=1.
\label{eq:elementary-pa-normalization}
\end{equation}
The function $p_a$ is the conditional spatial distribution assigned to one admissible Schr\"odinger--Nagasawa continuation inside the resolved information element, without introducing an additional hidden particle.  A convenient self-similar representation is $p_a(x,T)=\ell_0^{-d}f_a[(x-X_a)/\ell_0,T]$, where $\ell_0$ is the elementary information-resolution length.

For each possibility define the Fisher score~\cite{HallReginatto2002}
\begin{equation}
\bm s_a
=
\nabla\ln p_a.
\label{eq:Fisher-score}
\end{equation}
The Fisher score is tied to the BSM drift through Eq.~\eqref{eq:Fisher-variation}: the same functional generates the Bohm contribution of the intrinsic Schr\"odinger--Nagasawa operator. The score measures the local translation sensitivity of the density entering that operator.

Define the collective score at this internal resolution by
\begin{equation}
\bm S_N=\sum_{a=1}^N\bm s_a.
\label{eq:collective-score}
\end{equation}
The corresponding quadratic Fisher content is
\begin{align}
I_F^{(N)}
&\equiv
\mean{|\bm S_N|^2}_{\rm loc}
\nonumber\\
&=
\sum_{a=1}^N\mean{\bm s_a^2}_{\rm loc}
+
\sum_{a\neq b}\mean{\bm s_a\cdot\bm s_b}_{\rm loc}
\nonumber\\
&\equiv
I_{F,1}^{(N)}+I_{F,2}^{(N)}.
\label{eq:Fisher-decomposition}
\end{align}
Here $\langle\cdots\rangle_{\rm loc}$ denotes the normalized local average with respect to the one- or two-possibility weight appropriate to the term. In particular, $\langle A_a\rangle_{\rm loc}=\int\dd^dx\,p_a(x)A_a(x)$ for a one-possibility quantity. The one- and two-possibility weights are assumed to be compatible marginals of the same local joint description.  Within this closure, $I_F^{(N)}$ measures the collective quadratic Fisher-score content without requiring identification with the Fisher information of an arbitrary $N$-body joint distribution.

We make two explicit closure assumptions. First, equivalent elementary possibilities have the same marginal Fisher norm,
\begin{equation}
\mean{\bm s_a^2}_{\rm loc}
=
I_0,
\qquad
I_0\equiv\frac{1}{\ell_0^2}.
\label{eq:elementary-I0}
\end{equation}
Cauchy--Schwarz then gives
\begin{equation}
\left|\mean{\bm s_a\cdot\bm s_b}_{\rm loc}\right|
\leq I_0.
\label{eq:pair-CS}
\end{equation}
This bounds the coherent covariance that one pair can contribute.

The physical projection is instead supplied by the reciprocal forward--backward weight.  A pair contributes coherently only insofar as its two continuations belong to the organized $F/B$ overlap measured by the bounded fraction $f_{\rm FB}(T)$ of Sec.~\ref{sec:avrami}.  In the captured sector the corresponding scores are taken to align with the same collective mode.  The minimal homogeneous closure is therefore
\begin{equation}
I_{F,2}^{(N)}(T)
=
N(N-1)I_0
f_{\rm FB}(T).
\label{eq:IF2-muFB}
\end{equation}
The linear dependence on $f_{\rm FB}$ is a closure of the self-similar information model, not an identity of the 2PI equations.  The 2PI calculation supplies the physical correlator $C_{\rm FB}$ from which $f_{\rm FB}$ is extracted. The Fisher construction then uses this reciprocal occupation fraction to weight the pair-score covariances.

The total Fisher content becomes
\begin{equation}
I_F^{(N)}(T)
=
NI_0
+
N(N-1)I_0f_{\rm FB}(T).
\label{eq:IF-interpolation}
\end{equation}
Before reciprocal capture,
\begin{equation}
f_{\rm FB}\simeq0,
\qquad
I_F^{(N)}\simeq NI_0,
\end{equation}
whereas reciprocal saturation gives
\begin{equation}
f_{\rm FB}\longrightarrow1,
\qquad
I_F^{(N)}\longrightarrow N^2I_0.
\label{eq:Fisher-saturated}
\end{equation}
Within this closure, the $N^2$ scaling results from saturation of the pair-score bound after projection onto the physical reciprocal overlap.

\subsection{Resolution-independent information velocity}

At internal resolution $N$, the intrinsic Schr\"odinger--Nagasawa spatial coefficient of an elementary possibility is $\hbar/(4m_0)$.  Combining it with the $1/N$ internal-time rescaling gives the effective diffusivity entering the self-similar information current,
\begin{equation}
D_N^{(0)}=\frac{\hbar}{4Nm_0}.
\label{eq:DN-definition}
\end{equation}

The local information current associated with the $a$th possibility is
\begin{equation}
\bm j_a
=
-D_N^{(0)}\nabla p_a,
\label{eq:information-current}
\end{equation}
and the corresponding local transport velocity is
\begin{equation}
\bm v_a
=
\frac{\bm j_a}{p_a}
=
-D_N^{(0)}\nabla\ln p_a
=
-D_N^{(0)}\bm s_a.
\label{eq:local-information-velocity}
\end{equation}
Consequently
\begin{equation}
v_F^2(T)
=
\left[D_N^{(0)}\right]^2I_F^{(N)}(T).
\label{eq:Fisher-velocity-v2}
\end{equation}
Using Eq.~\eqref{eq:IF-interpolation},
\begin{equation}
v_F^2(T)
=
\frac{\hbar^2I_0}{16m_0^2}
\left[
\frac1N+
\left(1-\frac1N\right)f_{\rm FB}(T)
\right].
\label{eq:vF-interpolation}
\end{equation}

A resolution-independent saturated fixed point exists if the elementary mass--length product
\begin{equation}
m_0\ell_0=\kappa_I
\label{eq:invariant-product}
\end{equation}
is invariant under refinement.  The saturated information-velocity scale is then
\begin{equation}
c_\star=\frac{\hbar}{4\kappa_I}.
\label{eq:cstar-v2}
\end{equation}
Equation~\eqref{eq:vF-interpolation} becomes
\begin{equation}
v_F^2(T)
=
c_\star^2
\left[
\frac1N+
\left(1-\frac1N\right)f_{\rm FB}(T)
\right].
\label{eq:vF-cstar-interpolation}
\end{equation}
Cauchy--Schwarz therefore gives the intrinsic bound
\begin{equation}
v_F^2(T)\leq c_\star^2,
\label{eq:Fisher-bound}
\end{equation}
and at reciprocal saturation
\begin{equation}
f_{\rm FB}(T)\longrightarrow1
\qquad\Longrightarrow\qquad
v_F(T)\longrightarrow c_\star.
\label{eq:Fisher-saturation-cstar}
\end{equation}
The limiting speed is thus fixed by the intrinsic Schr\"odinger--Nagasawa/Fisher construction and by saturation of the reciprocal pair weight.  A generic Brownian identity $v^2=D^2I_F$ would not by itself select this fixed point: the BSM construction additionally uses the Bohm--Fisher relation, the mass-resolution scaling and the $F/B$ capture fraction.  The correlation-dependent factor $\zeta$ does not enter $c_\star$ but controls how the dressed relative sector realizes the same limiting speed.

\subsection{Dressed renewal clock and fixed-point matching}

The Fisher construction of
the previous subsection is formulated in terms of the resolved
distributions of possible continuations and their captured pair
correlations.  It therefore provides a global information-current
velocity,
\begin{equation}
v_F(T),
\end{equation}
whose saturated value is the resolution-independent scale
\begin{equation}
v_F(T)\longrightarrow c_\star
\qquad
\text{for}
\qquad
f_{\rm FB}(T)\longrightarrow1.
\label{eq:Fisher-saturated-speed-reminder}
\end{equation}

By contrast, the 2PI infrared reduction describes the same organized
reciprocal sector through its local dressed kernel,
\begin{equation}
\widehat{\mathcal K}_{\rm rel}^{\rm IR}(q,T)
=
h_{\rm eff}(T)q^2+M(T),
\qquad
h_{\rm eff}(T)=\zeta(T)h_0,
\end{equation}
with $M(T)>0$ on the localizing branch.  We retain
\begin{equation}
\Gamma_B(T)\equiv M(T)
\label{eq:effective-branching-frequency}
\end{equation}
as the dressed local renewal frequency of the organized reciprocal
sector.  The complete $M(T)$ contains the stationary, cubic and quartic
2PI contributions of Eq.~\eqref{eq:total-branching-mass}.

The corresponding dressed renewal duration is
\begin{equation}
\tau_B(T)=\frac{1}{M(T)}
\label{eq:branching-duration}
\end{equation}
and the spatial transfer coefficient associated with the same relative
kernel is
\begin{equation}
\mathcal D_{\rm tr}(T)
=
4h_{\rm eff}(T)
=
\zeta(T)\frac{\hbar}{m}.
\label{eq:transfer-diffusivity}
\end{equation}
Thus
\begin{equation}
\ell_B(T)
=
\sqrt{\mathcal D_{\rm tr}(T)\tau_B(T)},
\label{eq:branching-transfer-length}
\end{equation}
and the corresponding local renewal-transfer velocity is
\begin{equation}
v_B^2(T)
=
\mathcal D_{\rm tr}(T)M(T)
=
4h_{\rm eff}(T)M(T).
\label{eq:branching-transfer-velocity}
\end{equation}

No equality between $v_F(T)$ and $v_B(T)$ is assumed during the
transient formation regime.  The former follows the global capture of
reciprocal weight, whereas the latter is built from the instantaneous
dressed infrared response kernel.

The two velocities introduced above characterize the same reciprocal
organization at different statistical levels.  At reciprocal saturation, however, both constructions refer to the same
stationary organized sector.  The Fisher description has then exhausted
the available pair-information weight, while the 2PI coefficients have
reached their dressed fixed point.  A single saturated reciprocal state
should therefore possess a unique long-wavelength information-transfer
scale.  We consequently formulate the fixed-point consistency condition
\begin{equation}
v_{B,\infty}
=
v_{F,\infty}
=
c_\star.
\label{eq:Fisher-renewal-fixed-point-matching}
\end{equation}
This relation is not imposed as a microscopic identity throughout the
evolution.  It is a matching condition between the global Fisher
description and the local 2PI description of the same saturated
reciprocal state.

It is then useful to introduce the intrinsic mass clock
\begin{equation}
\Gamma_m
\equiv
\frac{mc_\star^2}{\hbar}
\label{eq:intrinsic-mass-clock}
\end{equation}
which is defined independently of the 2PI dressing. At this stage,
$\Gamma_m$ is a frequency scale constructed from $m$ and $c_\star$.
Its identification with a collective gap is an additional matching
condition examined in Sec.~\ref{sec:relativistic-fixed-point}. Since
\begin{equation}
4h_0=\frac{\hbar}{m},
\end{equation}
one has
\begin{equation}
c_\star^2
=
4h_0\Gamma_m.
\label{eq:bare-clock-speed}
\end{equation}

Let
\begin{equation}
\zeta_\infty
=
\lim_{T\to\infty}\zeta(T).
\end{equation}
Taking the saturated limit of
Eq.~\eqref{eq:branching-transfer-velocity} and imposing
Eq.~\eqref{eq:Fisher-renewal-fixed-point-matching} gives
\begin{equation}
4h_{\rm eff,\infty}M_\infty
=
4h_0\Gamma_m.
\end{equation}
Using $h_{\rm eff,\infty}=\zeta_\infty h_0$, this becomes
\begin{equation}
\zeta_\infty M_\infty
=
\Gamma_m.
\label{eq:dressed-clock-matching}
\end{equation}
Hence
\begin{equation}
M_\infty
=
\Gamma_{B,\infty}
=
\frac{\Gamma_m}{\zeta_\infty}
=
\frac{mc_\star^2}{\zeta_\infty\hbar}.
\label{eq:renormalized-branching-fixed-point}
\end{equation}

The intrinsic clock $\Gamma_m$ belongs to the Fisher fixed point and is not
renormalized by the 2PI covariance.  The local 2PI description instead
redistributes the same saturated transport scale between spatial
mobility and renewal frequency,
\begin{equation}
h_0
\longrightarrow
h_{\rm eff,\infty}
=
\zeta_\infty h_0,
\qquad
\Gamma_m
\longrightarrow
M_\infty
=
\frac{\Gamma_m}{\zeta_\infty}.
\end{equation}
Their product is therefore preserved,
\begin{equation}
4h_{\rm eff,\infty}M_\infty
=
4h_0\Gamma_m
=
c_\star^2.
\end{equation}
Thus $\zeta_\infty$ dresses the internal spatial and temporal
organization of the localized reciprocal sector without redefining
either the intrinsic information speed $c_\star$ or the conserved mass
parameter $m$.

The saturated renewal duration and transfer length are consequently
\begin{align}
\tau_{B,\infty}
=
\frac{1}{M_\infty}
=
\frac{\zeta_\infty\hbar}{mc_\star^2}
,
\label{eq:tauB-infty}
\\
\ell_{B,\infty}
=
\sqrt{\mathcal D_{\rm tr,\infty}\tau_{B,\infty}}
=
\frac{\zeta_\infty\hbar}{mc_\star}
.
\label{eq:lB-infty}
\end{align}

The screened pole itself has
\begin{equation}
\xi_\infty
=
\sqrt{\frac{h_{\rm eff,\infty}}{M_\infty}}
=
\frac{\zeta_\infty}{2}
\frac{\hbar}{mc_\star},
\qquad
\ell_{B,\infty}=2\xi_\infty.
\label{eq:xi-dressed-compton}
\end{equation}
Thus the localization scale is Compton-like but need not equal the bare
Compton length.  The 2PI correlation dressing supplies the factor
$\zeta_\infty$, while the factor $1/2$ follows from the convention
$h_0=\hbar/(4m)$.
The transfer length defines the information-resolution wave number $q_I$.
\begin{equation}
q_I
=
\frac{1}{\ell_{B,\infty}}
=
\frac{mc_\star}{\zeta_\infty\hbar}
\label{eq:qI}
\end{equation}
and therefore
\begin{equation}
q_I c_\star\tau_{B,\infty}=1.
\end{equation}

At saturation, further microscopic branching no longer extends the
captured reciprocal measure.  With the forward--backward capture
fraction of Sec.~\ref{sec:avrami},
\begin{equation}
\Gamma_{\rm new}(T)
=
\left[1-f_{\rm FB}(T)\right]M(T).
\label{eq:new-information-rate}
\end{equation}
Hence
\begin{equation}
f_{\rm FB}(T)\longrightarrow1
\qquad\Longrightarrow\qquad
\Gamma_{\rm new}(T)\longrightarrow0,
\end{equation}
while
\begin{equation}
M(T)\longrightarrow M_\infty
\end{equation}
remains finite.  The late-time regime is therefore a renewal fixed point:
microscopic renewal persists at a finite dressed frequency while no new
reciprocal weight remains to be captured.

The saturated dressed scales are summarized by
\begin{equation}
M_\infty
=
\frac{\Gamma_m}{\zeta_\infty},
\qquad
\tau_{B,\infty}
=
\frac{\zeta_\infty\hbar}{mc_\star^2},
\qquad
\ell_{B,\infty}
=
\frac{\zeta_\infty\hbar}{mc_\star}.
\label{eq:branching-cutoff-triplet}
\end{equation}
The mass scaling remains intrinsic, while $\zeta_\infty$ controls the
spatial and temporal resolution of the correlated relative sector rather
than renormalizing the definition of $c_\star$ or the conserved mass
parameter $m$.
\section{Global collective poles and two-sector propagation}
\label{sec:global-collective-poles}

\subsection{From the Born--Oppenheimer hierarchy to the global field}

The two-point kernel combines two response legs. Their frequency difference
describes a slow beat, while their sum retains a faster internal oscillation.
The following calculation identifies these two frequency combinations within
the instantaneous adiabatic closure.

The previous sections used the Born--Oppenheimer hierarchy to organize
the two frequency scales contained in the exact factorization of the
localized connected correlator. The combination
\begin{equation}
\varepsilon_x+\varepsilon_y
\end{equation}
sets the fast internal scale and determines the relative structure of
the cluster, whereas
\begin{equation}
\varepsilon_x-\varepsilon_y
\end{equation}
sets the slow scale associated with its collective propagation.

This interpretation is controlled by the hierarchy
\begin{equation}
\left|
\varepsilon_x-\varepsilon_y
\right|,
\ |\Omega|
\ll
\left|
\varepsilon_x+\varepsilon_y
\right|.
\label{eq:BO-global-hierarchy}
\end{equation}
The relative sector fixes the internal profile and correlation length,
while the slow sector transports the resulting correlated structure in
the collective variables $(Q,\Omega)$.

We now use the same reciprocal correlator from a global viewpoint.  Both
frequency combinations are retained simultaneously.  Since $C_{UU}$ and
$C_{WW}$ have the same response denominator, their pole positions are common, while the physical forward--backward kernel is reconstructed from their residues,
\begin{equation}
C_{\rm FB}^{\rm eq}(q,Q;\Omega,T)
\simeq
\frac14\left(C_{UU}^{\rm eq}-C_{WW}^{\rm eq}\right).
\end{equation}
The relative momentum $q$ remains the internal momentum of the correlator,
while $(Q,\Omega)$ describe its collective propagation.  The center-of-mass
response is obtained after projection over $q$.

\subsection{Exact factorization of the global denominator}

We now restore the diagonal coefficient $E$ in the frequency factor. After integration over the relative frequency, the equal-time correlator has the general form
\begin{equation}
C_{WW}^{\rm eq}(q,Q;\Omega,T)
=
\sum_{\sigma=\pm1}
\frac{
\mathcal R_\sigma(q,Q;\Omega,T)
}{
\mathcal D_\sigma(q,Q;\Omega,T)
},
\label{eq:global-eq-correlator}
\end{equation}
where $\mathcal R_\sigma$ contains the effective noises and the
frequency residues. The remaining denominator is
\begin{equation}
\mathcal D_\sigma
=
M_\sigma^2+\varepsilon_x^2,
\label{eq:global-Dsigma}
\end{equation}
with
\begin{align}
M_\sigma
&=
E(q,T)(s_++s_-)
-
\ii
\left[
\frac{\Omega}{2}
+
\sigma \varepsilon_y
\right],
\label{eq:global-Msigma}
\\
\varepsilon_x
&=
\sqrt{P_x},\qquad\qquad\qquad
\varepsilon_y=
\sqrt{P_y},
\label{eq:global-pxpy}
\end{align}
and
\begin{equation}
P_x=B_xC_x,
\qquad
P_y=B_yC_y.
\label{eq:global-PxPy}
\end{equation}

The denominator factorizes exactly as
\begin{align}
\mathcal D_\sigma
&=
\left\{
E(q,T)(s_++s_-)
-
\frac{\ii}{2}
\left[
\Omega
-
2\sigma(\varepsilon_x-\varepsilon_y)
\right]
\right\}
\nonumber\\
&\quad\times
\left\{
E(q,T)(s_++s_-)
-
\frac{\ii}{2}
\left[
\Omega
+
2\sigma(\varepsilon_x+\varepsilon_y)
\right]
\right\}.
\label{eq:global-factorization}
\end{align}

Introducing the shifted collective frequency
\begin{equation}
\overline{\Omega}
=
\Omega
+
2\ii E(q,T)(s_++s_-),
\label{eq:shifted-global-frequency}
\end{equation}
the four linear poles are organized into the two quadratic kernels
\begin{align}
\mathcal K_-(q,Q;\Omega,T)
&=
\overline{\Omega}^{\,2}
-
4(\varepsilon_x-\varepsilon_y)^2,
\label{eq:global-Kminus-exact}
\\
\mathcal K_+(q,Q;\Omega,T)
&=
\overline{\Omega}^{\,2}
-
4(\varepsilon_x+\varepsilon_y)^2.
\label{eq:global-Kplus-exact}
\end{align}

In the conservative limit $E=0$, the corresponding squared
frequencies are
\begin{equation}
\begin{aligned}
\omega_-^2(q,Q,T)
&=
4
\left[
\varepsilon_x(q,Q,T)-\varepsilon_y(q,Q,T)
\right]^2,
\\
\omega_+^2(q,Q,T)
&=
4
\left[
\varepsilon_x(q,Q,T)+\varepsilon_y(q,Q,T)
\right]^2.
\end{aligned}
\label{eq:global-two-frequencies}
\end{equation}

At zero collective momentum,
\begin{equation}
P_x(q,0,T)
=
P_y(q,0,T)
=
P(q,T),
\end{equation}
and therefore
\begin{equation}
\varepsilon_x(q,0,T)
=
\varepsilon_y(q,0,T)
=
\sqrt{P(q,T)}.
\end{equation}
The two branches satisfy
\begin{align}
\omega_-^2(q,0,T)
&=
0,
\label{eq:global-massless-branch}
\\
\omega_+^2(q,0,T)
&=
16P(q,T).
\label{eq:global-massive-branch}
\end{align}

The difference branch is massless at $Q=0$. The sum branch carries the
finite internal frequency
\begin{equation}
\omega_{\rm int}(q,T)
=
4\sqrt{P(q,T)}.
\label{eq:global-internal-frequency}
\end{equation}

\subsection{Two-sector form of the reciprocal correlator}

In the conservative limit, the two diagonal channel correlators can be written
with the same pole families,
\begin{align}
C_{WW}^{\rm pole}
&=
\frac{2\Omega Z_-^{W}}{\Omega^2-\omega_-^2}
+
\frac{2\Omega Z_+^{W}}{\Omega^2-\omega_+^2},
\label{eq:global-two-sector-W}
\\
C_{UU}^{\rm pole}
&=
\frac{2\Omega Z_-^{U}}{\Omega^2-\omega_-^2}
+
\frac{2\Omega Z_+^{U}}{\Omega^2-\omega_+^2}.
\label{eq:global-two-sector-U}
\end{align}
The regular pieces are collected separately and remain regular at both pole
families.  After removing the common causal normalization, the reduced
$Q=0$ residues are
\begin{align}
\widehat Z_-^{W}
&=
N_{WW}^{\rm eff}
+
\frac{C}{B}N_{UU}^{\rm eff},
&
\widehat Z_+^{W}
&=
N_{WW}^{\rm eff}
-
\frac{C}{B}N_{UU}^{\rm eff},
\label{eq:global-reduced-ZW}
\\
\widehat Z_-^{U}
&=
N_{UU}^{\rm eff}
+
\frac{B}{C}N_{WW}^{\rm eff},
&
\widehat Z_+^{U}
&=
N_{UU}^{\rm eff}
-
\frac{B}{C}N_{WW}^{\rm eff}.
\label{eq:global-reduced-ZU}
\end{align}
For the parity-even reciprocal projection, the physical pole weights are
therefore
\begin{equation}
Z_\pm^{\rm FB}
=
\frac14\left(Z_\pm^{U}-Z_\pm^{W}\right).
\label{eq:global-ZFB-definition}
\end{equation}
The pole contribution to the forward--backward kernel is
\begin{align}
C_{\rm FB}^{\rm pole}(q,Q;\Omega,T)
&=\frac{2\Omega Z_-^{\rm FB}(q,Q,T)}
{\Omega^2-\omega_-^2(q,Q,T)}
\nonumber\\
&\quad+\frac{2\Omega Z_+^{\rm FB}(q,Q,T)}
{\Omega^2-\omega_+^2(q,Q,T)}.
\label{eq:global-two-sector-exact}
\end{align}
Thus the global gapless and gapped sectors are properties of the same physical
reciprocal observable.  The $U$ and $W$ projections determine how much weight
that observable carries on each pole family.

\subsection{Simultaneous long-wavelength expansion}

The collective dispersion is obtained by expanding the two pole
products simultaneously. Exchange symmetry between the two legs gives
\begin{align}
P_x(q,Q,T)
&=
P(q,T)
+
\alpha(q,T)Q
+
\beta(q,T)Q^2
+
\mathcal O(Q^3),
\label{eq:global-Px-expansion}
\\
P_y(q,Q,T)
&=
P(q,T)
-
\alpha(q,T)Q
+
\beta(q,T)Q^2
+
\mathcal O(Q^3).
\label{eq:global-Py-expansion}
\end{align}
The coefficients are defined by
\begin{align}
\alpha(q,T)
&=
\left.
\partial_QP_x(q,Q,T)
\right|_{Q=0},
\label{eq:global-alpha-definition}
\\
\beta(q,T)
&=
\left.
\frac{1}{2}
\partial_Q^2P_x(q,Q,T)
\right|_{Q=0}.
\label{eq:global-beta-definition}
\end{align}

The odd coefficient $\alpha$ describes the splitting of the two legs,
whereas the even coefficient $\beta$ describes their common
collective curvature. Expanding the square roots gives
\begin{align}
\varepsilon_x-\varepsilon_y
&=
\frac{
\alpha(q,T)
}{
\sqrt{P(q,T)}
}
Q
+
\mathcal O(Q^3),
\label{eq:global-root-difference}
\\
\varepsilon_x+\varepsilon_y
&=
2\sqrt{P(q,T)}
\nonumber\\
&\quad+
\left[
\frac{
\beta(q,T)
}{
\sqrt{P(q,T)}
}
-
\frac{
\alpha^2(q,T)
}{
4P^{3/2}(q,T)
}
\right]
Q^2
+
\mathcal O(Q^4).
\label{eq:global-root-sum}
\end{align}

The difference branch therefore has the infrared form
\begin{equation}
\omega_-^2(q,Q,T)
=
v_-^2(q,T)Q^2
+
\mathcal O(Q^4),
\label{eq:global-minus-dispersion}
\end{equation}
with
\begin{equation}
v_-^2(q,T)
=
\frac{
4\alpha^2(q,T)
}{
P(q,T)
}.
\label{eq:global-vminus}
\end{equation}

The sum branch becomes
\begin{equation}
\omega_+^2(q,Q,T)
=
16P(q,T)
+
v_+^2(q,T)Q^2
+
\mathcal O(Q^4),
\label{eq:global-plus-dispersion}
\end{equation}
with
\begin{equation}
v_+^2(q,T)
=
16\beta(q,T)
-
\frac{
4\alpha^2(q,T)
}{
P(q,T)
}.
\label{eq:global-vplus}
\end{equation}
The two coefficients obey
\begin{equation}
v_+^2(q,T)+v_-^2(q,T)
=
16\beta(q,T).
\label{eq:global-velocity-sum}
\end{equation}

The massless branch is controlled by the antisymmetric splitting of the two legs. The gapped sum branch depends additionally on their common collective curvature.
Where $v_-^2(q,T)>0$, the gapless dispersion is linear at small
collective momentum, $|\Omega|\propto|Q|$. In the notation
$|\Omega|\propto|Q|^z$, this corresponds to the dynamical exponent $z=1$.

\subsection{Local collective coefficients}

For local 2PI coefficients without explicit $Q$ dependence, we define
\begin{align}
B(q,T)
&=
-\left[h_{\rm eff}(T)q^2+M(T)\right],
\nonumber\\
C(q,T)
&=
h_{\rm eff}(T)s(q)-D(q,T)s^2(q),
\label{eq:global-local-BC}
\end{align}
with
\begin{equation}
s(q)=k_n^2-q^2,
\qquad
P(q,T)=B(q,T)C(q,T).
\label{eq:global-local-P}
\end{equation}
The linear coefficient is
\begin{equation}
\alpha(q,T)
=
-h_{\rm eff}q\,[B(q,T)+C(q,T)].
\label{eq:global-alpha-local}
\end{equation}
The local quadratic curvature is
\begin{align}
\beta(q,T)
&=
B(q,T)
\left[-\frac{h_{\rm eff}}4
+\frac{D(q,T)}2\left(k_n^2+q^2\right)\right]
\nonumber\\
&\quad
-\frac{h_{\rm eff}}4C(q,T)
+h_{\rm eff}^2q^2.
\label{eq:global-beta-kin}
\end{align}

\subsection{Infrared propagation equation}

Substituting the long-wavelength dispersions into
Eq.~\eqref{eq:global-two-sector-exact} gives
\begin{equation}
\begin{aligned}
C_{\rm FB}^{\rm pole}(q,Q;\Omega,T)
&=
\frac{2\Omega Z_-^{\rm FB}(q,T)}
{\Omega^2-v_-^2(q,T)Q^2}
\\
&\quad+
\frac{2\Omega Z_+^{\rm FB}(q,T)}
{\Omega^2-16P(q,T)-v_+^2(q,T)Q^2}
\\
&\quad+
\mathcal O(Q^4).
\end{aligned}
\label{eq:global-CFB-two-sector-IR}
\end{equation}
The first term is the massless difference sector,
\begin{equation}
\mathcal K_-(q,Q,\Omega,T)
=
\Omega^2-v_-^2(q,T)Q^2,
\label{eq:global-Kminus-IR}
\end{equation}
and the second is the gapped sum sector,
\begin{equation}
\mathcal K_+(q,Q,\Omega,T)
=
\Omega^2-16P(q,T)-v_+^2(q,T)Q^2.
\label{eq:global-Kplus-IR}
\end{equation}
At fixed relative momentum, the reciprocal pole contribution satisfies
\begin{align}
\mathcal K_-\mathcal K_+
C_{\rm FB}^{\rm pole}
&=
2\Omega
\left[
Z_-^{\rm FB}\mathcal K_+
+
Z_+^{\rm FB}\mathcal K_-
\right].
\label{eq:global-CFB-field-equation}
\end{align}
The homogeneous propagation condition remains
\begin{equation}
\mathcal K_-(q,Q,\Omega,T)
\mathcal K_+(q,Q,\Omega,T)
=0.
\label{eq:global-homogeneous-equation}
\end{equation}
The two allowed collective branches are therefore unchanged by passing from
the channel projections to $C_{\rm FB}$, since this step only recombines their physical residues.


\subsection{One-mode Born--Oppenheimer projection of the collective kernels}

The $q$-resolved kernels above still retain the internal
coordinate of the localized reciprocal object.  A collective field that does
not resolve this internal coordinate is obtained by projecting the
\emph{inverse response operator}, rather than by averaging already inverted
propagators.  Let $\varphi_\xi(q,T)$ denote the normalized localized internal Born--Oppenheimer (BO) profile
\begin{equation}
\int\frac{\dd q}{2\pi}\,
|\varphi_\xi(q,T)|^2=1.
\label{eq:BO-internal-mode-normalization}
\end{equation}
The internal profile is chosen to represent the screened relative sector
identified in Sec.~\ref{sec:relative-localization}. Its use as a single
projector specifies the one-mode approximation made here.
At leading one-mode Born--Oppenheimer order we define
\begin{equation}
\mathcal K_\pm^{\rm BO}(Q,\Omega,T)
=
\left\langle
\varphi_\xi
\left|
\mathcal K_\pm(q,Q,\Omega,T)
\right|
\varphi_\xi
\right\rangle_q .
\label{eq:BO-projected-kernel-definition}
\end{equation}
In the conservative sector $E=0$, the simultaneous small-$Q$
expansion therefore gives
\begin{align}
\mathcal K_-^{\rm BO}
&=
\Omega^2-c_-^2(T)Q^2+\mathcal O(Q^4),
\nonumber\\
\mathcal K_+^{\rm BO}
&=
\Omega^2-\Omega_g^2(T)-c_+^2(T)Q^2+\mathcal O(Q^4),
\label{eq:BO-projected-collective-kernels}
\end{align}
with the \emph{same} internal projector in both sectors,
\begin{equation}
\begin{aligned}
c_-^2(T)
&=
\left\langle v_-^2(q,T)\right\rangle_{\xi,T},
\\
c_+^2(T)
&=
\left\langle v_+^2(q,T)\right\rangle_{\xi,T},
\\
\Omega_g^2(T)
&=
16\left\langle P(q,T)\right\rangle_{\xi,T},
\end{aligned}
\label{eq:BO-projected-collective-coefficients}
\end{equation}
where
$\langle F\rangle_{\xi,T}\equiv
\langle\varphi_\xi|F|\varphi_\xi\rangle_q$.
This is a one-mode projection of the inverse kernel.  Couplings to internal modes orthogonal to
$\varphi_\xi$ constitute the higher Born--Oppenheimer corrections to
this reduction.  The derivative coefficient $D(q,T)$ is retained in
$P$, $\alpha$, $\beta$, and hence in $v_\pm$ before the projection. The leading BO regime assumes only the conservative condition $E=0$ and the
bandwise hierarchy of Eq.~\eqref{eq:infrared-derivative-hierarchy}.

\subsection{Physical interpretation}

The same reciprocal connected kernel supports two collective responses.
The difference of the two leg frequencies generates a massless propagation
sector, while their sum generates a gapped sector with
gap
\begin{equation}
4\sqrt{P(q,T)}.
\end{equation}
The latter is inherited from the internal dynamics that structures the
cluster in the Born--Oppenheimer representation.

Because the relative profile has a finite internal spectral width, the
collective response is not fixed by the single value $q=0$.  The fixed-$q$
pole families derived above are retained, while the common one-mode
Born--Oppenheimer projection of Eq.~\eqref{eq:BO-projected-kernel-definition}
defines the collective parameters of both branches from the same localized internal BO profile.
\section{Discussion}
\label{sec:discussion}

\subsection{From the frozen hierarchy of Ref.~\cite{dumonteil_branching_2026} to a self-consistent
connected theory}

Ref.~\cite{dumonteil_branching_2026} introduced the statistical separation underlying branching
stochastic mechanics in terms of two stochastic reciprocal fields
\(\Phi_F\) and \(\Phi_B\). In the common reciprocal basis, the relevant
joint observable is the centered signed kernel
\(C_{\rm FB}(x,y)=\mathbb E_\omega[\psi_F(x)\psi_B(y)]\). On the
anticorrelated branch its negative diagonal defines the organized connected weight, while its off-diagonal dependence carries the reciprocal
relative structure.
The one-sector genealogical covariance is a distinct object describing
common ancestry inside either marginal superprocess.  Within the
frozen-rate hierarchy the Bohm/Fisher rate was prescribed, while the
joint reciprocal section of Ref.~\cite{dumonteil_branching_2026} identified the diagonal-preserving
channel in which a finite relative response could appear.
The reciprocal closure of that reference introduced an anti-correlated
forward--backward covariance and isolated a trace-free shape sector.
When supplied with a positive reciprocal screening coefficient $\mu_{\rm FB}$, this
sector is screened over the characteristic length
\begin{equation}
\xi_{\rm FB}^2
\sim
\frac{D_{\rm eff}}{\mu_{\rm FB}}.
\label{eq:discussion-paperI-screening}
\end{equation}
Here $D_{\rm eff}$ is the effective relative diffusion coefficient of the closure in Ref.~\cite{dumonteil_branching_2026} and $\mu_{\rm FB}$ is its reciprocal screening coefficient. This established the structural possibility of a localized connected
 sector while holding the extended Schr\"odinger field fixed at the
first-moment level within that closure. The strength of the reciprocal covariance, the
value of the screening coefficient, and the autonomous propagation of the resulting
correlated object were left to a self-consistent response-field
formulation.

The present paper develops that formulation. The fluctuating
Bohm/Fisher functional is promoted from a prescribed branching rate
to a nonlinear feedback acting on the connected hierarchy. The cubic
response--correlation loop and the quartic Hartree contraction jointly dress
the crossed response channel, while the Hartree covariance also renormalizes
the relative stiffness. Their normalization and central-time dependence are
fixed by the self-consistent response and covariance blocks. The two papers
therefore address successive levels of the same construction:
\begin{equation}
\begin{aligned}
\text{Ref.~\cite{dumonteil_branching_2026}:}\quad&
\begin{gathered}
\text{existence and structure of the}\\
\text{connected overlap sector},
\end{gathered}
\\[2pt]
\text{this work:}\quad&
\begin{gathered}
\text{self-consistent feedback,}\\
\text{saturation, and propagation}.
\end{gathered}
\end{aligned}
\label{eq:discussion-paper-sequence}
\end{equation}

\subsection{Response-field construction and 2PI closure}

Starting from the stochastic Schr\"odinger--Nagasawa pair, the MSRJD
construction~\cite{Martin1973,Janssen1976,DeDominicis1976} introduces the
physical and response fields required to formulate the connected stochastic
dynamics.  The confined quadratic theory gives a useful exact benchmark at
fixed modal cutoff: the retarded response remains oscillatory, whereas the
equal-time connected covariance grows secularly under continuous branching
noise and the derivative Bohm/Fisher vertices enhance the high-mode
contributions.  The resulting temporal and ultraviolet accumulation motivates
a self-consistent resummation of response and correlation functions.

The 2PI effective action provides such a closure
~\cite{Cornwall1974,Berges2004,AartsBerges2001,Bode2022}.  Bode's MSR--2PI
construction for classical stochastic processes gives the general
nonequilibrium framework in which response functions and second cumulants are
evolved self-consistently.  Here the same strategy is specialized to the
Bohm/Fisher interaction generated by BSM.  The primitive branching covariance
is evaluated on the prescribed Schr\"odinger--Nagasawa background and retained
as the quadratic noise kernel, while $\Gamma_2$ is truncated to the
Bohm/Fisher drift vertices.  This background-noise truncation isolates the
feedback mechanism studied in this work without freezing the stochastic sector:
the auxiliary response--response block is kept during the functional variation
so that the Bohm $HCC$ topology generates a dressed physical noise before the
MSRJD condition $H=0$ is imposed.  The field-dependent multiplicative-noise
vertex of the complete process of Ref.~\cite{dumonteil_branching_2026} is a systematic extension not included
in the present closure.

The cubic Bohm vertex generates the three retarded families
\begin{equation}
D(q,T),\qquad E(q,T),\qquad \mu_{2PI}(q,T),
\end{equation}
and its off-shell $HCC$ completion generates the corresponding colored-noise
correction.  The quartic Bohm vertex adds the Hartree coefficients $a_4(T)$
and $M_4(T)$.  The coefficient $D$ controls the higher-gradient continuation
of the relative kernel and $E$ dresses its diagonal dynamics.  The local equal-time connected $ww$ covariance entering the Hartree contraction defines
\begin{equation}
\zeta(T)
=
1+2C_{ww}(r=0,T)
=
1+2\langle w^2\rangle_c,
\end{equation}
so that
\begin{equation}
h_{\rm eff}(T)
=
h_0\zeta(T).
\end{equation}
The dimensionless factor $\zeta(T)$ provides the dynamic stiffness
renormalization while preserving the inverse-mass scaling of the
Schr\"odinger--Nagasawa coefficient.

The Hartree contribution $M_4$ is an equal-time covariance contraction and
$\mu_{2PI}$ is generated by a mixed response--correlation loop.  Together with
the stationary reference contribution they define the intrinsic local
frequency
\begin{equation}
M(T)=M_4(T)-\mu_{2PI}(T)-h_{\rm eff}(T)k_n^2.
\label{eq:discussion-total-M}
\end{equation}
This is the dressed local renewal frequency of the 2PI relative sector.  Its time dependence is fixed by the dressed response and covariance, while the separate terms remain useful diagnostics of the 2PI feedback.

\subsection{Relative localization and bounded formation}

The Born--Oppenheimer factorization separates the slow collective coordinate
from the internal relative structure of the reciprocal kernel.  In the leading
finite-well residue reduction, the $U$ and $W$ channels share a common
stochastic contribution that cancels from
$C_{\rm FB}\simeq(C_{UU}-C_{WW})/4$.  The remaining reciprocal weight contains
two distinct relative sectors: an extended reference-mode contribution with
poles at $q=\pm k_n$, and a screened contribution controlled by
\begin{equation}
h_{\rm eff}(T)q^2+M(T).
\end{equation}
The latter defines
\begin{equation}
\xi(T)=\sqrt{\frac{h_{\rm eff}(T)}{M(T)}}.
\label{eq:discussion-xi-instantaneous}
\end{equation}
The higher-gradient coefficient $D$ is retained in the finite-reference
$U$-residue denominator and modifies the extended/short-distance continuation
without changing this screened pole or the definition of $\xi$.
The Born diagonal normalization belongs to the collective factor of the
Born--Oppenheimer separation.  With a normalized relative factor
$F_{\rm rel}(0,T)=1$,
\begin{equation}
C_{\rm FB}(X,r,T)
\simeq
-\rho_{\rm BSM}(X,T)F_{\rm rel}(r,T),
\end{equation}
 and the reciprocal saturation matching condition is
$\rho_{\rm BSM}(X,T)\to R^2(X)$.  The $k_n$ pole therefore retains the
extended relative reference structure, while the massive denominator adds the
new finite relative scale. Neither relative term alone is identified with the Born profile.

The temporal formation is described directly through the reconstructed reciprocal kernel.  Its bounded fraction $f_{\rm FB}(T)$ is fitted by a single Avrami law, while $U$ and $W$ remain internal diagnostics of the covariance decomposition.  The 2PI theory fixes the dressed spatial coefficients and the evolving renewal frequency, whereas the Avrami representation describes the bounded formation of the physical forward--backward weight in post-processing.

At late times,
\begin{equation}
M(T)\longrightarrow M_\infty,
\qquad
\xi(T)\longrightarrow
\xi_\infty
=
\sqrt{\frac{h_{\rm eff,\infty}}{M_\infty}}.
\end{equation}
The finite value of $M_\infty$ is interpreted through the branching clock and
Fisher scaling developed in Sec.~\ref{sec:selfsimilar}.

\subsection{Self-similarity and the saturated information scales}

The mass-resolution construction examines the same intrinsic Schr\"odinger--Nagasawa information measure at different internal resolutions,
\begin{equation}
m=Nm_0.
\label{eq:discussion-mass-resolution}
\end{equation}
The parameter $N$ labels a refinement of the measure-valued description rather than a number of physical constituents.  Under this refinement the elementary internal time decreases as $1/N$ while the bare spatial coefficient increases as $N$.
The elementary possibility $a$ carries a normalized local distribution $p_a$ and score $\bm s_a=\nabla\ln p_a$.  The reciprocal capture fraction $f_{\rm FB}$ extracted from $C_{\rm FB}$ selects the pairs that contribute coherently to the score covariance.  As $f_{\rm FB}\to1$, the pair contribution saturates the Cauchy--Schwarz bound and the Fisher content grows from order $N$ to order $N^2$.  Combining this scaling with the invariant elementary product
\begin{equation}
m_0\ell_0=\kappa_I
\end{equation}
gives the intrinsic, resolution-independent information speed
\begin{equation}
c_\star=\frac{\hbar}{4\kappa_I}.
\end{equation}
Introducing the intrinsic mass clock $\Gamma_m=mc_\star^2/\hbar$, the saturated dressed renewal matching gives
\begin{equation}
M_\infty
=\frac{\Gamma_m}{\zeta_\infty},
\qquad
\tau_{B,\infty}=\frac{\zeta_\infty\hbar}{mc_\star^2},
\qquad
\ell_{B,\infty}=\frac{\zeta_\infty\hbar}{mc_\star},
\label{eq:discussion-saturated-triplet}
\end{equation}
with
\begin{equation}
\xi_\infty=\frac{\zeta_\infty}{2}\frac{\hbar}{mc_\star}.
\end{equation}
The Hartree dressing therefore changes the internal renewal and localization scales while leaving the intrinsic $c_\star$ and the mass parameter $m$ unchanged.

The self-similar resolution construction does not by itself determine how the distinct pole sectors of the global connected correlator share this velocity or how the collective gap reconstructs the intrinsic mass clock.  Those relations are formulated below as fixed-point conditions of the Born--Oppenheimer-projected response.

\subsection{Global propagation of the reciprocal connected field}

The Born--Oppenheimer representation separates the internal relative structure
from its slow collective propagation.  The global representation retains both frequency factors of the instantaneous adiabatic kernel and reconstructs its two pole families.  Since the
$UU$ and $WW$ projections have common denominators, the physical reciprocal
kernel carries the same two branches with residues
\begin{equation}
Z_\pm^{\rm FB}
=
\frac14\left(Z_\pm^U-Z_\pm^W\right).
\end{equation}
In the conservative infrared regime,
\begin{align}
C_{\rm FB}^{\rm pole}(q,Q;\Omega,T)
&=
\frac{2\Omega Z_-^{\rm FB}(q,T)}
{\Omega^2-v_-^2(q,T)Q^2}
\nonumber\\
&\quad+
\frac{2\Omega Z_+^{\rm FB}(q,T)}
{\Omega^2-16P(q,T)-v_+^2(q,T)Q^2}
\nonumber\\
&\quad+
\mathcal O(Q^4).
\label{eq:discussion-two-pole-correlator}
\end{align}
The difference sector is massless at $Q=0$, whereas the sum sector carries the
internal gap
\begin{equation}
\omega_{\rm int}(q,T)=4\sqrt{P(q,T)}.
\label{eq:discussion-internal-gap}
\end{equation}
Their propagation coefficients are
\begin{align}
v_-^2(q,T)
&=
\frac{4\alpha^2(q,T)}{P(q,T)},
\label{eq:discussion-vminus}
\\
v_+^2(q,T)
&=
16\beta(q,T)-\frac{4\alpha^2(q,T)}{P(q,T)}.
\label{eq:discussion-vplus}
\end{align}
The coefficient $\alpha$ measures the antisymmetric splitting of the two legs,
while $\beta$ measures their common collective curvature.  The pole locations
are therefore inherited from the common response kernel, while the reciprocal
observable determines their physical weights through the difference of the
$U$ and $W$ residues.

\subsection{Reciprocal saturation and the relativistic fixed point}
\label{sec:relativistic-fixed-point}

The emergent relativistic interpretation is supported by several structures obtained independently in the present construction: the reciprocal forward--backward dynamics, the separation into a gapless and a gapped pole, the appearance of a finite intrinsic speed through Fisher saturation, and a finite correlation-dressed localization scale.  The purpose of this discussion is to state how these elements fit together while separating derived results from fixed-point matching.

The forward and backward sectors are the two temporal orientations of one reciprocal process.  Each marginal process separately has diffusive Gaussian support and therefore does not define a causal cone.  Their connected product carries the physical reciprocal weight.  In the reciprocal basis,
\begin{equation}
U=\psi+\psi^\dagger,
\qquad
W=\psi^\dagger-\psi,
\end{equation}
so that
\begin{equation}
4\psi\psi^\dagger=U^2-W^2.
\end{equation}
This indefinite two-sided structure belongs to reciprocal field space and is not by itself a spacetime Lorentz symmetry.  The backward field is a terminal reciprocal weight rather than a signal propagating into the past: it assigns physical compatibility to the field of forward possibilities.  The normalized fraction $f_{\rm FB}$ used in the Fisher construction is precisely the reduced measure of this organized overlap.  A possible continuation that carries no reciprocal weight does not contribute to the saturated pair information.

The fixed-$q$ analysis gives two response families of the same reciprocal kernel, with conservative spectral product
\begin{equation}
P(q,T)=B(q,T)C(q,T)
\end{equation}
and full $D(q,T)$ dependence retained in $C(q,T)$.  The difference family is gapless at $Q=0$, while the sum family carries the internal frequency $4\sqrt{P(q,T)}$.  Their collective coefficients are defined with the same localized Born--Oppenheimer projector,
\begin{equation}
c_\pm^2(T)
=\left\langle v_\pm^2(q,T)\right\rangle_{\xi,T},
\qquad
\Omega_g^2(T)=16\left\langle P(q,T)\right\rangle_{\xi,T}.
\label{eq:projected-gap-definition}
\end{equation}
The conservative BO reduction assumes $E=0$ and
$|D(q,T)s^2(q)|\ll|M(T)+h_{\rm eff}(T)k_n^2|$ over the internal band carrying $\varphi_\xi$.

Fisher scaling independently defines the intrinsic information velocity $c_\star$ and, together with the mass parameter, the intrinsic mass clock
\begin{equation}
\Gamma_m=\frac{mc_\star^2}{\hbar}.
\end{equation}
The 2PI theory separately produces the Hartree dressing $\zeta$, the dressed renewal coefficient $M$, and the projected gap $\Omega_g$.  The saturated soft-transfer condition gives $M_\infty=\Gamma_m/\zeta_\infty$.  The relation between the projected collective gap and these local quantities is not derived by the minimal infrared kernel alone and must therefore be stated as an additional fixed-point matching condition of the complete Born--Oppenheimer projection.

We hypothesize that the gapless and gapped poles describe two collective responses of the same saturated reciprocal structure. At Fisher saturation the captured scores are maximally aligned. We further impose, as a fixed-point matching hypothesis, that both collective branches inherit the same limiting spatial transport scale. The gapped branch additionally reconstructs the intrinsic mass clock rather than the dressed local renewal clock.

The resulting full-theory fixed-point conditions are
\begin{equation}
c_{+,\infty}=c_{-,\infty}=c_\star,
\quad
\Omega_{g,\infty}=\Gamma_m=\frac{mc_\star^2}{\hbar},
\quad
M_\infty=\frac{\Gamma_m}{\zeta_\infty}.
\label{eq:relativistic-fixed-point-matching}
\end{equation}
Equivalently, the gap matching predicts
\begin{equation}
\Omega_{g,\infty}=\zeta_\infty M_\infty.
\label{eq:gap-renewal-zeta-matching}
\end{equation}
Equation~\eqref{eq:gap-renewal-zeta-matching} is a fixed-point prediction to be tested against the complete frequency-dependent and internally projected 2PI response. It is not inferred solely from the local Lorentzian denominator $h_{\rm eff}q^2+M$. In particular, any ordering between the two projected velocities obtained after the further minimal specialization $D=E=k_n=0$ is a property of that minimal intrinsic closure only and does not constrain the complete BO-projected theory, where the finite-reference structure, derivative dressing, self-consistent memory corrections, and internal projection are retained.  The common-cone relation is therefore introduced only as a full-theory fixed-point matching condition.

Under these conditions the collective infrared kernels reduce to
\begin{align}
\mathcal K_{-,\ast}(\Omega,Q)
&\simeq
\Omega^2-c_\star^2Q^2,
\nonumber\\
\mathcal K_{+,\ast}(\Omega,Q)
&\simeq
\Omega^2-c_\star^2Q^2-\Gamma_m^2.
\label{eq:fixed-point-kernels}
\end{align}
Defining $\mathcal E=\hbar\Omega$ and $p_{\rm cm}=\hbar Q$, the gapped pole obeys
\begin{equation}
\mathcal E^2
=
p_{\rm cm}^{\,2}c_\star^2
+m^2c_\star^4.
\end{equation}
The Hartree factor therefore does not renormalize the conserved mass parameter appearing in the collective Klein--Gordon gap.  Instead it dresses the internal localization and renewal scales that connect the intrinsic mass clock to the 2PI relative dynamics.

The microscopic Gaussian support remains noncompact in the presence of the backward weight. The causal cone refers only to the infrared poles of the connected response.  The role of reciprocity is instead to determine which possible continuations carry physical pair weight.  Fisher saturation fixes the limiting speed only after this $F/B$ projection, and the relativistic cone arises when that intrinsic information speed is matched by the two collective pole branches.  Establishing a strict microscopic front velocity would require control of the full retarded kernel, not only its low-frequency expansion.

The saturated dressed renewal and transfer scales obey
\begin{equation}
\tau_\star=\frac{1}{M_\infty}=\frac{\zeta_\infty\hbar}{mc_\star^2},
\qquad
\ell_\star=\frac{\zeta_\infty\hbar}{mc_\star},
\qquad
\frac{\ell_\star}{\tau_\star}=c_\star.
\end{equation}
Writing $\lambda_{C,\star}=\hbar/(mc_\star)$, one has
\begin{equation}
\ell_\star=\zeta_\infty\lambda_{C,\star},
\qquad
\xi_\infty=\frac{\ell_\star}{2}=\frac{\zeta_\infty}{2}\lambda_{C,\star}.
\end{equation}
These are correlation-dressed internal scales associated with a physical mass $m$.

The relativistic interpretation therefore rests on the convergence of independent indicators: reciprocal $F/B$ saturation, the infrared crossover toward $z=1$, Fisher saturation of the intrinsic speed, an isolated projected massive pole, and a common fixed-point cone.  Quantitatively, the decisive checks become
\begin{equation}
\frac{c_{+,\rm pole}^2}{c_{-,\rm pole}^2}\longrightarrow1,
\qquad
\frac{\Omega_g}{\Gamma_m}\longrightarrow1,
\qquad
\frac{\Omega_g}{M}\longrightarrow\zeta_\infty.
\end{equation}

\subsection{Statistical hierarchy and the quantum-to-classical transition}

The preceding construction suggests a physical hierarchy organized by
 statistical order. Within the imposed background projection, the first
moment is the extended Schr\"odinger--Nagasawa
mean field, which propagates the distributed compatibility weight of the
possible continuations and reproduces Schr\"odinger dynamics. In the present nonrelativistic formulation the first finite collective
propagation cone appears in the connected sector, while the first-moment field
remains extended.

The connected second moment is the first level at which reciprocal organization,
a finite relative length, and a collective propagation kernel appear together.
Localization is carried by correlations between possible continuations while the
 first-moment Schr\"odinger field is held fixed by the closure. In this hierarchy the
quantum-to-classical transition is associated with the organization of higher
statistical moments of the same underlying dynamics.

This hierarchy suggests an interpretation of measurement. The one-point
sector may retain the extended quantum field, while the connected two-point
sector provides a candidate carrier of a localized correlated event. Establishing
this interpretation dynamically requires an explicit detector coupling and lies
beyond the present field-theoretical construction.

A future many-body formulation could test whether this statistical hierarchy
extends to entangled states. It leaves open the possibility that nonlocal quantum
correlations, including EPR-type structure, remain encoded in an extended
first-moment wave-functional sector while a finite collective cone characterizes
localized connected excitations at higher statistical order. Testing this possibility for EPR-type correlations requires a many-body
extension of the present one-body theory. The current construction identifies
the separation of statistical levels on which such a question can be posed.

The reciprocal forward--backward structure is essential to this
separation. Already in Ref.~\cite{dumonteil_branching_2026}, a real stationary mode cannot be assigned
a single directed momentum while preserving its vanishing current. Its
spectral scale can instead be resolved into the two counter-propagating
components
\begin{equation}
p_+
=
+\hbar k_n
\qquad
p_-
=
-\hbar k_n
\end{equation}
whose total momentum flux vanishes in the standing state. In the present
connected theory, the same reciprocal structure appears dynamically at
the level of the two frequency branches: their difference and sum produce
the gapless and gapped quadratic propagation kernels. The relativistic
structure discussed above thus appears as a possible collective property of the
organized reciprocal sector, emerging from its two frequency branches.

The reciprocal fixed point defined by the matching conditions in the previous subsection gives a
particularly economical closure of this picture. If this fixed point is
shown to be dynamically attractive, the same saturated connected sector
would simultaneously support the common propagation cone, the massive
Klein--Gordon branch, and the associated correlation-dressed de Broglie scales. If the
resolution-independent information velocity is further identified with
the observed invariant speed,
\begin{equation}
c_\star
=
c
\end{equation}
relativistic propagation and the corresponding de Broglie scaling would therefore emerge
together in the connected sector rather than being introduced as
independent microscopic assumptions.

A final structural freedom is the elementary source intensity $\nu_2$.
The Schr\"odinger--Nagasawa operator fixes the propagation of the first
moment but does not determine the normalization with which new connected
possibilities are injected into the higher-moment hierarchy. The
parameter $\nu_2$ therefore controls the elementary feeding scale of the
connected sector and affects its formation time and fluctuation amplitude
without modifying the underlying Schr\"odinger mean equation. In this
sense, $\nu_2$ acts as a creation-scale degree of freedom of the branching
representation. Determining whether its value is purely
representational, is selected by the coupling to an external measuring
environment, or obeys a more fundamental universality condition remains
an open question.

\subsection{Status of the construction and open problems}

The derived results and fixed-point matching conditions can now be summarized
separately. The Schr\"odinger--Nagasawa rewriting and the MSRJD response-field
construction follow from the specified stochastic dynamics. The free connected
correlators and their secular growth are exact within the quadratic theory at
fixed cutoff, while the two-loop 2PI theory is a causal self-consistent
truncation.  The total coefficient $M(T)$ is the dressed renewal frequency of
the intrinsic relative kernel and the Hartree correction is encoded by
$h_{\rm eff}=h_0\zeta$.  Fisher saturation instead fixes the bare resolution-independent speed $c_\star=\hbar/(4\kappa_I)$ from the intrinsic Schr\"odinger--Nagasawa possibilities weighted by the reciprocal capture fraction $f_{\rm FB}$.  The intrinsic mass clock is $\Gamma_m=mc_\star^2/\hbar$, whereas the saturated dressed renewal frequency is $M_\infty=\Gamma_m/\zeta_\infty$.
The Avrami law is used only as an effective post-processing description of the
bounded forward--backward fraction $f_{\rm FB}$ reconstructed from
$C_{\rm FB}$, while $U$ and $W$ remain internal 2PI diagnostics.  The common cone
and the projected gap remain fixed-point conditions.  In particular the gap
condition is $\Omega_{g,\infty}=\Gamma_m=\zeta_\infty M_\infty$, whose
dynamical realization must be tested in the complete two-particle response.

The first technical problem is the complete solution of the
nonequilibrium 2PI memory equations, including the cutoff dependence and
the collective curvature of $D$, $E$, and $\mu_{2PI}$
~\cite{AartsBerges2001}. Such a calculation would determine the damping
and residues of the collective poles and the full internal-momentum
weight entering the $q$ integration.

The second problem concerns the microscopic capture dynamics. The
effective Avrami parameters should ultimately be obtained from the
branching process itself rather than from the reduced kinetic law. This
includes the dependence of the capture efficiency on the reciprocal
organization and the role of $\nu_2$ in the formation time.

The third problem is the dynamical stability of the reciprocal
fixed-point manifold. The saturated construction identifies the conditions under which the projected
sum sector joins the derived soft-sector cone and reconstructs the intrinsic
mass gap. Their attraction under the full nonequilibrium dynamics remains to be
established. A direct test is to determine whether the $Q^2$
curvature of the projected plus-sector response approaches
$c_\star^2$ while its intercept approaches $\Gamma_m^2$, equivalently whether
$\Omega_g/M\to\zeta_\infty$. This question
can be addressed numerically in a future implementation of the full
self-consistent 2PI evolution even if a closed analytic attraction
theorem remains unavailable. Such a calculation would provide a direct
nonperturbative test of the simultaneous relativistic and de~Broglie
matching predicted by the saturated construction.
Finally, connecting the localized connected sector to actual measurement
outcomes requires coupling the theory to a detector and deriving how a
local interaction nucleates, amplifies, and records the correlated
structure. A genuine many-body formulation is likewise required before
the proposed separation between nonlocal first-moment quantum structure
and causal propagation of localized connected events can be tested on
entangled systems. Spin, fermionic statistics, and gauge degrees of
freedom constitute further extensions of the present scalar response
theory.

The present work closes the main structural problem left by Ref.~\cite{dumonteil_branching_2026} at the
level of the field-theoretical architecture. The Schr\"odinger--Nagasawa mean
field remains the extended first-moment sector, while the self-consistent
Bohm/Fisher feedback generates a localized reciprocal sector and two collective
propagation branches at second order. The main unresolved question is their
dynamical fixed-point selection. If attraction toward the reciprocal common-cone
manifold is established, the same connected dynamics would support relativistic
propagation together with correlation-dressed internal localization and transfer
scales while preserving the standard mass normalization of the collective gap
and the extended Schr\"odinger field at first order. More broadly, this hierarchy suggests a
framework in which quantum nonlocal structure, localized measurement events, and
relativistic causal propagation could occupy distinct statistical levels of a
single branching dynamics.

\appendix
\section{Normalization of the $U/W$ basis and Wigner frequencies}
\label{app:UW-normalization}

The main text uses the unnormalized combinations
$U=\psi+\psi^\dagger$ and $W=\psi^\dagger-\psi$.  With
$X^T=(U,W)$ and $\widetilde X^T=(\widetilde U,\widetilde W)$,
\begin{equation}
\Psi=A X,
\qquad
\widetilde\Psi=A\widetilde X,
\qquad
A
=
\frac{1}{2}
\begin{pmatrix}
1&-1\\
1&1
\end{pmatrix}.
\label{eq:UW-inverse-transformation}
\end{equation}
For the reciprocal response block of Eq.~\eqref{eq:KS-hat},
\begin{equation}
A^T\widehat{\mathcal K}_S A
=
\frac{1}{2}\widehat{\mathcal K}_S.
\label{eq:UW-kernel-normalization}
\end{equation}
Thus the time derivative and the off-diagonal Schr\"odinger--Nagasawa
operator acquire the same factor $1/2$.  In particular,
\begin{equation}
\frac{1}{2}\frac{\hbar}{2m}
=
\frac{\hbar}{4m}
\equiv h_0.
\label{eq:h0-from-UW}
\end{equation}
The prefactor $1/2$ in the four-component matrix
Eq.~\eqref{eq:K2PI-matrix} has a different origin: it is the standard
symmetrization of the quadratic bilinear form.

For the two-time Fourier transform, let $\varpi_x$ and $\varpi_y$ be the
frequencies conjugate to the original leg times.  With
$T=(t_x+t_y)/2$ and $\tau=t_x-t_y$,
\begin{equation}
\ee^{-\ii\varpi_x t_x}\ee^{+\ii\varpi_y t_y}
=
\ee^{-\ii\Omega T}\ee^{-\ii\nu_{\rm W}\tau},
\label{eq:two-time-Fourier-convention}
\end{equation}
where
\begin{equation}
\Omega=\varpi_x-\varpi_y,
\qquad
\nu_{\rm W}=\frac{\varpi_x+\varpi_y}{2}.
\label{eq:collective-frequency-convention}
\end{equation}
The reduced $U/W$ kernel uses
\begin{equation}
\omega_x=\frac{\varpi_x}{2},
\qquad
\omega_y=\frac{\varpi_y}{2},
\qquad
\nu=\frac{\nu_{\rm W}}{2},
\label{eq:reduced-leg-frequencies}
\end{equation}
so that
\begin{equation}
\omega_x=\nu+\frac{\Omega}{4},
\qquad
\omega_y=\nu-\frac{\Omega}{4}.
\label{eq:frequency-quarter-factor}
\end{equation}
Accordingly, $\Omega$ remains the physical collective frequency used in
the dispersion relations of the main text.

\section{Operator expansion of the MSRJD action}
\label{app:operators}
\subsection{Fisher variation and Bohm response kernel}

For $R=\sqrt{\rho}$,
\begin{align}
\delta I_F
&=
\int\dd^dx
\left[
-2\frac{\nabla^2\rho}{\rho}
+\frac{(\nabla\rho)^2}{\rho^2}
\right]\delta\rho
\nonumber\\
&=
-4\int\dd^dx\,
\frac{\nabla^2R}{R}\,\delta\rho .
\label{eq:app-Fisher-variation}
\end{align}
This gives Eq.~\eqref{eq:Fisher-variation}. The linear response of
$\mathfrak b[\rho]\equiv2Q[\rho]/\hbar$ is the distributional kernel
\begin{equation}
K(x,z)
=
-\frac{Q[\rho](x)}{\hbar\rho(x)}\delta(x-z)
-\frac{\hbar}{2mR(x)}
\nabla_x^2
\left[
\frac{\delta(x-z)}{R(x)}
\right].
\label{eq:app-Kernel}
\end{equation}
Its action on a test fluctuation $f$ is
\begin{equation}
(Kf)(x)
=
-\frac{Q[\rho](x)}{\hbar\rho(x)}f(x)
-\frac{\hbar}{2mR(x)}
\nabla_x^2\left[\frac{f(x)}{R(x)}\right].
\label{eq:app-K-test}
\end{equation}
Equations~\eqref{eq:app-Kernel} and \eqref{eq:app-K-test} apply on the
support $\rho>0$, whereas nodal backgrounds require the corresponding
regularized distribution.

\subsection{Similarity-transformed quadratic matrix}

Define
\begin{equation}
\mathcal M_S
=
\begin{pmatrix}
\ee^{-S/\hbar}&0\\
0&\ee^{S/\hbar}
\end{pmatrix}.
\label{eq:app-MS}
\end{equation}
\begin{equation}
\delta\Phi=\mathcal M_S\Psi,
\qquad
\widetilde\Phi=\mathcal M_S^{-T}\widetilde\Psi.
\end{equation}
Then
\begin{equation}
\widehat{\mathcal K}_S
=
\mathcal M_S^{-1}\mathcal K\mathcal M_S,
\qquad
\widehat\Gamma_S^{(0)}
=
\mathcal M_S^{-1}\widehat\Gamma^{(0)}
\mathcal M_S^{-T}.
\label{eq:app-similarity}
\end{equation}
The needed identities include
\begin{align}
\ee^{S/\hbar}\partial_t\ee^{-S/\hbar}
&=
\partial_t-\frac{\partial_tS}{\hbar},
\nonumber\\
\ee^{S/\hbar}\nabla^2\ee^{-S/\hbar}
&=
\nabla^2-\frac{2}{\hbar}\nabla S\cdot\nabla
+\frac{(\nabla S)^2}{\hbar^2}
-\frac{\nabla^2S}{\hbar}.
\label{eq:app-similarity-identities}
\end{align}
The conjugate identities carry the opposite signs in the terms linear
in $S$. Using
\begin{equation}
\partial_tS
=
-V-Q_0-\frac{(\nabla S)^2}{2m}
\label{eq:app-HJ}
\end{equation}
cancels the diagonal $V+Q_0$ contributions and gives
Eqs.~\eqref{eq:KS-hat}--\eqref{eq:Dt-B0}.

\section{Representative quartic Bohm contribution and ultraviolet scaling}
\label{app:secular}

The complete quartic Bohm vertex is derived in the next appendix section.
For the fixed-order ultraviolet benchmark it is enough to retain one of its
allowed derivative structures.  On a locally uniform background,
Eq.~\eqref{eq:Bohm-quartic-UW} contains
\begin{equation}
S_{B,\rm der}^{(4)}
\supset
\frac{h_0}{4R^2}
\int\dd t\,\dd^dx\;
\widetilde U\,W\,\nabla^2(W^2).
\label{eq:app-representative-quartic}
\end{equation}
This term is representative for power counting because it contains the two
spatial derivatives carried by the quartic Bohm interaction.  A Wick
contraction of the two internal $W$ legs gives a response-kernel correction
of the form
\begin{equation}
\Pi(t)\sim
\frac{h_0}{R^2}
\sum_q k_q^2 C_{WW}^{qq}(t)
\label{eq:app-Wick-contraction}
\end{equation}
up to bounded modal overlaps and external-mode factors.  The free confined
covariance has the secular behavior
\begin{equation}
C_{WW}^{qq}(t)
=4\nu_2\mathcal R_{qq}^{(n)}t+\mathcal O(1).
\end{equation}
For a bounded smooth reference profile,
$\mathcal R_{qq}^{(n)}=\mathcal O(1)$ at large mode index.  Since
$k_q^2=\mathcal O(q^2)$, a fixed modal cutoff $N_{\rm UV}$ therefore gives
\begin{equation}
\Pi(t;N_{\rm UV})
=\mathcal O\!\left(\nu_2tN_{\rm UV}^3\right).
\label{eq:app-UV-N3}
\end{equation}
The secular factor comes from the continuously accumulated covariance and
the cubic cutoff dependence from the two derivatives of the Bohm vertex.
This power-counting argument does not determine the sign or channel decomposition of the full dressed self-energy.

\section{Cubic and quartic Bohm vertices}
\label{app:quartic}

It is convenient to use
\begin{equation}
u=\frac{U}{2R},
\qquad
w=\frac{W}{2R},
\qquad
\rho
=
R^2+RU+\frac{U^2-W^2}{4}.
\label{eq:app-rho12}
\end{equation}
With
\begin{equation}
\mathcal L_R
=
\nabla^2+2\nabla\ln R\cdot\nabla,
\qquad
\mathcal A_R f\equiv\mathcal L_R(f/R),
\end{equation}
and
\begin{equation}
\mathcal B_R
\equiv
R\mathcal A_R
=
\nabla^2-\frac{\nabla^2R}{R},
\label{eq:app-BR-spectral}
\end{equation}
one has
\begin{equation}
\mathcal L_R(f/R)=R^{-1}\mathcal B_Rf.
\label{eq:app-LR-BR-identity}
\end{equation}
The notation \(\mathcal B_R\) is reserved below for this spectral operator.
In particular, the operator acting directly on the quadratic argument \(W^2\)
in the \(W;WW\) vertex is instead
\begin{equation}
\mathcal B_RR^{-1}=R\mathcal A_RR^{-1},
\end{equation}
since
\(R\mathcal D_R(W,W)=\mathcal B_R(W^2/R)\).
The cubic interaction in the response normalization of the main text is
\begin{equation}
S_{3,B}
=
-h_0\int\widetilde U W\,\mathcal L_R(U/R)
+
\frac{h_0}{2}\int R\widetilde W\,\mathcal L_R(W^2/R^2).
\label{eq:app-corrected-cubic}
\end{equation}
Equivalently, with $\kappa=h_0/2$, this is
Eq.~\eqref{eq:Bohm-cubic-UW}.  The complete quartic contribution is
\begin{equation}
\begin{aligned}
S_{4,B}
=
2h_0\int R\Big[
&\widetilde U
\left\{
w\mathcal L_R(w^2)
+
2uw\mathcal L_Ru
\right\}
\\
-&\widetilde W
\left\{
\mathcal L_R(uw^2)
+
w^2\mathcal L_Ru
\right\}
\Big].
\end{aligned}
\label{eq:app-corrected-quartic}
\end{equation}
For constant $R$ these vertices reduce to
\begin{align}
S_{3,B}
&=
\int
\left[
-\frac{h_0}{R}\widetilde U W\nabla^2U
+
\frac{h_0}{2R}\widetilde W\nabla^2(W^2)
\right],
\label{eq:app-cubic-uniform}
\\
S_{4,B}
&=
\frac{h_0}{4R^2}
\int
\left[
\widetilde U
\left\{
W\nabla^2(W^2)
+
2UW\nabla^2U
\right\}
\right.
\nonumber\\
&\hspace{28mm}
\left.
-
\widetilde W
\left\{
\nabla^2(UW^2)
+
W^2\nabla^2U
\right\}
\right].
\label{eq:app-quartic-uniform}
\end{align}

\subsection{Quartic Hartree contraction}

For a general local covariance define
\begin{gather}
b=\langle w^2\rangle_c,
\qquad
c=\langle uw\rangle_c,
\qquad
\mathbf j_w=\langle w\nabla w\rangle_c,
\nonumber\\
e_u=\langle u\mathcal L_Ru\rangle_c,
\qquad
e_w=\langle w\mathcal L_Rw\rangle_c,
\qquad
f_w=\langle|\nabla w|^2\rangle_c,
\nonumber\\
e_{wu}
=
\langle w\mathcal L_Ru\rangle_c.
\label{eq:app-quartic-moments}
\end{gather}
Contracting two physical legs of
Eq.~\eqref{eq:app-corrected-quartic} produces
\begin{equation}
\delta S_{2,4}
=
2h_0
\int
R
\left(
\widetilde U\,\ell_U
-
\widetilde W\,\ell_W
\right),
\end{equation}
where
\begin{align}
\ell_U
={}&
2c\mathcal L_Ru
+
2e_{wu}u
+
2b\mathcal L_Rw
+
4\mathbf j_w\cdot\nabla w
\nonumber\\
&\qquad
+
(4e_w+2f_w+2e_u)w,
\\
\ell_W
={}&
\mathcal L_R(bu)
+
b\mathcal L_Ru
+
2\mathcal L_R(cw)
+
2e_{wu}w.
\label{eq:app-quartic-ell}
\end{align}
No homogeneity assumption has been used at this stage.

For the local homogeneous and parity-even reduction used in the body,
the slowly varying scalar moments satisfy
\begin{equation}
\nabla b
=
\nabla c
=
\mathbf j_w
=
0.
\end{equation}
We introduce
\begin{equation}
\tau_u
=
\langle|\nabla u|^2\rangle_c,
\qquad
\tau_w
=
\langle|\nabla w|^2\rangle_c,
\qquad
\tau_{uw}
=
\langle\nabla u\cdot\nabla w\rangle_c.
\label{eq:app-quartic-gradient-moments}
\end{equation}
The diagonal Hartree coefficients are then
\begin{equation}
a_4
=
2h_0 b,
\qquad
M_4
=
2h_0(\tau_u+\tau_w),
\label{eq:app-a4-M4}
\end{equation}
while the crossed equal-time moments generate the additional diagonal
response contribution
\begin{equation}
d_4(q,T)
=
2h_0
\left[
c(T)s(q)-\tau_{uw}(T)
\right],
\qquad
s(q)=k_n^2-q^2.
\label{eq:app-d4}
\end{equation}
The complete local quartic Hartree correction therefore reads
\begin{equation}
\delta\mathcal K_4(q,T)
=
\begin{pmatrix}
d_4(q,T)
&
a_4(T)s(q)-M_4(T)
\\[1mm]
-a_4(T)s(q)
&
-d_4(q,T)
\end{pmatrix}.
\label{eq:app-quartic-Hartree-full}
\end{equation}

The crossed contribution is traceless: it appears with opposite signs
in the two diagonal response blocks.  Consequently, if the remaining
diagonal part of the inverse propagator is denoted by
$\mathcal A(\omega,q,T)$, the pole denominator has the structure
\begin{equation}
\Delta(\omega,q,T)
=
\Delta_0(\omega,q,T)
-
d_4^2(q,T),
\label{eq:app-d4-denominator}
\end{equation}
where $\Delta_0$ denotes the denominator obtained from the off-diagonal
quartic reduction.  In particular, no contribution linear in $d_4$
appears in the pole condition.

The analytic infrared reduction employed in the main text retains the
dominant diagonal covariance and gradient moments while treating the
crossed sector as subleading,
\begin{equation}
\frac{|c|}
{\sqrt{
\langle u^2\rangle_c
\langle w^2\rangle_c
}}
\ll1,
\qquad
\frac{|\tau_{uw}|}
{\sqrt{\tau_u\tau_w}}
\ll1.
\label{eq:app-crossed-Hartree-hierarchy}
\end{equation}
The coefficient $d_4$ is built entirely from the crossed $UW$ moments,
whereas the dominant Hartree corrections $a_4$ and $M_4$ are generated by
the diagonal $UU$ and $WW$ covariance sectors.  In the localized regime the
crossed correlators remain subleading, so $d_4$ is correspondingly smaller
than the diagonal Hartree corrections.  Moreover, in the pole determinant
its first contribution is quadratic, through $d_4^2$.  We therefore use
\begin{equation}
d_4(q,T)\simeq0
\label{eq:app-d4-IR-neglect}
\end{equation}
in the analytic factorization, yielding
\begin{equation}
\delta\mathcal K_4^{\rm IR}(q,T)
=
\begin{pmatrix}
0
&
a_4(T)s(q)-M_4(T)
\\
-a_4(T)s(q)
&
0
\end{pmatrix},
\label{eq:app-quartic-Hartree-IR}
\end{equation}
which is Eq.~\eqref{eq:quartic-Hartree-reduction} of the main text.

This approximation is local to the analytic reduction and does not remove
the crossed Hartree contribution from the complete 2PI dynamics.  The
numerical evolution retains $d_4$ together with the full crossed covariance
sector, where it remains subleading in the localizing regime.

The dimensionless local correlation dressing is
\begin{equation}
\zeta(T)
=
1+2b
=
1+2\langle w^2\rangle_c,
\end{equation}
so that
\begin{equation}
h_{\rm eff}(T)
=
h_0+a_4(T)
=
h_0\zeta(T).
\end{equation}
The factor $\zeta$ is therefore the equal-time Hartree dressing of the
correlated 2PI relative sector.  It is not inserted into the intrinsic
Schr\"odinger--Nagasawa/Fisher refinement, whose information velocity is
defined before the correlation-dependent 2PI dressing.

For constant $R$,
\begin{equation}
\begin{aligned}
a_4&=\frac{h_0}{2R^2}\langle W^2\rangle_c,\\
M_4&=\frac{h_0}{2R^2}
\left(\langle|\nabla U|^2\rangle_c
+\langle|\nabla W|^2\rangle_c\right),
\end{aligned}
\label{eq:app-a4-M4-uniform}
\end{equation}
while the crossed contribution becomes
\begin{equation}
d_4(q,T)
=
\frac{h_0}{2R^2}
\left[
\langle UW\rangle_c\,s(q)
-
\langle\nabla U\cdot\nabla W\rangle_c
\right].
\label{eq:app-d4-uniform}
\end{equation}
Thus the dominant Hartree dressing is controlled by the diagonal $ww$
covariance and the diagonal gradient energies, whereas the crossed
Hartree structure contributes only through the subleading traceless
correction $d_4$.

\section{Two-loop Bohm/Fisher closure and off-shell $HCC$ completion}
\label{app:2PI}

Before imposing the physical MSRJD condition, write
\begin{equation}
\mathbb G=
\begin{pmatrix}
C&G\\
G^T&H
\end{pmatrix},
\qquad
\mathbb H_0=
\begin{pmatrix}
0&\mathcal K_0^T\\
\mathcal K_0&-\mathcal N_0
\end{pmatrix},
\end{equation}
where $\mathcal N_0$ is the background-evaluated branching kernel in the
normalization of the $U/W$ basis.  The reduced interaction entering
$\Gamma_2$ is
\begin{equation}
S_{\rm int}^{\rm red}
=
S_B^{(3)}+S_B^{(4)},
\label{eq:app-reduced-interaction}
\end{equation}
with the cubic part written directly as the sum of the two physical response
vertices
\begin{equation}
S_B^{(3)}
=
\int_1\left[V_U(1)+V_W(1)\right],
\end{equation}
\begin{align}
V_U(1)
&=
-2\kappa\,
\widetilde U(1)W(1)\mathcal A_1U(1),
\label{eq:app-VU}\\
V_W(1)
&=
\kappa R(1)\,
\widetilde W(1)\mathcal D_1[W(1),W(1)].
\label{eq:app-VW}
\end{align}
Thus there are only two cubic Bohm vertices in the reduced theory: a
$\widetilde U\,U\,W$ vertex with the derivative acting on the $U$ leg, and a
$\widetilde W\,W\,W$ vertex with the external Bohm operator carried by the
$W$ response leg.

At two-loop order, pairings of $V_A(1)V_B(2)$ with
$A,B\in\{U,W\}$ generate the two contraction classes needed in the physical
two-point equations,
\begin{equation}
V_A(1)V_B(2)
\quad\Longrightarrow\quad
\begin{cases}
GGC &\longrightarrow\ \Sigma^R_{AB},\\[1mm]
HCC &\longrightarrow\ \delta\mathcal N^{BB}_{AB}.
\end{cases}
\label{eq:app-two-cubic-topologies}
\end{equation}
The $GGC$ contraction contains two response lines and one physical covariance
and dresses the retarded kernel.  The $HCC$ contraction contains the auxiliary
response--response line $H$ and two physical covariances.  Functional
variation is performed before imposing the physical condition $H=0$, so opening the $H$ line leaves a finite $CC$ contribution to the effective
noise.  The quartic Bohm vertex contributes separately through the one-loop
Hartree contraction derived in Appendix~\ref{app:quartic}.

This representation avoids introducing an additional cubic tensor notation:
all channel structure follows from the four pairings
$V_UV_U$, $V_UV_W$, $V_WV_U$, and $V_WV_W$.  Their explicit modal response
contractions and the diagonal-sector $HCC$ contributions are given in
Appendix~\ref{app:Sigma}.  The numerical contraction retains the complete
$C_{UU}$, $C_{UW}$, $C_{WU}$, and $C_{WW}$ matrix inside these pairings, and the mixed vertex pairings generate the crossed $UW$ and $WU$ noise blocks.

The exact branching covariance of Ref.~\cite{dumonteil_branching_2026} contains, in addition, the
field-dependent term displayed in Eq.~\eqref{eq:Gamma-background-vertex}.
Promoting that term to an MSRJD interaction would add a
$\widetilde X\widetilde X X$ branching vertex and corresponding mixed
Bohm--noise skeletons.  These terms are omitted under the background-noise/Bohm-feedback truncation defined in Sec.~\ref{sec:msrjd}, rather than canceled by causality.  The present $\Gamma_2$ therefore isolates
the renormalization generated by Bohm/Fisher feedback, while the elementary
branching covariance remains in $\mathbb H_0$.  Restoring the multiplicative
branching vertex is a systematic extension of the interaction functional.

\section{Self-energies and diagonal-sector analytic projection}
\label{app:Sigma}

The numerical self-consistent evolution retains all four physical covariance
blocks $C_{UU}$, $C_{UW}$, $C_{WU}$, and $C_{WW}$ in the GGC and HCC feedback.
The reciprocal construction of Ref.~\cite{dumonteil_branching_2026} privileges correlated and
anticorrelated stochastic combinations.  In the corresponding $U/W$ basis we
therefore use, for the analytic identification of the Bohm/Fisher coefficient
families, the hierarchy
\begin{equation}
|C_{UW}|,\ |C_{WU}|\ll |C_{UU}|,
\qquad
|C_{UW}|,\ |C_{WU}|\ll |C_{WW}|.
\label{eq:weak-crossed-covariance-hierarchy}
\end{equation}
The coupled dynamics generates crossed covariances, which remain in the self-consistent numerical GGC and HCC contractions despite their subleading role in the analytic channel hierarchy.  Equation~\eqref{eq:weak-crossed-covariance-hierarchy}
is used only to expose analytically the leading diagonal-covariance sector
that defines $D$, $E$, and $\mu_{2PI}$ below.  The numerical evolution therefore
provides the less projected realization of the same Bohm/Fisher closure.

In a real orthonormal basis $e_a$ of
$\mathcal B_R$,
\begin{equation}
\mathcal B_Re_a=s_ae_a,
\qquad
V_{apr}=\int\dd x\,\frac{e_a(x)e_p(x)e_r(x)}{R(x)},
\end{equation}
the two cubic Bohm vertices take the modal form
\begin{align}
V_U
&=
-h_0\sum_{a,p,r}
s_pV_{apr}\,
\widetilde U_aU_pW_r,
\label{eq:app-modal-VU}
\\
V_W
&=
\frac{h_0}{2}
\sum_{a,p,r}
s_aV_{apr}\,
\widetilde W_aW_pW_r.
\label{eq:app-modal-VW}
\end{align}
The symmetry of $V_{apr}$ under interchange of the two physical legs has
already been used in the first line.  These two expressions are the modal
counterparts of Eqs.~\eqref{eq:app-VU} and \eqref{eq:app-VW}. Their four possible pairings generate all response and noise self-energies below.
Defining
\begin{equation}
\mathcal I_{ab}[F_{pr}]
=h_0^2\sum_{p,r}V_{apr}V_{bpr}F_{pr},
\end{equation}
the leading diagonal-covariance pieces selected by
Eq.~\eqref{eq:weak-crossed-covariance-hierarchy} and used to identify the
causal cubic response families are
\begin{align}
\Sigma^{BB}_{UU;ab}
&=-s_b\,\mathcal I_{ab}[s_pG_{UU}(p)C_{WW}(r)],
\\
\Sigma^{BB}_{WW;ab}
&=-s_a\,\mathcal I_{ab}[s_pG_{WW}(p)C_{WW}(r)],
\\
\Sigma^{BB}_{WU;ab}
&=s_as_b\,\mathcal I_{ab}[G_{WU}(p)C_{WW}(r)],
\\
\Sigma^{BB}_{UW;ab}
&=\mathcal I_{ab}[s_p^2G_{UW}(p)C_{WW}(r)
-s_r^2G_{WU}(p)C_{UU}(r)].
\label{eq:app-response-reduced}
\end{align}
Thus the two-external-derivative family lies in the $WU$ entry, while the
family with no imposed external factor lies in $UW$.

For a locally uniform patch, $g=h_0/R$, $r=q-p$, and
\begin{align}
E_U(q;t,t')&=-g^2\int_p s(p)G_{UU}(p;t,t')C_{WW}(r;t,t'),
\\
E_W(q;t,t')&=-g^2\int_p s(p)G_{WW}(p;t,t')C_{WW}(r;t,t'),
\\
D(q;t,t')&=g^2\int_p G_{WU}(p;t,t')C_{WW}(r;t,t'),
\\
\mu_{2PI}(q;t,t')&=g^2\int_p\left[
s(p)^2G_{UW}(p;t,t')C_{WW}(r;t,t')
\right.\nonumber\\
&\hspace{15mm}\left.
-s(r)^2G_{WU}(p;t,t')C_{UU}(r;t,t')\right].
\label{eq:app-DEmu-corrected}
\end{align}
After the adiabatic projection and the exchange reduction
$E_U=E_W\equiv E$, these expressions produce
Eq.~\eqref{eq:2PI-cubic-reduced-selfenergy}.

Projecting the $HCC$ variation onto the same leading diagonal-covariance
sector gives
\begin{align}
\delta\mathcal N^{BB}_{UU;ab}
&=\mathcal I_{ab}[s_p^2C_{UU}(p)C_{WW}(r)],
\\
\delta\mathcal N^{BB}_{WW;ab}
&=\frac{s_as_b}{2}\mathcal I_{ab}[C_{WW}(p)C_{WW}(r)].
\label{eq:app-noise-reduced}
\end{align}
The crossed $HCC$ blocks generated by the mixed pairings
$V_UV_W$ and $V_WV_U$ are retained in the numerical evolution but are not
needed to define the two diagonal residue weights displayed here.
In the uniform Fourier patch this becomes
\begin{align}
\delta\mathcal N^{BB}_{UU}(q;t,t')
&=g^2\int_p s(p)^2
\nonumber\\
&\quad\times C_{UU}(p;t,t')C_{WW}(q-p;t,t'),
\\
\delta\mathcal N^{BB}_{WW}(q;t,t')
&=\frac{g^2s(q)^2}{2}\int_p
\nonumber\\
&\quad\times C_{WW}(p;t,t')C_{WW}(q-p;t,t').
\label{eq:app-noise-uniform}
\end{align}
The external factor $s(q)^2$ belongs to the $WW$ noise correction, whereas the $UU$ noise correction contains only internal derivatives.  These
noise self-energies modify $C=G\mathcal N_{\rm eff}G^T$ while the physical
response--response covariance remains zero.

\section{Derivation of the relative--collective factorization}
\label{app:relative-collective-factorization}

This appendix derives the intrinsic slow-pole factorization used in
Sec.~\ref{sec:relative-localization}.  The finite-$k_n$ Born--Oppenheimer
residue decomposition is given in the main text, while the intrinsic projection below isolates the screened pole that controls the collective velocity.  The stationary finite-well reference
frequency has already been included in $M(T)$ through
Eq.~\eqref{eq:total-branching-mass}.  The analytic slow-pole projection is
therefore written with $k_n=0$ and
\begin{equation}
D(q,T)=E(q,T)=0.
\label{eq:app-DE-zero}
\end{equation}
With
\begin{equation}
q_x=q+\frac Q2,
\qquad
q_y=q-\frac Q2,
\end{equation}
the spatial blocks are
\begin{align}
B_x&=-\left(h_{\rm eff}q_x^2+M\right),
&B_y&=-\left(h_{\rm eff}q_y^2+M\right),
\nonumber\\
C_x&=-h_{\rm eff}q_x^2,
&C_y&=-h_{\rm eff}q_y^2.
\label{eq:app-blocks}
\end{align}
Their products are
\begin{align}
P_x&=h_{\rm eff}q_x^2\left(h_{\rm eff}q_x^2+M\right),
\label{eq:app-Px}\\
P_y&=h_{\rm eff}q_y^2\left(h_{\rm eff}q_y^2+M\right),
\label{eq:app-Py}
\end{align}
where all coefficients are evaluated at the same central time $T$.  Define
$\varepsilon_x=\sqrt{P_x}$ and $\varepsilon_y=\sqrt{P_y}$.  After integration over the relative
frequency, the denominator associated with the pole $\sigma=\pm1$ is
\begin{equation}
\mathcal D_\sigma
=P_x-
\left(\frac\Omega2+\sigma \varepsilon_y\right)^2.
\label{eq:app-Dsigma}
\end{equation}

\subsection{Expansion of the two spatial legs}
Define
\begin{equation}
P(q,T)
=
h_{\rm eff}(T)q^2\left[h_{\rm eff}(T)q^2+M(T)\right]
\label{eq:app-Ps}
\end{equation}
and $\varepsilon(q,T)=\sqrt{P(q,T)}$.  Exchange symmetry gives
\begin{equation}
\varepsilon_x-\varepsilon_y
=
Q\,\partial_q\varepsilon(q,T)+\mathcal O(Q^3).
\label{eq:app-p-difference}
\end{equation}
The denominator factorizes as
\begin{equation}
\mathcal D_\sigma
=
\left[\varepsilon_x-\frac\Omega2-\sigma \varepsilon_y\right]
\left[\varepsilon_x+\frac\Omega2+\sigma \varepsilon_y\right].
\label{eq:app-factorization}
\end{equation}
Under
\begin{equation}
|\Omega|,\ |\varepsilon_x-\varepsilon_y|\ll \varepsilon(q,T),
\label{eq:app-adiabatic-hierarchy}
\end{equation}
the slow factor is
\begin{equation}
\mathcal D_\sigma^{\rm slow}
\simeq
\varepsilon(q,T)[\Omega_{\rm d}(q,Q,T)-\sigma\Omega],
\label{eq:app-D-slow}
\end{equation}
where
\begin{equation}
\Omega_{\rm d}=2(\varepsilon_x-\varepsilon_y)=v(q,T)Q+\mathcal O(Q^3)
\end{equation}
and
\begin{equation}
v(q,T)
=
2h_{\rm eff}q
\frac{2h_{\rm eff}q^2+M(T)}
{\sqrt{h_{\rm eff}q^2[h_{\rm eff}q^2+M(T)]}}.
\label{eq:app-v-explicit}
\end{equation}
Hence
\begin{equation}
v^2(q,T)
=
4h_{\rm eff}
\frac{[2h_{\rm eff}q^2+M(T)]^2}
{h_{\rm eff}q^2+M(T)}.
\label{eq:app-v-squared}
\end{equation}

\subsection{Channel residues and reciprocal reconstruction}
For the diagonal effective-noise projection used in the analytic residue
algebra, direct multiplication of $G\mathcal NG^T$ gives positive
$d_xd_y^\ast$ contributions in both diagonal covariances.  At $Q=0$, the reduced pole residues are
\begin{align}
\widehat Z_-^{W}
&=
N_{WW}^{\rm eff}+\frac{C}{B}N_{UU}^{\rm eff},
&
\widehat Z_+^{W}
&=
N_{WW}^{\rm eff}-\frac{C}{B}N_{UU}^{\rm eff},
\\
\widehat Z_-^{U}
&=
N_{UU}^{\rm eff}+\frac{B}{C}N_{WW}^{\rm eff},
&
\widehat Z_+^{U}
&=
N_{UU}^{\rm eff}-\frac{B}{C}N_{WW}^{\rm eff}.
\end{align}
The reciprocal weights follow from
\begin{equation}
\widehat Z_\pm^{\rm FB}
=
\frac14\left(\widehat Z_\pm^U-\widehat Z_\pm^W\right).
\end{equation}
In the intrinsic projection,
\begin{equation}
B=-\left(h_{\rm eff}q^2+M\right),
\qquad
C=-h_{\rm eff}q^2,
\end{equation}
and the screened denominator has the normalized shape
\begin{equation}
F_{\rm loc}(q,T)
=
\frac{M(T)}{h_{\rm eff}(T)q^2+M(T)},
\end{equation}
with
\begin{equation}
\xi^2(T)=\frac{h_{\rm eff}(T)}{M(T)}.
\end{equation}
The finite-$k_n$ extended reference contribution is restored before this
intrinsic projection, as shown in
Eq.~\eqref{eq:BO-ZFB-two-relative-sectors}.  Combining the conjugate slow
poles yields the collective factor
$2\Omega/[\Omega^2-\Omega_{\rm d}^2]$ used in the main text.

\section{Numerical implementation of the reduced self-consistent 2PI closure}
\label{app:numerical-2pi}

The numerical figures use the reduced causal two-loop closure in the
accompanying code. Space is discretized on the open Dirichlet grid
\begin{equation}
x_j=-L+j\Delta x,\qquad \Delta x=\frac{2L}{N_x+1},
\qquad j=1,\ldots,N_x.
\label{eq:app-grid}
\end{equation}
Each $U/W$ block is $N_x\times N_x$, and the combined matrices are
$2N_x\times2N_x$.  The distributed reference settings are dimensionless,
with $\hbar=m=L=1$, $n=1$, $\nu_2=100$, $N_x=60$,
$\Delta T=10^{-4}$ and $N_t=100$, giving $T_f=N_t\Delta T=10^{-2}$.  The initial connected covariance is zero,
and the bare-noise block prescription injects the same local
$4\nu_2R(x)\delta(x-y)$ covariance in $U$ and $W$.  Production runs may
override $N_x$, $N_t$, the memory depth, or the saving cadence. The settings used for each figure are stored with the corresponding result.

\paragraph{Evolution and memory.}
The bare generator $A_0$ is constructed for the fixed background, and
the instantaneous bare covariance $\mathcal Q_0$ is evaluated in the time
loop. This notation distinguishes it from the background Bohm potential
$Q_0=Q[R^2]$ in the analytical sections. A finite sliding memory stores
the two-time matrices $G(T,T-\tau_\ell)$ and $C(T,T-\tau_\ell)$.
The local cubic response and HCC noise projections use
\begin{align}
\Sigma_{R,3}^{\rm loc}(T)&\simeq
\Delta T\sum_{\ell=1}^{N_{\rm mem}(T)}
\Sigma_{R,3}[G(T,T-\tau_\ell),C(T,T-\tau_\ell)],\nonumber\\
\Sigma_K^{\rm loc}(T)&\simeq
\Delta T\sum_{\ell=1}^{N_{\rm mem}(T)}
\Sigma_K[C(T,T-\tau_\ell)]
\nonumber\\
&\qquad\qquad\qquad\qquad\qquad\qquad+\frac{\Delta T}{2}\Sigma_K[C(T,T)],
\label{eq:app-discrete-memory}
\end{align}
In Eq.~\eqref{eq:app-discrete-memory} the displayed $\Sigma_K$ memory
sum denotes the Bohm--Bohm $HCC$ contribution.  The numerical noise self-energy
used in the production evolution is
\begin{equation}
\Sigma_K^{\rm loc}
=
\Sigma_{K,BB}^{\rm loc}
+
\Sigma_{K,BN}^{\rm loc},
\label{eq:app-numerical-K-BB-BN}
\end{equation}
where $\Sigma_{K,BN}^{\rm loc}$ is the leading mixed Bohm--branching-noise
correction generated by the multiplicative branching vertex.  It is accumulated
only over positive retarded lags, with no $\tau=0$ endpoint in the It\^o convention, whereas the regular $BB$ $HCC$ endpoint retains the half
quadrature weight shown above.
Here $\tau_\ell=\ell\Delta T$.  The final term is the enabled half-weight $BB$ HCC endpoint, which is a regular covariance contraction rather than an equal-time response loop. Positive lags use uniform weights.  $N_{\rm mem}(T)$ is the number of saved
positive lags available at central time $T$, capped by the configured memory
depth, which is 100 lags in the distributed reference settings. The quartic Hartree term is instantaneous and is added
separately. Thus
\begin{align}
A_{\rm eff}=A_0+s_R(\Sigma_{R,3}^{\rm loc}+\Sigma_{R,4}),
\nonumber\\
\mathcal Q_{\rm eff}=\mathcal Q_0+s_K\Sigma_K^{\rm loc},
\label{eq:app-numerical-dressed-blocks}
\end{align}
The equal-time covariance advanced by the numerical solver obeys
\begin{equation}
\partial_T C(T)
=
A_{\rm eff}(T)C(T)
+
C(T)A_{\rm eff}^{\mathsf T}(T)
+
\mathcal Q_{\rm eff}(T).
\label{eq:app-covariance-evolution}
\end{equation}
The supplied settings use $s_R=-1$ and $s_K=+1$.  Here
$\Sigma_K^{\rm loc}$ is the numerical notation for the correlation/noise
self-energy denoted $\Sigma_C$ in the analytic sections, and
$\mathcal Q_{\rm eff}$ is the equal-time noise covariance entering
Eq.~\eqref{eq:app-covariance-evolution}.  Bare-source contact subtraction and
positive-semidefinite projection of the effective noise are disabled in this
reference configuration.

Within a time step the Dyson--Picard iteration updates the trial memory
and mixes both self-energies, with an adaptive trust bound. Hartree is
evaluated from the previous equal-time covariance and held fixed during
that iteration. The stopping monitor is the relaxed response increment
$\|\Sigma_R^{j+1}-\Sigma_R^j\|_F/
\max(\|\Sigma_R^{j+1}\|_F,\|A_0\|_F,\epsilon)$,
where $\epsilon=10^{-30}$ is a numerical floor preventing division by zero. The default tolerance is $10^{-10}$, with a cap of 500 iterations.
Reaching the cap does not by itself certify a fixed point, and this
monitor is not an independent residual test of the noise equation.

\paragraph{Hartree projection and distinct cutoffs.}
The production Hartree prescription, \texttt{correlation\_cutoff}, takes the full equal-time covariance $C(T,T)$.  Its $UU$, $UW$, $WU$, and $WW$ blocks are divided by $4R(x)R(y)$, symmetrized central row/column cuts are formed, and their finite-interval cosine spectra are used for the local quartic moments.
The spectra are integrated up to
\begin{equation}
q_H(T)=\min\!\left[\frac{\pi}{\Delta x},
\sqrt{\frac{|\mu_{2PI}(q_0,T-\Delta T)|}{h_{\rm eff}(T-\Delta T)}}\right],
\end{equation}
where $q_0$ is the saved diagnostic momentum nearest zero, with the
implemented small-coefficient guards. The resulting central moments define
scalar $a_4$ and $M_4$, which build the Hartree response
matrix through a central coarse-grained reduction of an inhomogeneous covariance. The resulting quartic self-energy is not fully position-dependent.
The cutoff regularizes only this instantaneous quartic contraction: it does not replace the propagated covariance by a fitted residual and does not modify the covariance entering the cubic $GGC$ or $BB/BN$ noise loops.  Those loops retain the complete covariance blocks, including $UW$ and $WU$, and the reference $HCC/BN$ loop has no additional correlation cutoff.
The response projection also uses the configured smooth spatial bulk window.

The saved diagnostic momentum grid extends to
\begin{equation}
q_{\rm grid,max}=\min(0.8\pi/\Delta x,40k_n)
\end{equation}
by default.  The three scales have different roles. The Hartree contraction uses $q_H$, the displayed BO profiles use $q_p=\min(q_{\rm grid,max},p/\xi_M)$, and
\begin{equation}
q_{\rm KG}=\xi_M^{-1}
\end{equation}
selects the localized internal band used in the collective projection.  The
supplied profile setting is $p=3$, while the collective setting corresponds to
$p_{\rm KG}=1$.  Both post-processing windows use a cosine taper over their
last $20\%$.  These scales are distinct from the fixed modal regulator of the
2PI loops and from the information-resolution scale $q_I$ of
Sec.~\ref{sec:selfsimilar}.

\paragraph{Saved coefficient diagnostics.}
Finite-grid bulk projections of the stored loops give the diagnostic
coefficients
\begin{align}
D(q,T)=2\kappa^2 I_{UW}(q,T),\nonumber\\
E(q,T)=2\kappa^2 I_{UU}(q,T),\nonumber\\
\mu_{2PI}(q,T)=2\kappa^2 J_{WU}(q,T).
\label{eq:app-numerical-DEM}
\end{align}
In translationally invariant shorthand their structures are
\begin{align}
I_{UU}&\sim\sum_\ell\int\dd k\,s(k)G_{UU}(k)C_{WW}(q-k),\nonumber\\
I_{UW}&\sim\sum_\ell\int\dd k\,G_{WU}(k)C_{WW}(q-k),\nonumber\\
J_{WU}&\sim\sum_\ell\int\dd k\,s(k)^2
\bigl[G_{WU}(k)C_{UU}(q-k)\nonumber\\
&\qquad\qquad\qquad\qquad\qquad-C_{WW}(k)G_{UW}(q-k)\bigr],
\label{eq:app-loop-projection-shorthand}
\end{align}
with the stored positive time lags and their weights. The actual
implementation uses finite-grid operators and bulk projections. Its
enabled diagnostic response-contact option subtracts the instantaneous
identity from the diagonal response blocks for these projections, while retaining it in the propagated response.
Figure~\ref{fig:2pi-coefficients} samples the time traces at the saved
$q_0$ nearest zero.

The coefficient-based lengths are
\begin{equation}
\xi_\mu=\sqrt{\frac{h_{\rm eff}}{|\mu_{2PI}(q_0,T)|}},\qquad
\xi_M=\sqrt{\frac{h_{\rm eff}}{M}},
\label{eq:app-Deltaq-width}
\end{equation}
where a length is used only when its defining ratio is positive and
finite. The $\xi_\mu^{-1}$ guide in the coefficient figure is not an
exponential fit. The dotted spatial envelope instead uses $\xi_M$.
Independent fitted widths, when available, are exported for comparison.

\paragraph{Relative profiles and common-component extraction.}
Relative slices are reconstructed at $x=X+r/2$, $y=X-r/2$ with
$X=x_{\rm ref}$, where $x_{\rm ref}$ is the fixed collective reference
position used for the profile extraction (the reference run uses
$x_{\rm ref}=0$).  We denote by $r_{\rm available}=\max|r|$ the largest
relative separation available on that reconstructed slice.  The stored
reciprocal observable is
\begin{equation}
C_{\rm FB}=\tfrac14(C_{UU}+C_{UW}-C_{WU}-C_{WW}).
\end{equation}
At equal time the crossed difference is antisymmetric under $r\to-r$ and
therefore drops from the parity-even profile used for the BO comparison.  The
crossed blocks themselves remain in the covariance evolution and are retained in the complete numerical self-energies.

For the one-dimensional publication profiles the central lattice sample
is replaced by its neighbors' mean, and a finite-interval cosine
reconstruction applies the profile window. If $\xi_M$ is unavailable,
$\xi_\mu$ supplies the window scale when possible. The inverse transform
of $N_{UU}^{\rm eff}+N_{WW}^{\rm eff}$ provides the common template.
The common and channel-specific templates are separately normalized
before estimating their coefficients. In $UU$, a two-template
least-squares fit separates the common and extended reference shapes.  The
extended BO template uses the denominator
$(q^2-k_n^2)[h_{\rm eff}-D(q,T)s(q)]$ of
Eq.~\eqref{eq:BO-ZU-finite}. The BO comparison is made in the conservative sector and on the
band where $|D|q^4/M\ll1$, while $D$ remains in this denominator as required by
Eq.~\eqref{eq:infrared-derivative-hierarchy}.  In $WW$, the common coefficient is fitted over
$1.8\xi_M\leq|r|\leq3\xi_M$, falling back to the usable noncentral interval
if fewer than five tail samples are available. The remaining $WW$
profile is then compared with the screened BO basis. No fitted common
template is subtracted from the reciprocal panel.

This fitted real-space extraction is not identical to directly
subtracting $N_{UU}^{\rm eff}+N_{WW}^{\rm eff}$ with its algebraic
coefficient in the reduced spectral residue. Separate direct-residue
figures implement that algebraic operation, both in the minimal formula
and with $D$ retained. They preserve the full effective noise, including
its bare part, in the non-common numerators. Neither diagnostic changes
the covariance supplied to the evolution.

Figure~\ref{fig:a-transition} independently rescales the available early,
middle and final curves to absolute peaks $1/3$, $2/3$ and $1$, so its vertical levels do not measure kinetic growth. The final BO comparison
is separately normalized. Figure~\ref{fig:cfb-cluster-formation} instead
starts from the full reciprocal map, subtracts the median over
$|r|\geq0.75r_{\rm available}$ at each time, and uses one final-time
absolute-peak scale over the displayed interval for the entire surface.
One overall sign makes the final central value positive. Its inset
uses the earliest profile snapshot at the actual stored time $T_i$,
usually the third stored sample, and an adjustable cosine comparison.
The interactive companion uses a $0.65r_{\rm available}$ tail threshold,
normalizes over the full stored relative interval and retains the
original reciprocal sign. Its vertical scale consequently differs from that of the static surface.

\paragraph{Avrami diagnostics.}
The numerical Avrami observable is reconstructed directly from the forward--backward correlator.  With the fixed display sign,
\begin{equation}
A_{\rm FB}(T)=\max_r C_{\rm FB}(r,T),
\end{equation}
and the normalized saturation estimator used in Fig.~\ref{fig:avrami} is
\begin{equation}
f_{\rm FB}(T)
=
\frac{A_{\rm FB}(T)-A_{\rm FB}(0)}
{A_{{\rm FB},{\rm tail}}-A_{\rm FB}(0)}.
\end{equation}
Here $A_{{\rm FB},{\rm tail}}$ is the median of the final eight samples on the saturated plateau of the checkpoint used for the figure.  This normalization provides the kinetic capture fraction used in the Avrami and Fisher reductions.

The fit $1-\exp[-\gamma(T-T_0)^n]$ uses the stated window $0.01<f_{\rm FB}<0.985$, with nonnegative $\gamma$ and the numerical exponent bounds reported by the plotting script.  The optional BSM-MC reference is normalized independently and affinely aligned to the $C_{\rm FB}$ conversion interval only to compare kinetic shapes, without assuming parameter-free equality of amplitudes or microscopic time scales.  The $U/W$ channel amplitudes may still be exported as internal residue diagnostics, but they are not used as Avrami formation variables in the present analysis.

\bibliographystyle{apsrev4-2}
\bibliography{references}

\end{document}